\documentclass[aps,prd,reprint,
amsmath,amssymb,
]{revtex4-2}
\usepackage{acro}[=v2] 
\usepackage{graphicx}
\usepackage{dcolumn}
\usepackage{bm}
\usepackage{hyperref}
\usepackage{float}
\usepackage{enumitem}
\usepackage[skip=10pt plus1pt, indent=10pt]{parskip}
\usepackage{lipsum}
\usepackage{xcolor}
\newcommand{\PW}{1.00}         

\usepackage{empheq}
\newcommand{\mnras}{MNRAS}             
\newcommand{\aj}{AJ}                   
\newcommand{\aap}{A\&A}                
\newcommand{\apjl}{ApJ}                
\newcommand{\apjs}{ApJS}               
\newcommand{\jcap}{J.~Cosmology Astropart. Phys.} 
\newcommand{\na}{New~Astron.}          

\makeatletter
\newcommand*{\hyperlinkcite}[1]{\hyper@link{cite}{cite.#1}}
\makeatother
\newcommand{\CPaperSDR}[1]{\hyperlinkcite{Shapiro2022}{SDR22}}
\newcommand{\CPaperHWM}[1]{\hyperlinkcite{Hartman2022a}{HWM22}}
\newcommand{\CPaperFR}[1]{\hyperlinkcite{Foidl2022}{FR22}}
\newcommand{\mPSI}{P_\text{SI}}
\newcommand{\PSI}{$\mPSI$}
\newcommand{\mPsigma}{P_{\sigma}}
\newcommand{\Psigma}{$\mPsigma$}
\newcommand{\mRTF}{R_\text{TF}}
\newcommand{\RTF}{$\mRTF$}
\newcommand{\mldb}{\lambda_\text{deB}}
\newcommand{\ldb}{$\mldb$}
\begin{document}

\preprint{APS/123-QED}

\title{Halo formation and evolution in SFDM and CDM: \break New insights from the fluid approach}

\author{Horst Foidl$^{1}$}
\email{horst.foidl@outlook.com}
\author{Tanja Rindler-Daller$^{1,2}$}%
\email{tanja.rindler-daller@univie.ac.at}
\author{Werner W. Zeilinger$^{1}$}
\email{werner.zeilinger@univie.ac.at}
\affiliation{%
	$^{1}$Institut f\"ur Astrophysik, Universit\"atssternwarte Wien,
	Fakult\"at f\"ur Geowissenschaften, Geographie und Astronomie,
	University of Vienna, T\"urkenschanzstraße 17, A-1180 Vienna, Austria\\
	$^{2}$Wolfgang Pauli Institut, Oskar-Morgenstern-Platz 1, A-1090 Vienna, Austria
}%
%
%
%
\date{\today}
\begin{abstract}
	We present dark matter (DM)-only simulations of halo formation and evolution in scalar field dark matter (SFDM) cosmologies in the Thomas-Fermi regime, also known as ``SFDM-TF'', where a strong repulsive two-particle self-interaction (SI) is included. This model is a valuable alternative to cold dark matter (CDM), with the potential to resolve the ``cusp-core'' problem of the latter. In general, SFDM behaves like a quantum fluid. Previous literature has presented fluid approximations for SFDM-TF in 1D and 3D, respectively, as well as numerical DM-only simulations of SFDM-TF halo formation, whose results are in agreement with earlier analytic expectations, that a core-envelope halo structure arises; a central region close to a ($n=1$)-polytropic core, surrounded by a CDM-like (i.e. Navarro–Frenk–White (NFW)-like) halo envelope. While those previous results are generally in mutual agreement, discrepancies have been also reported. Therefore, we perform dedicated 3D cosmological simulations of the halo infall problem for the SFDM-TF model, as well as for CDM and its corresponding CDM fluid approximation, where we implement both previous fluid approximations into the code RAMSES. We compare our findings with those previous simulations. Our results are very well in accordance with previous works and extend upon them, in that we can explain the reported discrepancies. They are not due to the different fluid approximations, nor the geometry, but rather a result of different simulation setups. 
	Moreover, we find some interesting details, as follows. The evolution of both SFDM-TF and CDM halos follows a two-stage process. In the early stage, the central density in the halo rises, its profile becomes close to a ($n=1.5$)-polytropic core being dominated by an ``effective'' velocity-dispersion pressure \Psigma{} that stabilizes it against gravity. In fact, this pressure stems from random orbital motion in CDM, but from random wave motion in SFDM.
	Consecutively, for CDM halos, this core transitions into a steep central cusp whose slope is almost the same as the outer density slope, which is close to $\rho \propto r^{-3}$ as expected from NFW. Finally, however, the central profile makes another transition to a ``shallower cusp'', very close to the NFW behavior of $\rho \propto r^{-1}$. On the other hand, in the formation of the SFDM-TF halo, the additional pressure \PSI{} due to SI determines the second stage of the evolution. At the end, \PSI{} dominates in the central region, whose density follows closely a $(n=1)$-polytropic core. This core is enshrouded by a nearly isothermal envelope, i.e. the outskirts are similar to CDM at this point. We also encounter a new effect in our simulations, namely a late-time expansion of both polytropic core plus envelope, because the size of the almost isothermal halo envelope is affected by the ``external pressure'', which decreases with the expansion of the background universe. 
	Hence, the core size of SFDM-TF halos is not necessarily determined only by the parameters of the model, as our simulations reveal that an initial primordial core of $\sim 100$~pc -- demanded by power spectrum constraints  -- can evolve into a larger core of $\gtrsim 1$~kpc, after all, during halo evolution, even without feedback from baryons.
\end{abstract}
%
%
\maketitle
%
\newpage
\setcounter{tocdepth}{4}
{\parskip=0pt 
}
\newpage
%
\section{Introduction}\label{sectroduction}

%
The nature of the cosmological dark matter (DM) remains an open problem in physics and astronomy. The paradigm of collisionless, nonrelativistic -- ``cold'' -- dark matter (CDM) has been a core feature of the cosmological standard model for almost four decades, given its ability to explain important astronomical observations, notably on cosmological scales such as the cosmic web of structure, the matter power spectrum and the temperature anisotropies of the cosmic microwave background, as well as the elemental abundances in the wake of big bang nucleosynthesis. Nevertheless, the particle nature of CDM has not yet been identified, despite ongoing efforts to detect plausible candidates of CDM. These are foremost ``weakly interacting massive particles'', also known as WIMPs, that need to be heavier than the proton, or the QCD axion with a mass of about $\sim 10^{-5}$ eV/$c^2$. The observation of the bullet cluster, which shows a clear spatial separation of gas masses and sources of gravity, is a very convincing indication that DM can be considered a particle (\citet{Clowe2006}). This is one reason why the majority of the research community favors the particle hypothesis of DM.

However, apart from the hitherto non-detection of DM candidates, a closer analysis and comparison of theoretical predictions of CDM structure and galaxy formation to observations on scales of individual galaxies, notably dwarfs, have revealed discrepancies. These fare under the header of ``CDM small-scale problems'', which combines a set of different issues that are partly related to each other. In light of the work of this paper, we are particularly interested to highlight the cusp-core problem, which refers to the fact that observations of the central regions of DM-dominated galaxies seem to prefer (nearly-)constant density ``cores'' of $\sim$~kpc size, while CDM universally predicts density ``cusps'' of order $\rho \propto r^{-1}$. The question remains open what the reasons for the discrepancies are and explanations broadly include e.g. a lack of proper modeling of the full complexities of galaxy astrophysics, or the various complications that arise when data is compared to models. Yet, a large community in the field pursues the more radical idea that CDM has to be replaced by another DM model, altogether, supported by the fact that particle physics models have continuously come up with more and more candidates for DM, in general.     

In this paper, we will be concerned with a class of so-called scalar field dark matter (SFDM) models, whose constituents are bosons with masses $m \gtrsim 10^{-22}$ eV/$c^2$, i.e. typically much lower than the mass of the QCD axion. More precisely, we will consider SFDM in the Thomas-Fermi (TF) regime, where bosons interact via a strong repulsive two-particle self-interaction (SI). This SI is usually parametrized with a single coupling strength $g$ that describes the strength of the SI, and is a free parameter apart from $m$.
However, the TF regime refers to that parameter space $(m,g)$, for which the pressure due to the repulsive SI is the main source that stabilizes a system in hydrostatic equilibrium against gravitational collapse. 
This subclass of SFDM models has been dubbed ``SFDM-TF'' in \citet{Dawoodbhoy2021} and \citet{Shapiro2022} (henceforth abbreviated \CPaperSDR{}), and we will use this notation as well. On the other hand, the term ``SIBEC-DM'' has been used in \citet{Hartman2022} and \citet{Hartman2022a} (henceforth abbreviated \CPaperHWM{}).
Earlier literature on SFDM-TF, using various other names, include the works of \citet{Peebles2000,Goodman2000,Boehmer2007,RindlerDaller2012}, or \citet{2016PDU....14...84F}.

A much more extensively studied class of SFDM models, including various constraints already derived from comparison to astronomical data, is ``fuzzy dark matter (FDM)''. Here, the particle mass is the only free parameter, for SI is disregarded from the outset.
Both FDM and SFDM-TF can provide a cure to the small-scale problems referred to above, because the characteristic length scale of the models, related to their Jeans lengths, can be much larger than for CDM, resulting in a stronger suppression of structure formation in the former. Also, these length scales help to provide central density cores in halos made of FDM or SFDM-TF, respectively, possibly alleviating the cusp-core problem.
However, that ``success'' depends on the choice of the corresponding free DM model parameters, and these are subject to observational constraints on which we will elaborate shortly. 

The reason why structure formation in SFDM differs from that of collisionless CDM has its root in the fact that the bosons of SFDM form a Bose-Einstein condensate (BEC), such that SFDM has features similar to a quantum superfluid, albeit these features are most important on sub-cosmological scales.   
In general and regardless of the regime in SI, the set of equations of motion for nonrelativistic SFDM consists therefore of a nonlinear Schr\"odinger equation (NLSE) -- also called Gross-Pitaevskii (GP) equation --, and a Poisson (P) equation. The NLSE is the evolution equation for the complex, so-called ``wave function of the condensate'', whose modulus squared is proportional to the mass density of SFDM. Simulations of SFDM structure and halo formation have thus employed dedicated algorithms to solve the NLSE as a wave equation, coupled to the Poisson equation, and this has been the method of choice in simulations of FDM, in particular.
However, this approach is computationally very expensive, once expanding backgrounds in cosmologically representative volumes are considered, because it requires resolutions of the order of and below the comoving Jeans length, which shrinks with cosmic time [see Eqs. 65-71 in \CPaperSDR{}]. The proper de Broglie length reads $\mldb = h/mv$, where $v$ is a characteristic velocity depending on the environment, and $h$ is Planck's constant. Now, for ultralight bosons, $\mldb \lesssim (1$-$2)$~kpc is possible by the time a simulation run gets down to redshifts of $z \lesssim 6$. As a result, 3D cosmological halo formation simulations of ``FDM-only'' can often not be performed up to the present, or simulations have to be confined in ``small'' boxes of a few Mpc comoving on a side. Such simulations of FDM have been performed notably in the works by \citet{Schive_FDM,Schive_SFDM_constraints} and \citet{SchwabeFDM}, while box sizes up to 10 Mpcs have been used by \citet{May2021}. Baryons have been included in the first realistic FDM galaxy formation simulations of \citet{Mocz2020}, which are even more computationally expensive, as a result, so their simulation is stopped at $z = 5.5$.
One of the most important results of these simulations concerns the finding that forming FDM halos establish a core-envelope structure early on in their evolution, and this structure evolves but remains stable over time.
On the one hand, the ``solitonic cores'' of these halos have features similar to the hydrostatic equilibrium solutions of the equations of motion. The halo envelopes, on the other hand, exemplify a highly dynamical behavior which, however, has been found to be ``close'' to CDM in an averaged sense. For example, by averaging over the density in radial bins, in order to calculate radial density profiles, the exponents of that profile turn out to be close to that of the NFW profile (\citet{nfw}) of CDM halos. With respect to this overall halo structure, 3D simulations of FDM have basically confirmed previous results of analytical calculations and 1D simulations of FDM, see e.g. \citet{Guzman2003,Guzman2004}, or \citet{Zimmermann2021}.

In this paper, we study the opposite regime to FDM, namely SFDM-TF, and the characteristic scale for these models, which predominantly affects halo formation and structure, is related to the TF radius to be introduced in the next section. Suffice to say here that the proper TF radius reads $\mRTF \propto g/m^2$, i.e. it is a fixed quantity for given $m$ and $g$, or rather for a given combination of $g/m^2$. For SFDM-TF,  $\mRTF \gg \mldb$ for given $m$ and $g$. Thus, if we require that $\mRTF \sim 1$~kpc, as the characteristic scale to help alleviate the small-scale problems, and since simulations of SFDM-TF need to resolve scales of order \RTF{}, we see that resolving the even smaller \ldb{} becomes a computationally impossible task to achieve, by using a brute-force calculation of the cosmological NLSE+P (Gross-Pitaevskii-Poisson (GPP)) system of equations.    
Therefore, a new approach has been developed in \citet{Dawoodbhoy2021}, in an effort to overcome this issue, as follows.
It has been realized for a while that the NLSE+P (GPP) system of equations can be reformulated into a set of (quantum) hydrodynamical equations, in which the complex NLSE transforms into two real equations, the continuity and the momentum (Euler) equations (see next section). These hydrodynamical equations have been used in some of the cited literature above, especially in the study of SFDM-TF halo structure.
In fact, fluid approximations have been also considered previously in other contexts, or for different DM models. A fluid approach has been also considered in the works by \citet{Chavanis_2019, Chavanis2022}. Inspired by previous investigations, and in an attempt to model halo formation in SFDM-TF, \citet{Dawoodbhoy2021} have derived a fluid formulation by starting from a statistical description based on an appropriate phase space formulation of SFDM, in general. This way, the quantum physics on scales of \ldb{} and below is ``smoothed-over'' in a way that allows to model the large-scale effects due to it, without having to resolve the (arbitrary) small de Broglie scale \ldb{} itself.
By focusing on the spherically symmetric case, they were able to derive a set of fluid equations that include the respective pressures that arise from physics on scales of \RTF{} due to SI, as well as that due to ``quantum pressure'' on scales of \ldb{}.
We will summarize the most important equations in Sec. \ref{sec:fundeq}. As a result, hydro solvers can be used to model halo formation and halo structure in SFDM-TF, for which this particular fluid approximation is valid.  
Using a 1D Lagrangian hydro code in spherical symmetry, \citet{Dawoodbhoy2021} performed individual halo infall calculations for different choice of \RTF{}, in order to study halo formation and evolution of halo structure over time.
While the analysis in \citet{Dawoodbhoy2021} was limited to static backgrounds, serving as a kind of proof-of-principle that the formalism works, the extension towards halo formation and evolution in expanding backgrounds was presented in \CPaperSDR{}, starting their simulations from redshifts  $z \sim 3000$. There, a calculation of the linear regime of structure growth in SFDM-TF was also carried out, in order to embed and provide the necessary initial conditions for the subsequent nonlinear halo evolution.
Performing their simulations, both papers established for the first time that a core-envelope halo structure also arises in SFDM-TF models, confirming some analytical expectations of previous literature as e.g. expressed in \citet{RindlerDaller2014a}.
More precisely, the halo cores found were similar to the hydrostatic equilibria of the corresponding equations of motion for SFDM-TF (see next section), with core radii close to \RTF{} that are dominated by SI energy, while the halo envelopes were reminiscent of those of CDM halos, grown either in static or expanding backgrounds. 
The formalism laid out in \citet{Dawoodbhoy2021} and \CPaperSDR{}, and the analysis of the results, have revealed that the ``quantum pressure'' (due to the inherent quantum nature of SFDM) manifests itself as a kind of internal phase-space velocity dispersion, which gives rise to an effective velocity-dispersion pressure. It is this pressure that supports virialized SFDM halos -- FDM or SFDM-TF alike -- against gravitational collapse, in the same manner as does the collective velocity-dispersion pressure of collisionless particles of CDM in its fluid formalism.  It is the common nature of this pressure that explains why halo envelopes of SFDM as well as of CDM share similar features, especially with respect to the overall density profiles. As will become clear in the forthcoming, the energy due to this velocity-dispersion-mediated pressure is formally close to the ``thermal energy'' of a $\gamma=5/3$ gas.

One might argue that 1D simulations lack realism and miss out on important physical phenomena. However, they provide unprecedented resolution, unattainable to their counterparts in 3D. For example, halo cores \RTF{} of factors  $\sim 10^{-3}$-$10^{-2}$ smaller than the virial radius were resolved by \citet{Dawoodbhoy2021} and \CPaperSDR{}.  Such resolutions will not be achieved any time soon in 3D simulations of SFDM in general nor of SFDM-TF in particular. Hence, there is added value in performing 1D simulations, despite the progress of newer 3D simulations, also carried out in this paper, as will become clear in the forthcoming.

Indeed, the issue of resolution has actually become more serious in light of some recent findings, as follows. By extending the SFDM-TF halo formation studies of \citet{Dawoodbhoy2021} and by performing
their semi-analytic linear structure formation calculation, \CPaperSDR{} found that the (unconditional) halo mass
function in SFDM-TF exhibits a cutoff at a higher halo mass scale than the corresponding one for FDM, albeit the subsequent falloff toward smaller masses is much shallower than for FDM.
As a result, even the formation of SFDM-TF halos of Milky-Way size is suppressed, compared to CDM, if their primordial cores are large, $\mRTF \gtrsim 1$~kpc. Thus, it strongly appears that the model favors sub-kpc primordial cores, $\mRTF \lesssim 0.1$~kpc, questioning the ability of SFDM-TF to resolve the small-scale problems mentioned earlier. These conclusions were confirmed subsequently by \citet{Hartman2022} and \citet{Foidl2022}. Recent works (e.g. \citet{Dave2023}, \citet{Pils2022}) found SFDM halo core sizes $>1$~kpc to be compatible with observed SPARC (\citet{Lellietal}) rotation curves.

However, subsequently to \citet{Dawoodbhoy2021} and \CPaperSDR{}, 3D simulations of SFDM-TF halo formation have been performed in \CPaperHWM{} (they call the model SIBEC-DM), using the cosmological N-body and hydro code RAMSES of \citet{Teyssier2002}. Inspired by the fluid approximation of \citet{Dawoodbhoy2021}, which is valid for (and limited to) spherical symmetry, \CPaperHWM{} used a similar set of hydrodynamical equations for SFDM-TF, but applied in 3D. In doing so, however, some assumptions from 1D were carried over to the 3D setting, such as e.g. the skewlessness of the phase-space distribution function. Normally, the N-body module in RAMSES takes care of the collisionless CDM dynamics, while the hydro module is used to evolve the baryons. Now, \CPaperHWM{} modified the hydro equations within RAMSES accordingly in order to perform cosmological SFDM-TF halo formation simulations, while disabling the N-body module.

Apart from 3D, the simulations of \CPaperHWM{} are also more realistic with respect to the larger box size and the adopted initial conditions, enabling to form several halos, of which some merge, and to follow their evolution over time. The computational challenges are high and, as a result, there are two shortcomings of these simulations. First, they have to stop their runs at a redshift of $z=0.5$, thus final snapshots of their evolved halos refer to this epoch. Second, and more severely, the chosen value for \RTF{} has to be larger than $\sim 1$~kpc (they chose values of $1$, $3$ and $10$~kpc), despite the fact that such model parameters have been ruled out by the previous work just mentioned.  Again, it is the demands on resolution which required high enough \RTF{}. 

The simulations of both groups, \citet{Dawoodbhoy2021}, \CPaperSDR{} and \CPaperHWM{}, share the limitation that baryons were not included. Thus, cosmological simulations of SFDM-TF with baryons have yet to be carried out.

Despite the differences in the scope and implementation, \CPaperHWM{} confirmed many of the results reported in \citet{Dawoodbhoy2021} and \CPaperSDR{}, in that they confirm the establishment of a core-envelope halo structure, with core radii remaining close to \RTF{} and a CDM-like/NFW-like profile of the halo envelope. However, in contrast to \citet{Dawoodbhoy2021} and \CPaperSDR{}, \CPaperHWM{} find that the cores, that are initially also dominated by SI energy, eventually ``thermalize'', such that the effective ``thermal energy''  due to the quantum-pressure sourced large-scale velocity-dispersion pressure will dominate throughout the halos. The authors attribute this finding to the possibility of mixing of fluids that are dynamically heated during collapse, a phenomenon that the previous 1D simulations of spherical infall of mass shells were not able to model correctly. 
Interestingly, though, the results of \CPaperHWM{} seem to fit better with the semi-analytical double-polytrope model of \citet{Dawoodbhoy2021} and \CPaperSDR{}, which was devised as a theoretical model to compare to their 1D simulations. 

This paper here was motivated by the question of the origin of the discrepancies of the results reported in \CPaperHWM{} versus those in \citet{Dawoodbhoy2021} and \CPaperSDR{}, because it had remained unclear whether these discrepancies are due to the different geometry and simulation setup, or due to the underlying equations of motion, after all.
Therefore, we took up the task of performing our own halo formation simulations using RAMSES in order to gain more insight into this question. In doing so, we implement both sets of equations into RAMSES, which allows us to switch between them in carrying out the simulations. Also, we probe realistic SFDM-TF parameters by choosing sub-kpc primordial core radii for \RTF{}. However, we do this at the expense of smaller box sizes, which means that our simulations are limited to single-halo collapse calculations, such as in \citet{Dawoodbhoy2021} and \CPaperSDR{}, albeit in 3D. While we can thus study halo structure and evolution in detail, up to the present, we cannot make any statements regarding halo statistics nor the physics of major merging.   

This paper is organized as follows: in Sec.\ref{sec:fundeq}, we briefly summarize the fundamental equations of motion of SFDM, followed by an exposition of the fluid approximations with their respective equations, which are at the heart of our approach in Sec.~\ref{sec:SFDMhydro}. Section~\ref{sec:ourINVESTIGATION} contains a discussion of the implementation of the fluid approximations into the cosmological code RAMSES, while the initial conditions (ICs) are described separately in Sec.~\ref{sec:ICsHalos}. Since the fluid approximations include the CDM regime, by neglecting the characteristic SI pressure of SFDM, we devote Sec.~\ref{sec:simsRAMSESCDM} to an analysis of CDM single-halo formation and evolution. The results of this section not only serve as a cross-check to validate that our code implementations are correct, but also confirm earlier findings in 1D pertaining to the CDM fluid approach, thus we discuss the evolution of halo density and pressure profiles that we find, in detail. As such, this section should be useful also to readers who are generally interested in CDM dynamics, independent of SFDM as an alternative model.
Finally, Sec.~\ref{sec:simsRAMSES} includes our main results concerning SFDM single-halo  formation and evolution, in the TF regime, using the fluid approximations.  We detail the comparison between the approaches used by \CPaperSDR{} vs \CPaperHWM{}, and put our results into perspective with theirs, as well as our explanations regarding the reported discrepancies. Section~\ref{sec:cosmosims} contains a brief report on our simulation run that goes beyond the single-halo dynamics, which is also a cross-check of our implementations when compared to previous work. Our conclusions and summary can be found in Sec.~\ref{sec:summary}.

\section{Fundamental equations}\label{sec:fundeq}

Before we introduce the fluid approximations in the next section, which are at the core of our analysis here, we briefly present the fundamental equations of motion of SFDM which underlie these approximations.
Regardless of SI regime, it is assumed that SFDM consists of a single species of bosons with particle mass $m$, whose dynamics can be described by a complex function $\psi(\mathbf{r}, t)$, which is basically the ``wave function of the condensate'' of the BEC, formed by these bosons.
As in standard CDM, galactic SFDM halos are nonlinear overdensities, compared to the background of a $\Lambda$SFDM universe which is assumed to be homogeneous and isotropic, just as in $\Lambda$CDM.

Individual SFDM halos in the nonrelativistic limit can be described by a NLSE, the GP equation (see \citet{Gross_BEC,Pitaevskii_BEC}, and applied to gravity in \citet{Kaup1968,Ruffini1969}),
\begin{equation} \label{eq:gross-pitaevskii}
	\begin{aligned}
		i \hbar \frac{\partial \psi(\mathbf{r}, t)}{\partial t} = & -\frac{\hbar^2}{2m} \Delta \psi(\mathbf{r}, t) + \left(m\Phi(\mathbf{r},t\right) + \\ & g|\psi\left(\mathbf{r}, t)|^2 \right) \psi(\mathbf{r}, t),
	\end{aligned}
\end{equation}
which is coupled to the Poisson equation
\begin{equation} \label{eq:poisson}
	\Delta \Phi(\mathbf{r},t) = 4\pi G m |\psi(\mathbf{r}, t)|^2.
\end{equation}
The full system is called Gross-Pitaevskii-Poisson (GPP) equations.
The Born assignment is used, such that $|\psi(\mathbf{r}, t)|^2 = n(\mathbf{r}, t)$ describes the number probability density of the bosons.
The two-boson contact SI is modeled as the third term on the right-hand side of (\ref{eq:gross-pitaevskii}), where $g$ is a constant coupling strength that determines whether the bosons interact attractively ($g < 0$), or repulsively ($g > 0$). Together with $m$, it is a free parameter of the SFDM model. $\Phi$ is the gravitational potential of the self-gravitating halo. In addition to the nonlinear SI term in the GP equation, the full GPP equations are nonlinear in any case due to their coupling, even if we set $g=0$. This case corresponds to FDM models, discussed in the Introduction.  

The assumption that all $N$ bosons within a given halo of volume $V$ can be described by $\psi$ then naturally leads to the normalization condition,
\begin{equation}
	\int_V |\psi|^2 = N.
\end{equation}

The literature has made extensive use of an equivalent representation of GPP, \eqref{eq:gross-pitaevskii} and \eqref{eq:poisson}, by transforming the GP equation into quantum hydrodynamic equations, pioneered in particularly by \citet{Bohm1952},\cite{Bohm1952a} and \citet{Takabayasi1954}.
Using the polar decomposition or Madelung transformation \citep{Madelung1927}, the wave function is decomposed into its phase and amplitude functions, 
\begin{equation} \label{eq:madelung}
	\psi(\mathbf{r}, t) = |\psi(\mathbf{r}, t)|e^{iS(\mathbf{r}, t)} = \sqrt{\frac{\rho(\mathbf{r}, t)}{m}} e^{iS(\mathbf{r}, t)}.
\end{equation}
$\rho(\mathbf{r}, t)$ is identified as the SFDM halo mass density and $S(\mathbf{r}, t)$ as the action (or phase) function. It is related to the associated bulk velocity of the halo as follows,
\begin{equation} \label{eq:bulk-velocity}
	\mathbf{v} = \frac{\hbar}{m} \nabla S.
\end{equation}
Making use of the ``hydrodynamic variables'' $\mathbf{v}$ and $\rho$, the complex GP equation is transformed into two real equations, which are interpreted as a continuity equation and an Euler-like momentum equation,
\begin{equation} \label{eq:continuity}
	\frac{\partial \rho}{\partial t} + \nabla \cdot (\rho \mathbf{v}) = 0,   
\end{equation}
\begin{equation} \label{eq:euler-eq}
	\frac{\partial \mathbf{v}}{\partial t} + \left( \mathbf{v} \cdot \nabla \right) \mathbf{v} = - \nabla Q - \nabla \Phi - \frac{1}{\rho} \nabla \mPSI,
\end{equation}
\begin{equation} \label{eq:poisson2}
	\Delta \Phi = 4 \pi G \rho.  
\end{equation}
They are supplemented by the Poisson equation (\ref{eq:poisson2}).
The involved quantities are well known in the field, namely the so-called quantum or Bohm potential,
\begin{equation} \label{eq:quantum-potential}
	Q = -\frac{\hbar^2}{2m^2} \frac{\Delta \sqrt{\rho}}{\sqrt{\rho}},
\end{equation}
and the SI pressure \PSI{}, which is of polytropic form with index\footnote{This notation shall not be confused with the number density $n(\mathbf{r},t)$; the distinction should be clear in context.} $n=1$ and polytropic constant $K = g/2m^2$,
\begin{equation} \label{eq:polytropic-pressure}
	\mPSI = K \rho^{1 + 1/n} = \frac{g}{2m^2} \rho^2.
\end{equation}
We will also use the notation $\gamma = 1+1/n$, i.e. the ($n=1$)-polytrope is equivalent to $\gamma = 2$. 

Now, let us consider the hydrostatic solution in the TF regime, where
we set $\hbar = 0$ and $\mathbf{v}=\mathbf{0}$ in \eqref{eq:euler-eq}, reducing it to
\begin{equation} \label{eq:TF-regime-diffeq}
	-\frac{1}{\rho} \nabla \mPSI = \nabla \Phi,
\end{equation}
i.e. only SI pressure balances gravity. In spherical symmetry, it can be shown that this equation leads to the well-known Lane-Emden equation for a ($n=1$)-polytrope, whose closed-form solution for the density profile is given by
\begin{equation} \label{eq:polytrope-density}
	\rho(r) = \rho_0 \frac{\sin (\pi r / R_\text{TF})}{\pi r / R_\text{TF}},
\end{equation}
with central density $\rho_0$, and $R_\text{TF}$ is the first zero of the density profile, which serves as the radius of the polytrope, see \citet{Goodman2000} and \citet{Peebles2000}. This is the so-called TF radius, which is given by
\begin{align} \label{eq:tfradius}
	R_\text{TF} = \pi \sqrt{\frac{g}{4 \pi G m^2}}
\end{align}
with the gravitational constant $G$. Despite the fact that two DM model parameters come in, we see that in the TF regime, the characteristic radius of interest is fully determined by a given choice of the \textit{single} parameter combination $g/m^2$. Also, the TF radius does not depend on the total mass of the polytrope. That is, in the context of SFDM-TF halos, where cores are close to these polytropes, their size does not depend on their (core) mass. Hence, once we specify a SFDM-TF model by fixing $g/m^2$, we get a fixed value for $R_\text{TF}$, independent of core mass. Generally, we stress that \RTF{} is subject to constraints from astronomical observations that are able to determine (upper) bounds on that radius, constraining SFDM-TF models, in turn. 
In practice, \RTF{} will be a free parameter in our simulations, though its choice will be informed by constraints just mentioned. In terms of \RTF{}, we can rewrite the associated SI pressure in (\ref{eq:polytropic-pressure}) as
\begin{equation} \label{eq:SIpressrtf}
	\mPSI = \frac{2G}{\pi}\mRTF^2 \rho^2,
\end{equation}
which is the form that will be used in our simulations to determine the halo SI pressure as a function of halo density, for fixed \RTF{}.

Whereas \PSI{} turns out to be easily modeled, the numerical treatment of the quantum potential $Q$, and its corresponding quantum pressure, is much more difficult. Therefore, it is useful to devise ``effective'' means to model the physics of $Q$ in a computationally feasible way. One approach will be discussed in the next section.

\section{Fluid Approximations for SFDM Halos}\label{sec:SFDMhydro}


It has been long realized that fluid approximations of DM dynamics constitute both, a helpful computational methodology, as well as a contribution to our physical understanding and comparison between different DM models. 

The idea of simulating collisionless CDM by means of fluid equations has been recurring in the literature, and was considered especially at earlier times when realistic N-body simulations were a bigger computational challenge than they are today. For instance, halo infall calculations in the fluid picture were presented by \citet{Teyssier1997}. If fluid approximations are limited to 1D (as they often are), they have the advantage of providing very high resolution at a much lower computational cost than other approaches.

To derive a fluid approximation, one usually starts with the equation of motion for the distribution function of the system.
For CDM, a collisionless Boltzmann equation for the phase-space distribution function is integrated to obtain a set of moment equations (momentum moments of the Boltzmann equation). Collisionless particles have an infinite set of moment equations, the BBGKY hierarchy (see \citet{Binney1987}). So, the hierarchy needs to be truncated at a certain order. By adopting some assumptions, one ends up with a closed set of fluid conservation equations for collisionless systems\footnote{The same approach has been adopted by \citet{Vorobyov2015} for the collisionless stellar component of dwarf galaxies.} such as CDM, the ``CDM fluid approximation''.  

\citet{Alvarez2003} have applied this formalism to study CDM halo formation. Adopting the universal mass assembly history (MAH) reported for CDM halos from N-body simulations, found by \citet{Wechsler2002}, and applying their fluid approximation, they have shown
that the resulting shape of the equilibrium halo profile and its evolution match those of CDM N-body simulations remarkably well.
We shall add that the notion of ``orbital crossing'' in collisionless N-body simulations is replaced by the appearance of a shock wave in the fluid picture which, upon gravitational collapse, will eventually form and propagate outward into the forming halo envelope. As such, it separates the interior of the eventually virialized halo from the material outside of the halo. This way, the location of the shock boundary provides a physical explanation for and interpretation as the halo virial radius.

\citet{Ahn2005} developed this model further in order to study halo formation in CDM, as well as in self-interacting DM (SIDM)\footnote{From a dynamics perspective, the main difference between CDM and SIDM is the addition of a finite repulsive particle scattering cross-section in the latter, which has a different physical origin than the repulsive SI in SFDM, however.} by adopting certain assumptions, namely spherical symmetry, skew-free velocity distribution and isotropic velocity dispersion. More precisely, \citet{Ahn2005} derived a fluid approximation for CDM halos in spherical symmetry from momentum moments $0, 1, 2$ of the collisionless Boltzmann equation, resulting in the following set of hydrodynamical equations:
\begin{subequations}\label{eq:cdmSDR}
	\begin{empheq}{align}
		\frac{\partial \rho}{\partial t} + \frac{1}{r^2} \frac{\partial(r^2 \rho v)}{\partial r} = 0,\label{eq:cdmSDRa}\\
		\frac{\partial v}{\partial t} + v \frac{\partial v}{\partial r} + \frac{1}{\rho} \frac{\partial}{\partial r} \left( \mPsigma \right) + \frac{\partial \Phi}{\partial r} = 0,\label{eq:cdmSDRb}\\
		\frac{\partial}{\partial t} \left( \frac{3 \mPsigma}{2 \rho} \right) + v \frac{\partial}{\partial r} \left( \frac{3 \mPsigma}{2 \rho} \right) + \frac{\mPsigma}{\rho r^2} \frac{\partial \left(r^2 v \right)}{\partial r} = 0,\label{eq:cdmSDRc}
	\end{empheq}
\end{subequations}
with the density $\rho$, the velocity $v$, the gravitational potential $\Phi$, and a pressure called $P_{\sigma}$ where $\sigma$ stands for velocity dispersion. It stems from the second moment equation, where a non-vanishing velocity-dispersion tensor gives rise to an overall pressure contribution $P_{\sigma}$ in the system. It is the only pressure contribution in a spherical system of collisionless particles. In fact, this set of equations has to be accompanied by an equation of state (EoS) which relates that pressure to density. \citet{Ahn2005} found that the nature of the velocity-dispersion pressure, under the assumed symmetries, demands that $P_{\sigma} \propto \rho^{\gamma}$ with $\gamma = 5/3$, valid for CDM and SIDM, indeed. 
For formal analogy to gas dynamics, we will often use the phrase ``thermal pressure'' for \Psigma{}, along with the corresponding ``thermal energy'' that comes with it, which will be introduced below.

Performing simulations using a Lagrangian 1D hydro code, \citet{Ahn2005} also confirmed that the CDM fluid approximation can explain very well certain universal properties, as found by N-body simulations, such as the NFW halo density profile, or temperature profiles of baryons in galaxy clusters; we refer to the in-depth account of these and similar results by \citet{Shapiro2004}. (A further comparison between fluid and N-body simulation results has been also performed by \citet{Koda2011}, in the framework of SIDM.)  \par

Inspired by this approach, \citet{Dawoodbhoy2021} engaged to derive a fluid approximation for SFDM in the TF regime (SFDM-TF), in a similar manner by starting from an equivalent equation of motion for the (quantum) Wigner phase-space distribution function.
In brief, it is possible in this TF regime to coarse-grain the original equations, using a smoothing scale that is much larger than the de Broglie length scale $\lambda_{deB}$, but smaller than the characteristic structure formation scale of order \RTF{}. This is a valid approach in SFDM-TF, because $R_\text{{TF}} \gg \lambda_{deB}$. As a result, the dynamics need not be resolved at scales of order $\lambda_{deB}$, which are the scales where the quantum potential $Q$ of Eq. \eqref{eq:quantum-potential} is at play. Promoting its local effects to a quantum-pressure tensor, in a similar form than what the velocity-dispersion tensor does for the collisionless case, it is possible to encode it through a macroscopic pressure contribution which is isotropic, if the same assumptions are adopted as before (sphericity, isotropy and skewless velocity distribution). The remarkable thing is that the so-encountered pressure has the same mathematical form as $P_{\sigma}$ above, i.e. the macroscopic velocity dispersion due to the presence of the quantum potential in SFDM is also modeled in this case via a $\gamma = 5/3$ law for \Psigma{}.  

Using their equations, \citet{Dawoodbhoy2021} performed halo formation simulations using their amended version of said 1D Lagrangian hydro code. Their analysis was restricted to simulations of single halos in a static background universe, whereas such single-halo infall simulations were subsequently extended to expanding backgrounds in \CPaperSDR{}; we display their Eqs. (22)--(24) here, using the same notation as before:
\begin{subequations}\label{eq:sfdmSDR}
	\begin{empheq}{align}
		\frac{\partial \rho}{\partial t} + \frac{1}{r^2} \frac{\partial(r^2 \rho v)}{\partial r} = 0,\label{eq:sfdmSDRa}\\
		\frac{\partial v}{\partial t} + v \frac{\partial v}{\partial r} + \frac{1}{\rho} \frac{\partial}{\partial r} \left( \mPsigma + \mPSI \right) + \frac{\partial \Phi}{\partial r} = 0,\label{eq:sfdmSDRb}\\
		\frac{\partial}{\partial t} \left( \frac{3 \mPsigma}{2 \rho} \right) + v \frac{\partial}{\partial r} \left( \frac{3 \mPsigma}{2 \rho} \right) + \frac{\mPsigma}{\rho r^2} \frac{\partial \left(r^2 v \right)}{\partial r} = 0.\label{eq:sfdmSDRc}
	\end{empheq}
\end{subequations}
Different from the CDM equations in (\ref{eq:cdmSDR}), we have in the momentum equation \eqref{eq:sfdmSDRb}, apart from \Psigma{}, the additional pressure \PSI{} from (\ref{eq:SIpressrtf}), which originates from the repulsive SI of SFDM-TF. However, \PSI{} does not contribute to the energy equation \eqref{eq:sfdmSDRc}, which is the same as for CDM.

Although the respective sets of equations for CDM vs SFDM-TF are derived from different equations of motion, the fluid approximations look very similar. This can be attributed to the same approximations on symmetry and EoS that enter in both approaches.  We also stress again that the fluid approximation for SFDM-TF is especially helpful, given the severe resolution issues pointed out in the Introduction, which we face in modeling SFDM-TF halo formation, particularly if sub-kpc core sizes are expected.  \par

Now, \citet{Dawoodbhoy2021} and \CPaperSDR{} used an amended 1D Lagrangian code, which was previously used by \citet{Ahn2005} for CDM and SIDM simulations of halo spherical infall and collapse, in order to calculate the formation of SFDM-TF halos and to investigate their structure, using the above set of fluid equations (\ref{eq:sfdmSDR}). In accordance with previous analytical calculations, they found that halos formed with a core close to a ($n=1$)-polytrope, surrounded by a halo envelope.
Unsurprisingly, the characteristic transition and slope of the latter differed, depending on whether halos formed in static or expanding backgrounds; however, in each case the halo envelopes would be close to those found in the CDM fluid approximation, i.e. ``CDM-like''.
In fact, generically, the SI pressure \PSI{} dominated in the core, whereas \Psigma{} dominated in the envelope. At a radius of approximately \RTF{} both pressures were nearly equal.\par
Soon after the work of \CPaperSDR{}, \CPaperHWM{} performed 3D simulations of halo formation in SFDM-TF. Given the same computational difficulties that prompted \citet{Dawoodbhoy2021} to come up with the fluid approximation in the first place, \CPaperHWM{} used a fluid approximation, as well. However, they used a different set of 3D fluid equations [their Eqs. (15)--(17)], which we display here as follows:
\begin{subequations}\label{eq:sfdmHEA}
	\begin{empheq}{align}
		\frac{\partial \rho}{\partial t} + \nabla \cdot \bm{j} = 0,\\
		\frac{\partial \bm{j}}{\partial t} + \nabla \left(P_{\sigma} + \mPSI \right) + \rho \left(\bm{v} \cdot \nabla\right)\bm{v} \:+\;\; \\ \nonumber
		\bm{v} \left(\nabla \cdot \rho \bm{v} \right) = - \rho \nabla \Phi,\\
		\frac{\partial E}{\partial t} + \nabla \cdot \left[ \left(E + P_{\sigma} + \mPSI\right)\bm{v}  \right] = -\bm{j} \cdot \nabla \Phi.\label{eq:sfdmHEAc}
	\end{empheq}
\end{subequations}
Note that \CPaperHWM{} uses the notation $P$, instead of $P_{\sigma}$, and $\phi$, instead of $\Phi$, while $\bm{j}$ is the 3D current. 
Comparing this set of equations with those in \eqref{eq:sfdmSDR}, we see that the SI pressure \PSI{} appears not only in the momentum equation, but also in the energy equation, because the latter includes the total energy $E$, not just the ``thermal energy'' due to \Psigma{}.
Furthermore, the momentum and the energy equation have a source term due to gravity on the right-hand side, different from Eqs. (\ref{eq:sfdmSDR}). 
Later in this paper, we will see that these differences in the fluid equations make no difference in the results, concerning halo evolution and halo structure. 
The form of the 3D equations in (\ref{eq:sfdmHEA}) is motivated by their close resemblance to the general fluid equations, which are implemented into the cosmological code RAMSES that is used by \CPaperHWM{} to perform their simulations. Yet, we stress that the form of (\ref{eq:sfdmHEA}) still assumes that the momentum distribution of the phase-space distribution function is isotropic, an assumption which is not valid in 3D in all phases of halo evolution. By enforcing spherical symmetry, rewriting of the energy equation in terms of the ``thermal pressure'' only, and neglecting the source terms on the right-hand side, Eqs. (\ref{eq:sfdmHEA}) reduce to Eqs. \eqref{eq:sfdmSDR}.   
\par

Finally, we comment that in the 1D simulations by \CPaperSDR{}, the effect of Hubble expansion is taken into account via appropriate initial conditions, while the 3D simulations in RAMSES of \CPaperHWM{} and our own take advantage of a prior transformation of the fluid equations, using super-comoving coordinates which were introduced by \citet{Martel1998}; we refer to this paper and \CPaperHWM{} for more details.\par

In accordance with the previous 1D results mentioned earlier, \CPaperHWM{} also found a core-envelope structure of SFDM-TF halos, but with different characteristics; in fact they report discrepancies compared to the results of \CPaperSDR{}, as follows:

\begin{itemize}[topsep=5pt,noitemsep,itemsep=5pt]
	\item \CPaperSDR{} found a clear dominance of the SI pressure \PSI{} in the core, throughout the entire simulation time of any given halo, whereas \CPaperHWM{} found that \PSI{} eventually does not dominate over $\mPsigma$ in the cores of the halos in their simulations.
	\item The final core radii in \CPaperHWM{} are larger than the primordial value given by \RTF{}.
\end{itemize}
The authors of \CPaperHWM{} suggest the following main reasons for the discrepancies:
\begin{itemize}[topsep=5pt,noitemsep,itemsep=5pt]
	\item There are naturally asymmetries in the 3D environment, while the 1D simulations cannot account for these.
	\item The cores get dynamically heated from the envelopes, due to some equilibration and mixing processes (e.g. infall of lobes of collapsing material), which lead to a higher \Psigma{} and thus prevents the dominance of \PSI{} in the cores, in turn.
\end{itemize}\par

Our main goal in this paper consists in determining the nature of the reported discrepancies by performing our own 3D simulations, using both sets of fluid equations for the sake of a consistent comparison.


%
\section{Implementing fluid approximations of SFDM-TF into RAMSES}\label{sec:ourINVESTIGATION}

The scope of our investigation demands that we implement both sets of equations \eqref{eq:sfdmSDR} and \eqref{eq:sfdmHEA} into a cosmological code and perform 3D halo formation simulations, in order to shed light onto the discrepancies between these previous papers. Since \CPaperHWM{} used RAMSES for their simulations, we will also use this code for this comparison. Furthermore, we investigate the implications for small-scale structure which is relevant for observations and models.\par
%

\subsection{Implementation by HWM22}\label{sec:differences}

In order to perform their 3D simulations, \CPaperHWM{} used an amended version of RAMSES, an adaptive mesh refinement (AMR) based code by \citet{Teyssier2002} \footnote{A public version of RAMSES is available for download at \href{https://bitbucket.org/rteyssie/ramses/src/master/}{https://bitbucket.org/rteyssie/ramses/src/master/}}. Moreover, they used the code MUSIC by \citet{Hahn2011} \footnote{MUSIC is publicly available for download at \href{https://bitbucket.org/ohahn/music/src/master/}{https://bitbucket.org/ohahn/music/src/master/}}, to generate the initial conditions (ICs). 
This way, they were able to form several halos within a 3D cosmological box with periodic boundary conditions. Several simulations were performed with different box sizes: one with box size of $2$~Mpc, four simulations with box size of $5$~Mpc and one simulation with box size of $15$~Mpc.

All simulation runs were started at an initial redshift of $z_{ini}=50$, a choice which is typical for CDM cosmological simulations.  As we will see, this is a shortcoming with respect to SFDM cosmologies, because it means that \CPaperHWM{} were restricted to models that have $\mRTF \gtrsim 1$~kpc.

Finally, the properties of the formed halos were analyzed at a lowest redshift of $z=0.5$, because the demands on CPU time prevented simulation runs up to the present $z=0$. In fact, the computational challenges with respect to resolution increase even more, the closer the evolution gets to the present time.  \par

In contrast, the simulations in \CPaperSDR{} of 1D halo infall were not limited as much by resolution, though their shortcoming is a less realistic environment, for only one single halo at a time was simulated. Their simulations started at matter-radiation equality $z_{eq} \sim 3000$ (or in terms of scale factor $a_{eq}$), right in order to account for the effects of early suppression of small-scale structure that were found from the linear growth regime of structure formation in $\Lambda$SFDM universes in \CPaperSDR{}, \citet{Hartman2022} and \citet{Foidl2022}. Sufficient spatial resolution was attained by using around 1000 mass shells in their Lagrangian code; yet halos with sub-kpc cores could be analyzed. 
The analysis of halo structure was carried out at the so-called formation time $a_f$ (respectively $z_f$) of halos; the definition of formation time can be found in their Eqs. (38) and (39). It is used to determine the earliest redshift at which the collapse of a halo of given mass shall occur, in accordance with \citet{Klyping2016_Massrelation}.\par

\subsection{Our implementation}\label{sec:implRAMSES}

As in \CPaperHWM{}, we perform SFDM-only cosmological 3D simulations using RAMSES, where we likewise modify and adapt the in-built RAMSES hydro module, which is usually used to model the baryons in cosmological simulations, in order now to calculate the SFDM-TF fluid dynamics, instead. At the same time, we disable the RAMSES N-body solver that would usually calculate the CDM dynamics. In modifying RAMSES, we also adapted the 1D Riemann solver in the RAMSES hydro module, for both sets of fluid approximations, such that we could switch between them accordingly.\par
More precisely, RAMSES is an AMR code that applies the Godunov method (\citet{Godunov1959}) to solve the hydrodynamic equations in conservative form [see Eqs. (8)--(10) in \citet{Teyssier2002}], which we display as follows:
\begin{subequations}\label{eq:hydroRAMSES}
	\begin{empheq}{align}
		\frac{\partial \rho}{\partial t} + \nabla \cdot \left(\rho \bm{u} \right) = 0,\\
		\frac{\partial }{\partial t} (\rho \bm{u}) + \nabla \cdot \left( \rho \bm{u} \otimes \bm{u} \right) + \nabla p = - \rho \nabla \Phi,\\
		\frac{\partial}{\partial t} \left( \rho e \right) + \nabla \cdot \left[ \rho \bm{u} \left(e + p/\rho\right)\right] = -\rho \bm{u} \cdot \nabla \Phi,
	\end{empheq}
\end{subequations}
with $\rho$ the density, $\mathbf{u}$ the velocity vector, $e$ the specific total energy, $p$ the specific pressure (which, in non-SFDM applications, is usually the thermal pressure of the baryonic gas), and $\Phi$ the gravitational potential (again, we use the notation $\Phi$, instead of $\phi$). 
Furthermore, RAMSES uses the second order MUSCL scheme (\citet{Leer1979}) to enable 3D simulations that utilize a 1D Riemann solver for Eqs.~\eqref{eq:hydroRAMSES}, which effectively reduces them to 1D. RAMSES provides a number of different Riemann approximations and an exact Riemann solver for configuration purpose.
We adapted the HLLC Riemann approximation\footnote{This Riemann approximation has been also used for the test problems in \citet{Teyssier2002}} to reflect the necessary modifications to Eqs.~\eqref{eq:hydroRAMSES}, in order to implement both sets of equations:  the 3D version of \eqref{eq:sfdmSDR} from \CPaperSDR{} and \eqref{eq:sfdmHEA} from \CPaperHWM{}. To this end, we introduced two additional variables in the state vector: first, $p_{\text{SI}}$ which is the specific form of the SI pressure defined in \eqref{eq:SIpressrtf}, and the second variable $p_{\sigma}$, the specific ``thermal pressure'' due to the (quantum) velocity dispersion, according to
\begin{equation} \label{eq:eosHEA}
	p_{\sigma} = (\gamma-1) \rho (e - \frac{1}{2} u - e_\text{SI}).
\end{equation}
Here, $e_\text{SI}$ represents the specific energy contribution originating from SI, while the overall expression within the second bracket then represents the specific internal energy; it is this energy that enters in the default caloric EoS which is used in RAMSES. Because $p_\text{SI}$ does not appear in the energy equation of \eqref{eq:sfdmSDR}, $e_\text{SI}$ is set to zero for this set of equations.\par
The analysis of the simulation data as well as part of the plots were done using the software package yt (\citet{Turk2011})\footnote{yt is publicly available for download at \href{https://yt-project.org}{https://yt-project.org}}.\par

\subsection{Verifying the CDM fluid approximation}\label{sec:compareNBody}

The implementation of the SFDM-TF fluid approximation into RAMSES lends itself to a comparison of a fluid vs an N-body cosmological simulation within the CDM framework first, as follows. The CDM fluid approximation, as derived by \citet{Ahn2005}, is recovered from Eqs.~\eqref{eq:sfdmSDR}, if we set $\mRTF=0$ which implies $\mPSI=0$. In this case, only the velocity-dispersion pressure \Psigma{} remains in the equations. Thus, the cosmological simulation of halo formation in the CDM fluid picture uses this set of fluid equations, whereas the comparison CDM simulation uses the standard N-body solver of RAMSES. For this comparison, we generated the ICs with MUSIC and applied the same cosmological parameters as those used for our forthcoming simulations, summarized in Table~\ref{tab:planck2018}.\par
\begin{figure*} [!htb]		
	{\includegraphics[width=\PW\columnwidth]{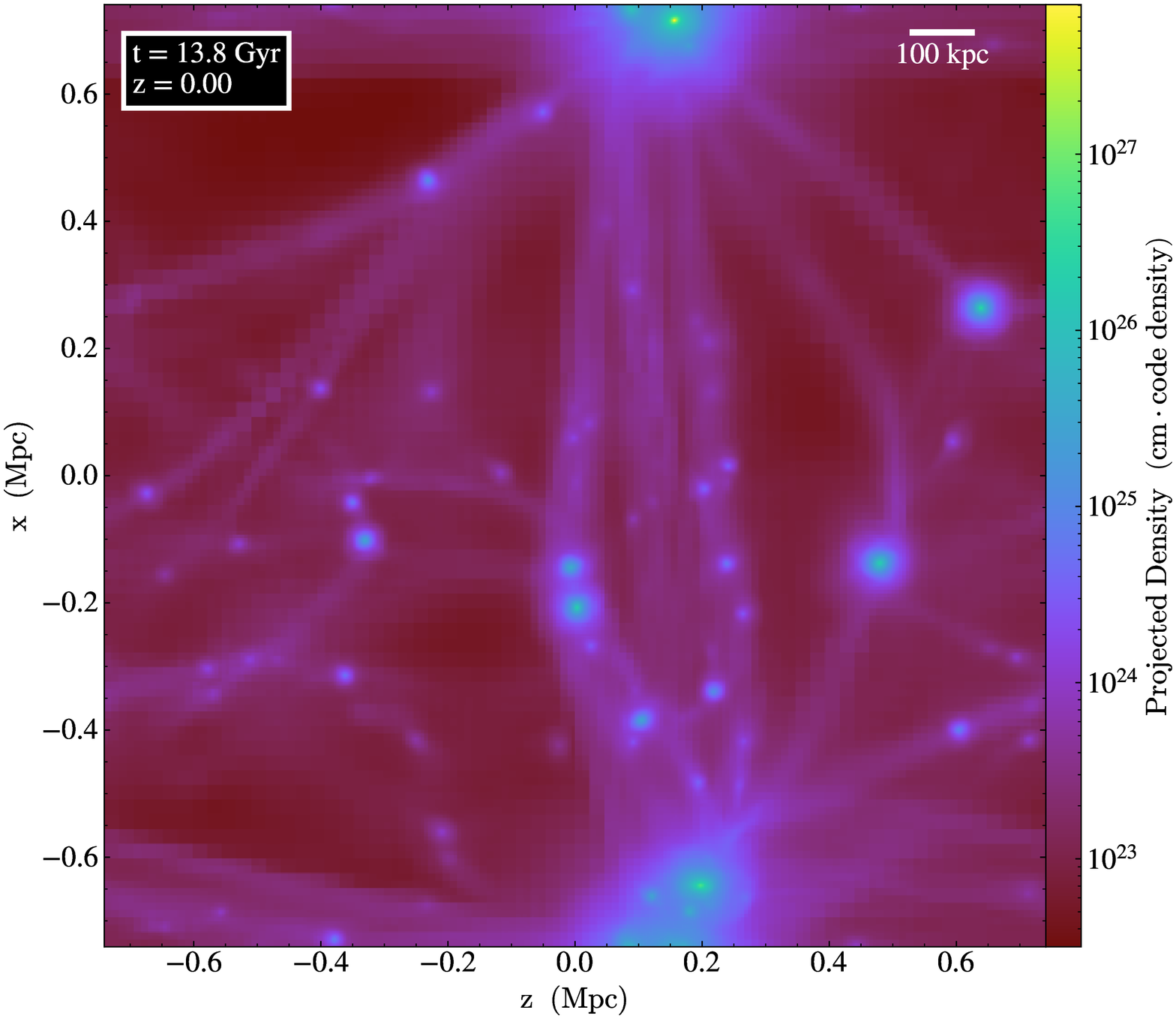}}
	{\includegraphics[width=\PW\columnwidth]{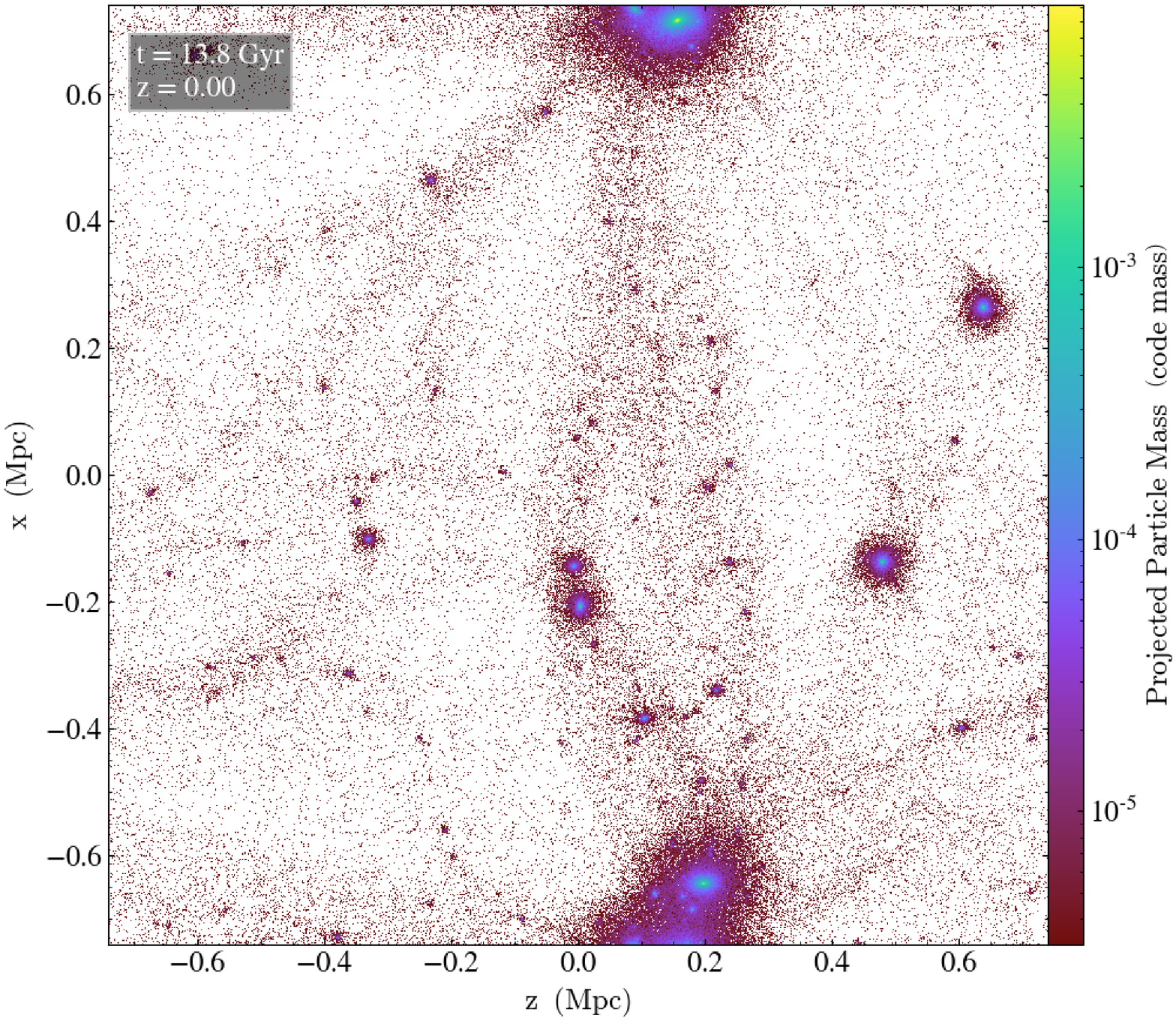}}
	\caption[Evolution $\Lambda$CDM Halo $\bm{z_{ini}=600}$]
	{\textbf{Evolution of $\Lambda$CDM halos in the fluid approximation (left-hand panel) vs an N-body simulation (right-hand panel). } Each simulation starts at ${z_{ini}=600}$; shown are the snapshots at the present $z=0$. The left-hand panel indicates the color-coded projected density distribution. The spatial resolution is determined by the AMR refinements, where every cell holds the average density of the spatial region. The right-hand panel displays the projected mass of the DM particles, where the location of the individual particles is not limited to the spatial resolution of the AMR grid. Owing to the periodic boundary conditions (BCs), the halo at the bottom of the plot coincides with that at the top, in each panel.
	}
	\label{fig:form-LCDM}
\end{figure*}
Figure~\ref{fig:form-LCDM} displays the results, where we compare the cosmological structure formation of CDM halos using the fluid approximation (left-hand panel) versus the N-body solver (right-hand panel), in a comoving box of $1$~Mpc size with periodic boundary conditions (BCs). In general, we can see that the overall structures and spatial extent of DM halos match quite well between both kinds of simulations, though there seems to be more substructure in the right-hand panel. Compared to this density field in the N-body case that shows individual DM particles, the smeared-out impression of the density field in the case of the fluid approximation comes about due to the different post-processing of the AMR grid, as the plot is built by individual AMR cells, i.e. it shows the projected average density in the cells, resulting in an overall lower spatial resolution. Apart from this reason, we cannot exclude the possibility that there is a genuine, physical ``wash-out'' effect that is inherent in the fluid approximation itself.   \par
To our knowledge, this is the first side-by-side comparison of a 3D cosmological simulation of CDM-only structure formation between the fluid picture and the N-body approach. Of course, a thorough quantitative comparison would be required to investigate the similarities and discrepancies between fluid and N-body simulation results of CDM, and there are only few such comparisons in the literature (see also previous citations to references, but they are mostly limited to the 1D fluid case). 
Since we focus in this paper on the single-halo infall problem in SFDM-TF, such a dedicated methodological comparison in the CDM regime is beyond the scope of this paper. 
Nevertheless, we feel safe to conclude that the implemented fluid approximation yields reasonable results in the CDM case, which makes us confident that the implementation works correctly. (We note that the same applies to the implementation by \CPaperHWM{}; setting the SI pressure to zero, $\mPSI = 0$, leads to the same CDM fluid approximation). 
In fact, we do present a detailed analysis of CDM single-halo evolution in Sec.\ref{sec:simsRAMSESCDM}.
But first, we dedicate the next section to details concerning initial conditions.

\section{ICs for halo formation}\label{sec:ICsHalos}

We performed SFDM-only cosmological simulations with the same set of cosmic parameters as used in our preceding work \citet{Foidl2022}, determined by \citet{Collaboration2020}, shown in Table~\ref{tab:planck2018}.\par
{\belowcaptionskip=0pt
	\begin{table}[!htbp] 
		\caption{Cosmological parameters from Planck 2018\label{tab:planck2018}}
		\begin{ruledtabular}
			\begin{tabular}{lrl}
				Parameter        & Value            & Comment \\
				\hline
				H$_{0}$          & $67.556$         & \\
				T$_{\text{CMB}}$~[K]    & $2.7255$         & \\
				N$_{\text{ur}}$   & $3.046$          & \\
				$\Omega_{\gamma}$ & $5.41867 \times$10$^{-5}$  & derived from T$_{\text{CMB}}$\\
				$\Omega_{\nu}$    & $3.74847 \times$10$^{-5}$  & derived from N$_{\text{ur}}$\\
				$\Omega_{b}$      & $0.0482754$      & \\
				$\Omega_{\text{SFDM}}$   & $0.263771$       & \\
				$\Omega_{\Lambda}$& $0.687762$       & \\
				$\Omega_{k}$      & $0$              & \\
				$\tau_{\text{reio}}$     & $0.0925$         & \\
				A$_{s}$          & $2.3 \times$10$^{-9}$       & \\
				n$_{s}$          & $0.9619$            & adiabatic ICs\\
			\end{tabular}
		\end{ruledtabular}
	\end{table}
}
We provide this table for the sake of completeness. Not all parameters are relevant for the generation of ICs and the run-time configuration of RAMSES.\par
\subsection{ICs for the cosmological box}\label{sec:ICsBox}

We generated the initial conditions (ICs) such that a perturbation overdensity is placed in the center of a cosmological box, whose size is sufficiently large to contain the matter to build up the halo of given (final) mass. 
%
We performed simulation runs in order to study the formation, evolution and structure of a halo of Milky-Way size with $10^{12}$~M$_{\odot}$, as well as a dwarf-galactic-size halo of $10^{9} $~M$_{\odot}$. In all simulations, we chose a fixed value of the primordial core radius of $R_\text{{TF}}=110$~pc, i.e. the SFDM-TF model parameters are the same in each simulation (see \citet{Foidl2022} for constraints on SFDM-TF parameters). This way, we can investigate whether a given primordial core affects halos of various final mass and size over the course of their evolution differently.  
To fix the starting point of our simulations, we used Eqs.~ (38) and (39) of \CPaperSDR{} to determine the earliest redshift at which the collapse could occur. Yet, we also performed simulations with a range of starting redshifts, $z_{ini}=3400;1000;600;200;50$ in order to investigate the impact of this choice for $z_{ini}$. 
It turned out that $z_{ini} = 600$ is roughly the latest possible time of collapse to form halos of the above mass, in accordance with \citet{Klyping2016_Massrelation}. In the consecutive simulations, we used thus $z_{ini} = 600$ as the fiducial starting redshift to compare our results, based on the different sets of equations that we implemented. 
In fact, our results do not change if we choose an earlier redshift than $600$, except for the higher demands on computation time.

The size of the simulation box is set to $4$~Mpc comoving on a side, applying periodic boundary conditions (BCs). The initial spatial resolution, sufficient for the initial density profile, was set to $\sim 25$~kpc comoving, which is increased dynamically by the AMR mechanism down to $\sim 0.1$~kpc comoving, in regions where necessary. The spatial resolution of the AMR grid, described as comoving, has to do with the fact that it is related to the scale factor, i.e. in the early stages of the evolution, where the spatial resolution is very important, it is significantly larger than in the late stages (see the fluctuations depicted in Figs.~\ref{fig:evol-var4}, \ref{fig:evol-var1-2} and \ref{fig:evol-var2-12}).
We first present and focus mostly on the results for the Milky-Way-type halo, while the results of and comparison with the dwarf-galactic-type halo is presented afterwards. This way, we can also put our results better into perspective in terms of \RTF{}. \par
\subsection{Initial density profiles}\label{sec:ICprofiles}

In \CPaperSDR{}, the choice of the initial density profile was crucial, because it determined the MAH of the (single) halo, which was the key to correctly model the cosmological environment. 
In contrast to that, RAMSES implements the cosmological environment ``for free'', so to speak. However, we wanted to investigate the potential influence of the initial density profile onto the final halo structure. Therefore, we tested several IC density profiles to this end. We chose (i) a tophat profile, which is often used in the literature to model halo infall; (ii) a linear density profile; (iii) and an ($\epsilon=1/6$) profile, which corresponds to a scale-free, spherical, linear perturbation with a growth rate of $dM/M \propto M^{-1/6}$ (see \CPaperSDR{}~and references therein).
\begin{figure} [!htb]
	{\includegraphics[width=0.9\columnwidth]{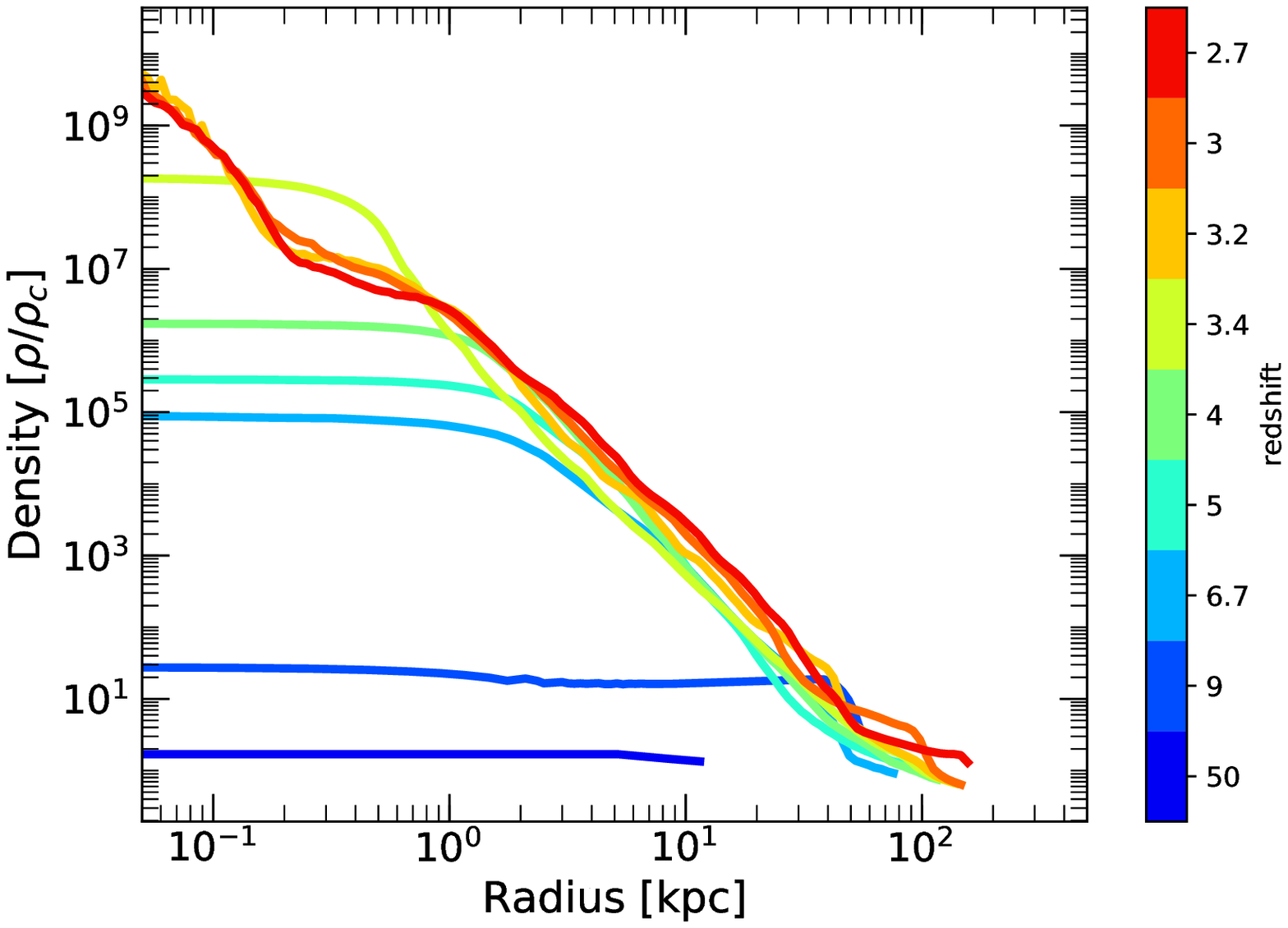}}
	{\includegraphics[width=0.9\columnwidth]{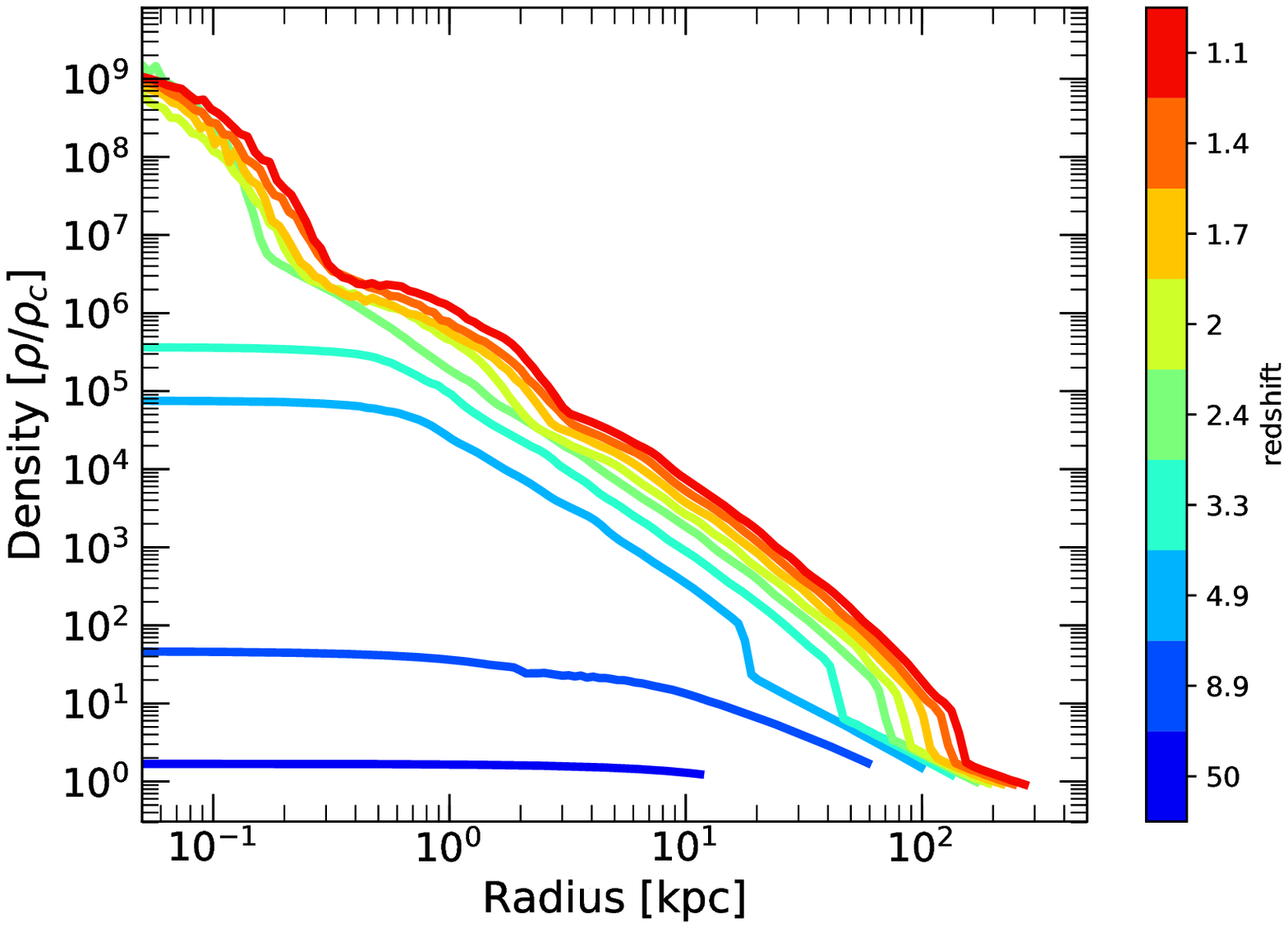}}
	{\includegraphics[width=0.9\columnwidth]{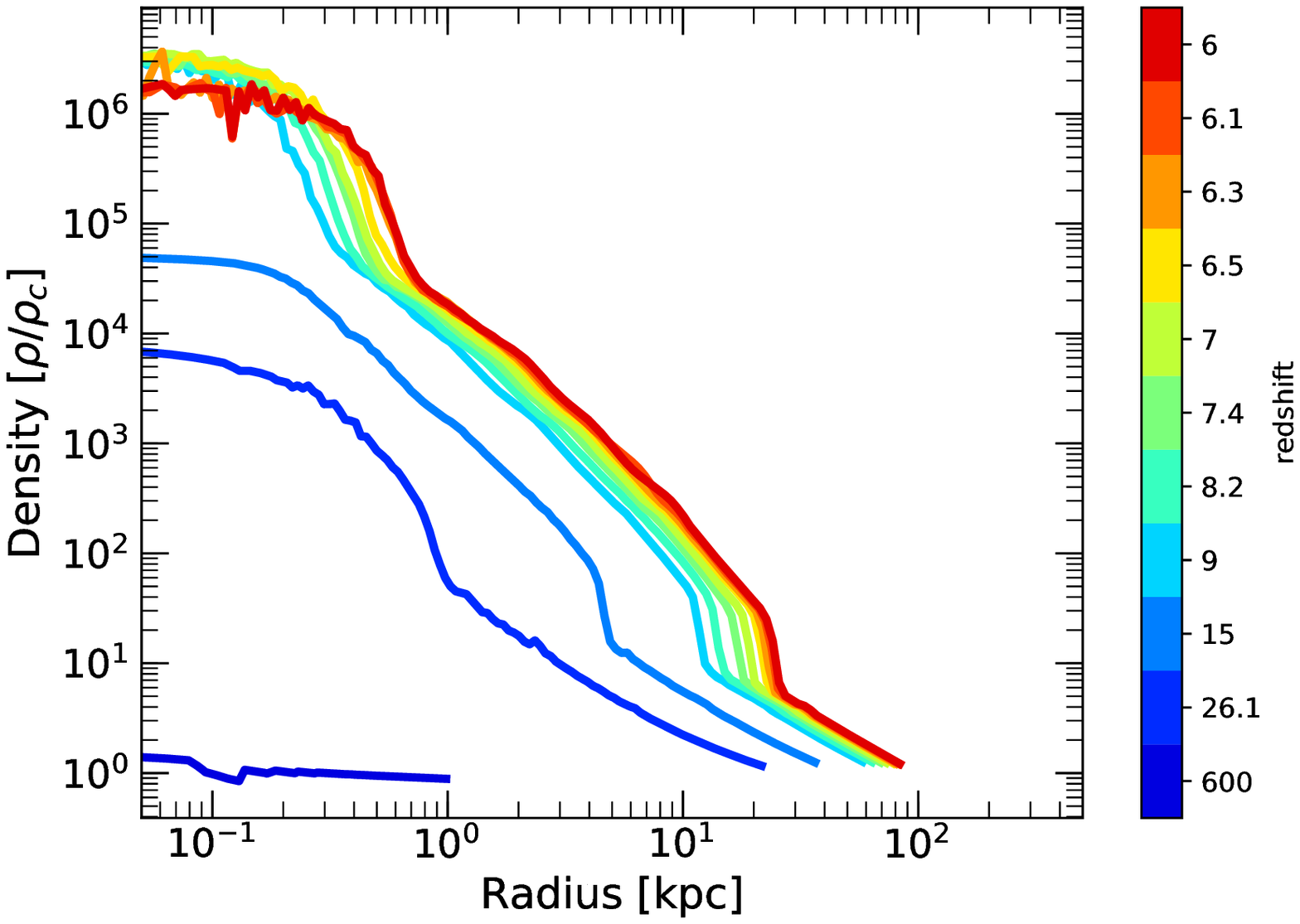}}
	\caption[Initial density profiles]
	{\textbf{SFDM-TF halo evolution for different choice of initial density profiles.} The solid lines display the density profile of a halo during its evolution, color-coded by redshift but note the different $z$-range between panels! Each panel displays halo density profiles over time, initiated by a tophat profile (top panel), a linear profile (middle panel), and a scale-free, spherical perturbation (bottom panel). In each case, we see that the halo density builds up smoothly but quickly toward the center, while a clear core-envelope halo structure evolves. The consecutive shock wave (recognizable as the steep gradient that forms after $z \sim 10$) propagates outward, transporting excess material from the inner regions to the outskirts of the halo.
	}%
	\label{fig:initial-profiles}%
\end{figure}%
Figure~\ref{fig:initial-profiles} displays the results of the impact of the different chosen initial density profiles onto the evolution of a SFDM-TF halo. Here and in the figures that follow, we normalize the halo density over the critical background density of the universe, denoted as $\rho_c$. The initial profiles were built with the same amount of overdensity at the center of the box and filling the box to its boundaries. We can see that the shape of the final halo profiles differ somewhat, for different initial profiles. Still, in each case we see a pileup of material in the halo center, which is separated from a milder decline in density in the halo outskirts.
This pileup is initially extremely pronounced for the tophat profile, compared to the linear profile and the spherical perturbation. When the outer boundary of the tophat profile, at $z \sim 3$, reaches the pileup of matter in the center, the infall of matter ceases and the profile develops an inner and an outer part with a milder decline there. The linear profile displays a much milder growth of the central pileup of matter and we can see the formation of a shock front moving outward. Such a shock front is also seen in the profile with the spherical initial perturbation. However, no shock front is formed in the case with the initial tophat profile.
After the rapid increase of the density in the center (for the linear profile and the spherical perturbation), the consecutive shock wave transports excess material from the inner regions to the outskirts of the halo. The shock front is recognizable as the steep density gradient that emerges at redshifts lower than $z \sim 10$, which itself propagates outward.\par
In the subsequent simulations, the location of this shock front at the present time, $z=0$, serves as a physical approximation for the halo virial radius, assuming such virialization is (approximately) achieved.\par
%

In \CPaperSDR{}, the shape of the initial profile served as a way to provide a realistic expanding background for their 1D simulations, especially in terms of mass infall. Our experiments here show that the cosmological environment of RAMSES is very robust in that we need not construct an ``artificial'' mass assembly history, from the chosen profile. However, as the ($\epsilon=1/6$) profile is the most realistic IC, in terms of the expanding cosmological background, we decided to use this profile in the subsequent simulations. In fact, this choice of initial profile leads to a clear core-envelope halo structure (see Fig.~\ref{fig:initial-profiles}), which was also found in the simulations of \CPaperSDR{} and \CPaperHWM{}, respectively.

Before we leave this section, we like to point out that the format of the IC files which RAMSES uses has no place to specify an initial pressure. Instead, RAMSES uses an approximation of the average temperature (normally due to the presence of baryons) to compute the initial pressure. We adopt this default behavior of RAMSES, in that we assign \Psigma{} to RAMSES' thermal pressure of Eq.~\eqref{eq:eosHEA} (since formally it is the same), while \PSI{} is determined according to \eqref{eq:SIpressrtf}. In contrast, \CPaperHWM{} introduced a free parameter $\zeta$, which is the ratio of \Psigma{} over \PSI{} at $z_{ini}$ and is supposed to be much smaller than one, in order to specify the initial value of their pressure. However, they find that their simulation results are rather insensitive to this parameter, a finding that is also in accordance with our comparison to their results.
%

\section{Formation and evolution of CDM halos}\label{sec:simsRAMSESCDM}

We start our series of simulations by studying first the formation of CDM halos in the CDM fluid approximation, where we set $\mRTF = 0$, implying $\mPSI=0$, in our SFDM-TF fluid equations, which implies standard CDM behavior. This way, we can also convince ourselves that the fluid approximation gives reasonable results when applied to the CDM regime, before we move on to the study of SFDM-TF halo formation.

\par
%
\subsection{The properties of the shock wave in CDM halos}\label{sec:shockwave}

In Sec.~\ref{sec:ICprofiles} we investigated the impact of the initial density profile on the structure of the forming halo. We could see that the collapsing matter concentrates in the center of the forming halo, where a shock wave builds up and transports excess material to the outskirts of the halo. As a result of this process, a nearly isothermal halo envelope forms.\par

\begin{figure} [!tb]		
	{\includegraphics[width=1.0\columnwidth]{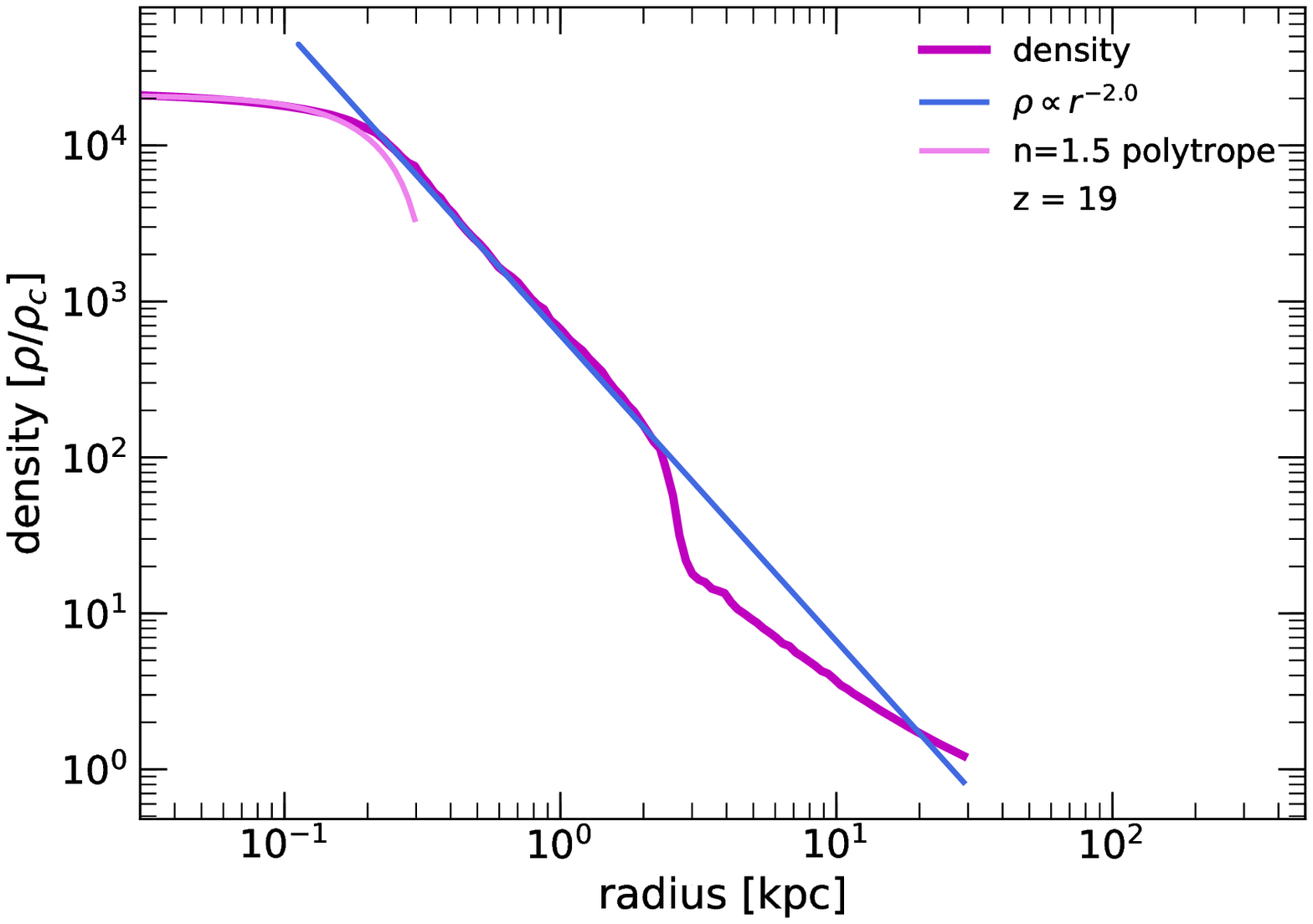}}
	{\includegraphics[width=1.0\columnwidth]{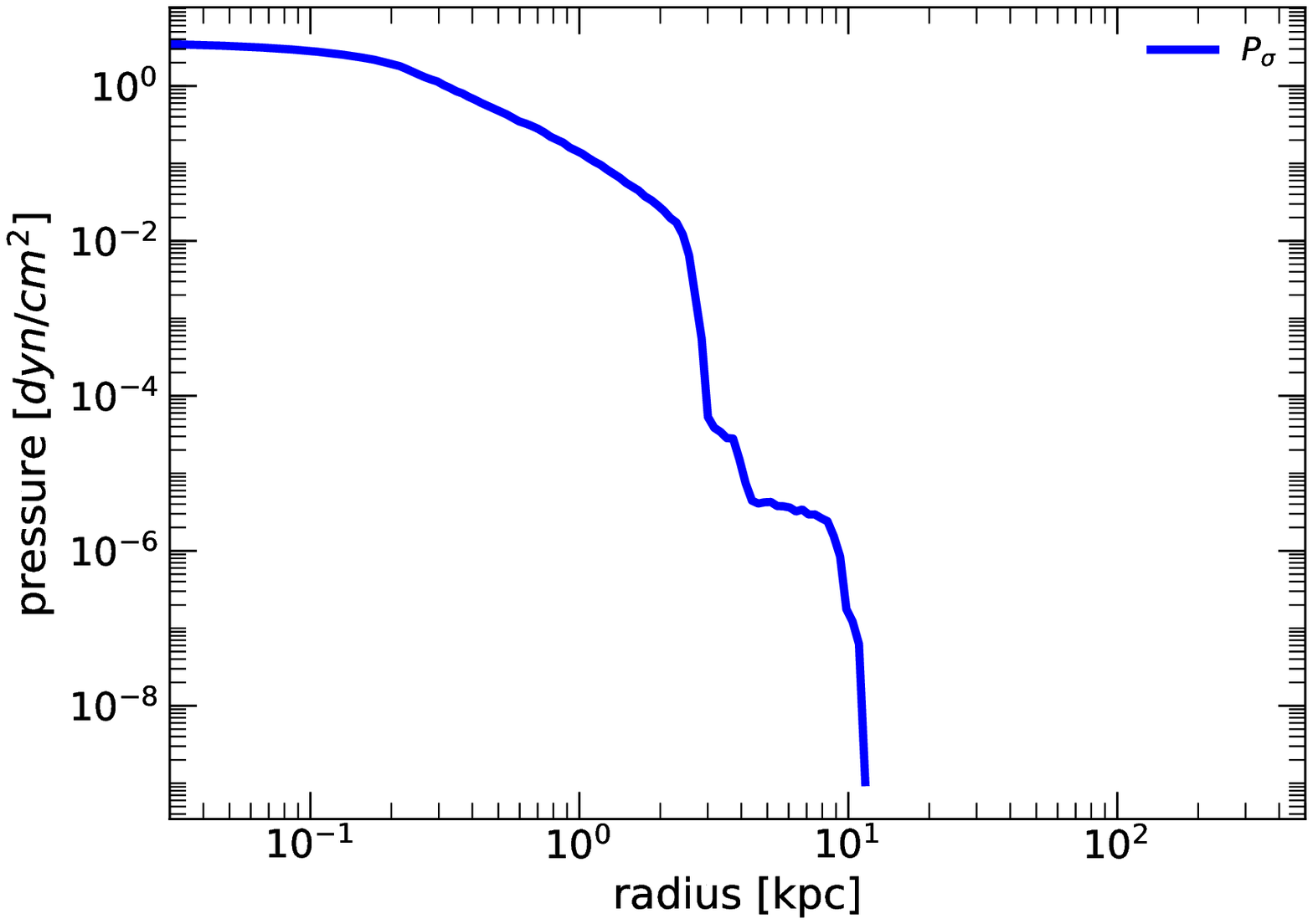}}
	\caption[shock front]
	{\textbf{Shock front in CDM halo formation}. The top panel displays the density profile during the formation of a CDM halo, shown at $z \sim 19$. The shock front is located between $\sim 2$-$3$~kpc. We see that immediately with the outward moving shock front an isothermal envelope establishes with a slope of $-2$; the shock is ``isothermal''. The inner region has the shape of a ($n=1.5$)-polytrope. The bottom panel displays the corresponding pressure profile of the halo at the same redshift. For CDM, \Psigma{} is the only pressure contribution.   Outside the shock front, \Psigma{} displays a steep falloff, as there is no significant velocity dispersion left.
	}
	\label{fig:shock-front}
\end{figure}

It is interesting to see that this isothermal structure forms immediately when the shock front moves outward, starting at $z \sim 25$, and is not a result of a relaxation following the formation process. So, we now take a look at the properties of the shock wave as it moves outward during the formation of the halo, as depicted in Fig.~\ref{fig:shock-front}.\par
The well-known Rankine-Hugoniot conditions (\citet{Rankine1870} and \citet{Hugoniot1998}) in \eqref{eq:RH-condition} relate hydrodynamic and thermodynamic quantities in front (subscripts $0$) and behind a shock front (subscripts $1$):
\begin{subequations}\label{eq:RH-condition}
	\begin{empheq}{align}
		\frac{\rho_1}{\rho_0} = \frac{\gamma + 1}{\gamma - 1} \label{eq:eq:RH-condition-d}\\
		\frac{T_1}{T_0} = \frac{\gamma - 1}{\gamma + 1} \frac{P_1}{P_0}, \label{eq:eq:RH-condition-p}
	\end{empheq}
\end{subequations}
where we use the general notation $P$ for the pressure and $T$ for the temperature. For CDM, $P$ corresponds to \Psigma{}, in fact. The adopted fluid approximation demands $\gamma = 5/3$ for CDM, and the same is true for SFDM-TF, or SIDM, concerning this effective pressure contribution from velocity dispersion; see Sec~\ref{sec:SFDMhydro}. By \eqref{eq:eq:RH-condition-d}, this implies that the passing of the shock front leads to an increase in density by a factor of $4$ across the shock.\par
However, the derivation of the Rankine-Hugoniot conditions assumes a shock front of zero thickness, which is clearly a simplification of reality. Indeed, in simulations such shock fronts have always finite size, and we can also see this in our simulations. In the example of Fig.~\ref{fig:shock-front} (top panel), the shock front extents to almost $1$~kpc. As mentioned already, we recognize that an isothermal halo envelope builds up, as soon as the shock front has moved through the forming halo. Hence, we can assume that $T_1 = T_0$.
Indeed, the actual shock front is close to an ``isothermal shock'', corresponding rather to the limit case $\gamma = 1$, 
as seen in the bottom panel of Fig.~\ref{fig:shock-front}, where the pressure across the shock changes by a large factor of $\sim 400$. \par

This means that the envelope is left in an isothermal state as the shock wave moves outward, prior to going through a phase of virialization. This confirms the results found by \citet{Dawoodbhoy2021} (see their Fig.~4) and \CPaperSDR{} (see their Fig.~2). This is remarkable, as they use a 1D Lagrangian code with artificial viscosity to handle the shock waves, in contrast to our 3D simulations which apply the Godunov method based on the Riemann problem. So, we have two different methodologies to model shocks which yield compatible results. Although artificial viscosity leads to a  ``smearing out'' of the shock front in the 1D simulations, this effect is compensated by the high spatial resolution attained in 1D. As a result, the spatial resolution of the shock front is comparable to the one gained from the AMR mechanism in our 3D simulations. However, we stress that the inner region of the halo is better spatially resolved in the 1D simulations.\par
\begin{figure} [!htb]		
	{\includegraphics[width=0.49\columnwidth]{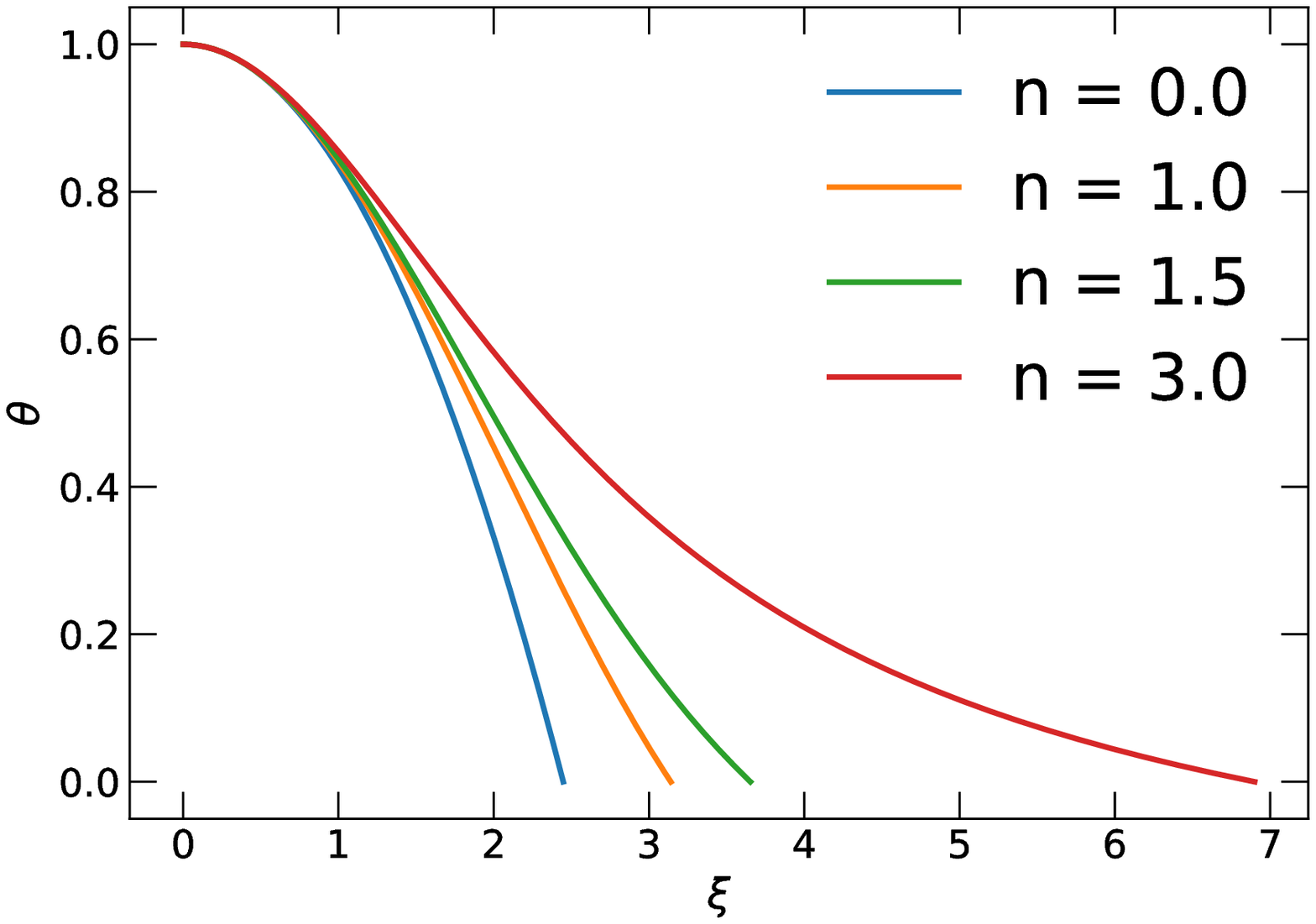}}
	{\includegraphics[width=0.49\columnwidth]{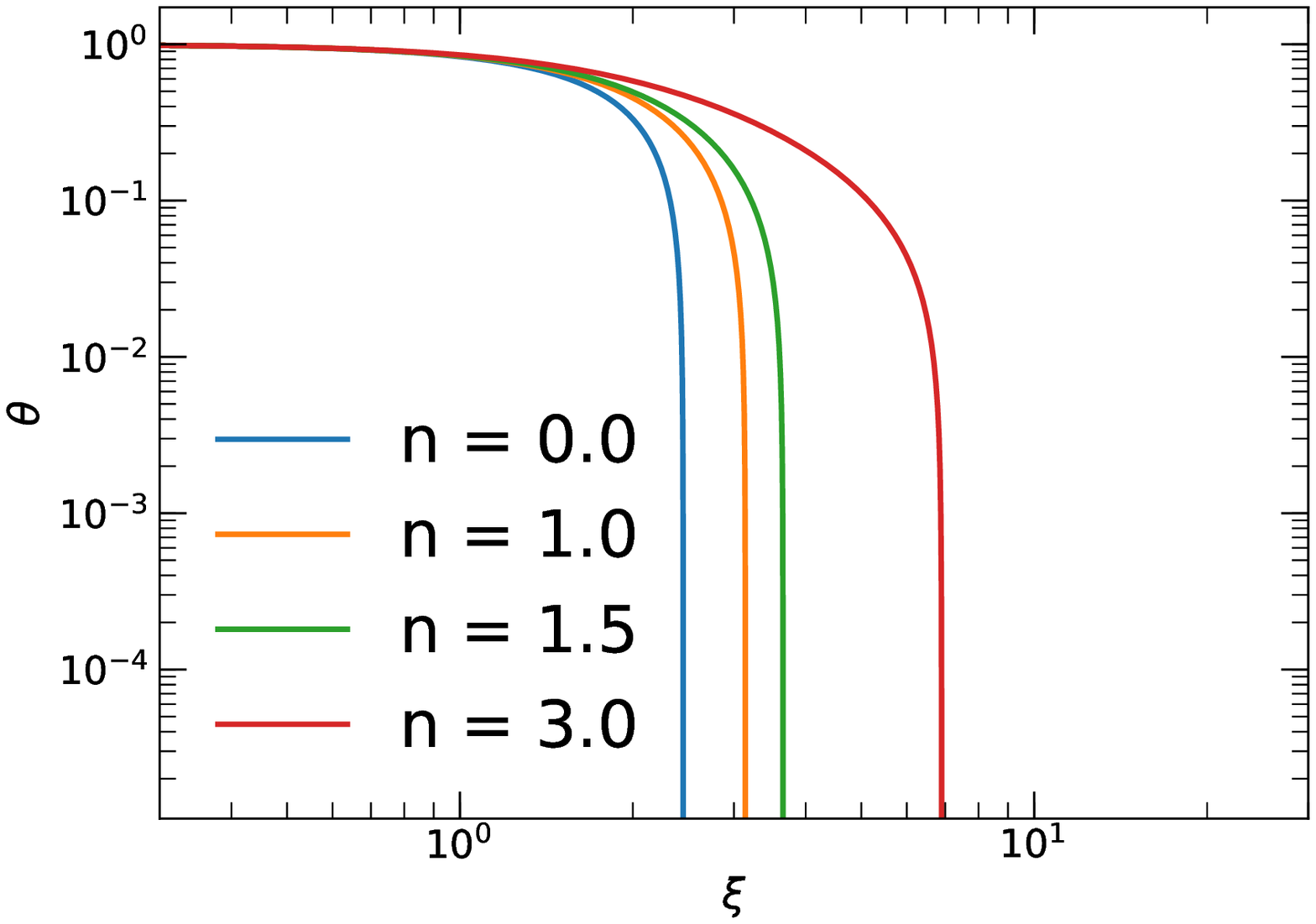}}
	\caption[Lane-Emden polytropes]
	{\textbf{Density profiles of Lane-Emden polytropes}. The left-hand panel displays polytropes in linear scale; the right-hand panel displays polytropes on log-log scale. Both panels display exemplary solutions to the Lane-Emden equation, with indices indicated in the legends. The $x$-axes display the dimensionless radius $\xi$ and the $y$-axes the dimensionless density $\theta$, with the same central density for each polytrope. The radius of the polytrope grows with increasing $n$.
	}
	\label{fig:polytropes}
\end{figure}

Moreover, we recognize in Fig.~\ref{fig:shock-front} (top panel) the formation of a core which is close to a ($n=1.5$)-polytrope. For this and later comparisons, we plot polytropes of different index $n$ in Fig.~\ref{fig:polytropes}, as an illustration.\par
More precisely, in the early stages of CDM halo formation ($z \sim 19$), the inner region is close to a ($n=1.5$)-polytrope (corresponding to $\gamma=5/3$), which balances the gravity of the accumulated matter that falls onto the halo. At $z \approx 7.5$, \Psigma{} is not sufficient to balance that gravity anymore and mass continues to pile up in the central region, steepening the slope of the density profile. Eventually, a cusp-like feature emerges similar to NFW profiles and expected for CDM halos, which we discuss in Sec.~\ref{sec:resultCDM}. 

We also show the corresponding pressure profiles \Psigma{} within the CDM halo in the bottom panels of Fig.~\ref{fig:shock-front} at $z \sim 19$, and in Fig.\ref{fig:formation-var4} at $z \sim 7.5$, respectively. For CDM, \Psigma{} is the only pressure contribution. At both redshifts, we see a steep falloff of this pressure outside the respective shock front, as a result of the corresponding drop in the velocity dispersion.

To put these results into perspective, we note already here that in the formation of a SFDM-TF halo, the additional pressure component \PSI{} related to $\gamma = 2$ plays an important role and changes the outcome as follows. \PSI{} builds up comparatively more slowly, but eventually dominates over \Psigma{} in the later stages of the evolution (see Sec.~\ref{sec:evolpressure}). As a result, this pressure component is able to balance gravity through the entire evolution of the halo, such that a halo core forms not only temporarily but remains in place and finally acquires a shape close to a ($n=1$)-polytrope (corresponding to $\gamma = 2$), in contrast to the cusp in CDM halos.
We discuss these findings in Sec.~\ref{sec:simsRAMSES}.\par
%

\subsection{CDM halo profiles}\label{sec:resultCDM}

\begin{figure} [!htb]		
	{\includegraphics[width=1.0\columnwidth]{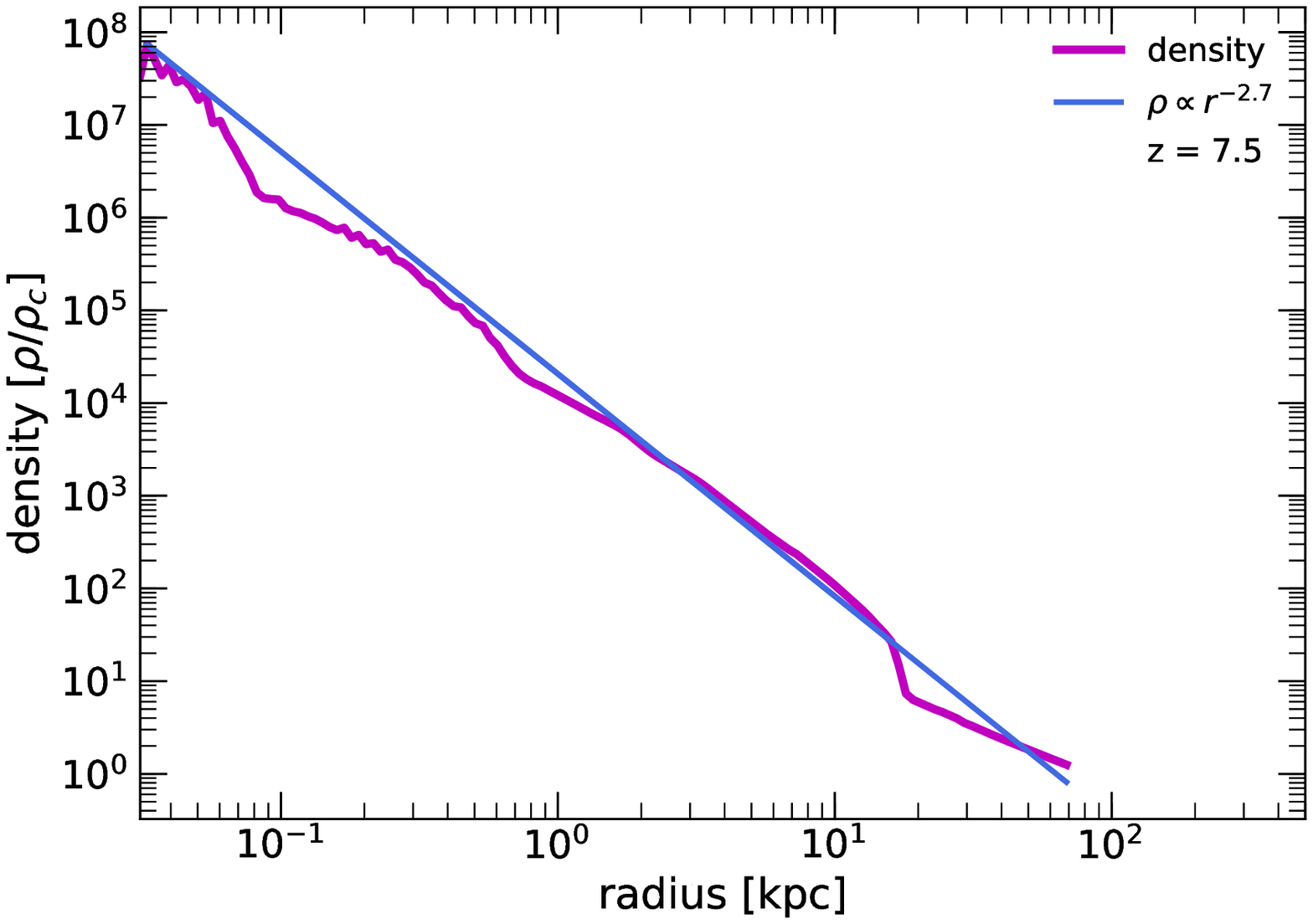}}
	\includegraphics[width=1.0\columnwidth]{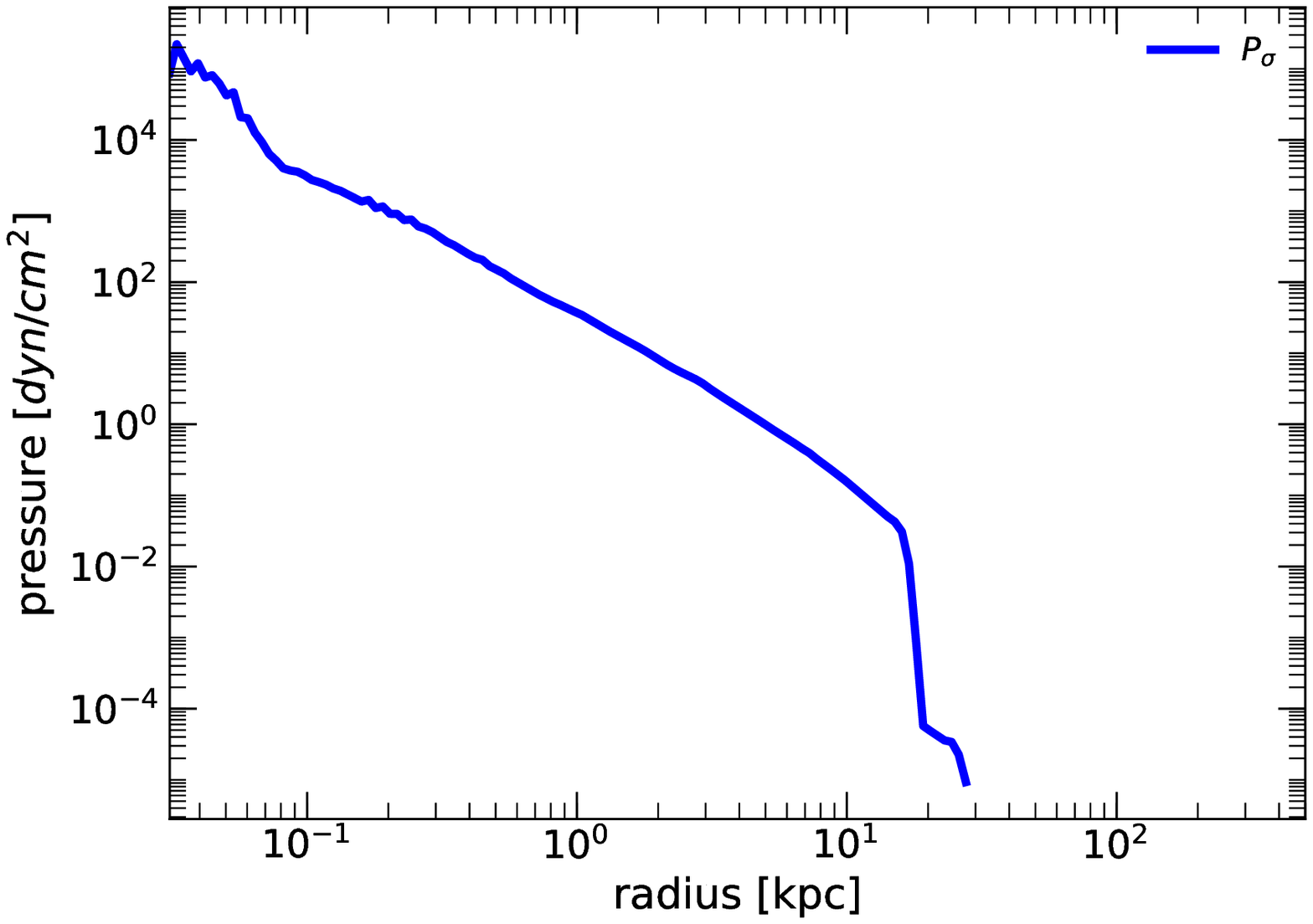}	
	\caption[Formation of a CDM Halo]
	{\textbf{Formation of a CDM halo of mass $10^{12} $~M$_{\odot}$}. Top panel: density profile at the formation time ($z \sim 7.5$). Toward the center the density profile develops a steep slope, i.e. a cusp, which is a characteristic feature of CDM halos. In the outskirts, the profile transforms to that of a nearly NFW-like envelope, with a slope of $\rho \propto r^{-2.7}$; its boundary is set by the outward moving shock front. Bottom panel: pressure profile \Psigma{} for the same halo at the same redshift as in the top panel.
	}
	\label{fig:formation-var4}
\end{figure}

\begin{figure} [!htb]%
	{\includegraphics[width=1.0\columnwidth]{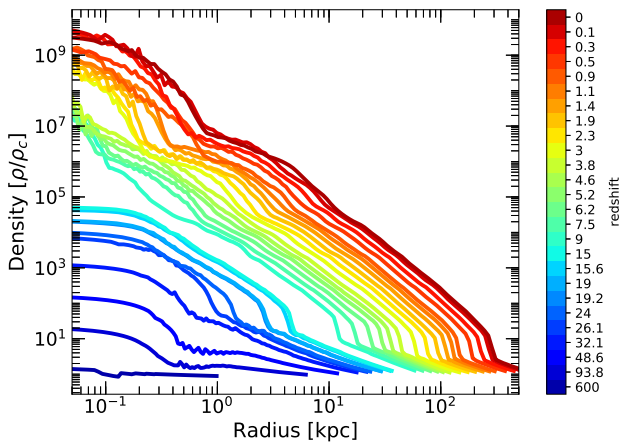}}
	\caption[Evolution CDM Halo]
	{\textbf{Evolution of a CDM halo of mass $10^{12} $~M$_{\odot}$.} The color-coded solid lines display the evolution of the halo density profile, beginning with the collapse of the initial ($\epsilon=1/6$) profile all the way to the present at $z=0$. We can see a constantly rising density in the central region. At $z \sim 25$, the shock wave begins to form and moves outward. After the halo formation time at $z\sim 7.5$, the inner ($n=1.5$)-polytropic core of the halo transitions to a cusp, steepening to $\rho \propto r^{-2.7}$. As time progresses, the halo keeps growing in mass by accreting further material from the surroundings. By $z=0$ the inner profile has flattened to $\rho \propto r^{-1.15}$, i.e. very similar to a NFW cusp of $r^{-1}$. 
	}%
	\label{fig:evol-var4}%
\end{figure}

In this subsection, we draw our attention to the density and pressure profiles as the CDM halo forms and evolves up to the present. In Fig.~\ref{fig:formation-var4}, we show a halo of mass $10^{12} $~M$_{\odot}$ at its formation time $z\sim 7.5$ (this is the same halo as shown in Fig.~\ref{fig:shock-front} at an earlier redshift). At this formation time, we can see that the earlier inner ($n=1.5$)-polytropic core -- which we see in Fig.~\ref{fig:shock-front} -- has transitioned into a steep cusp, because of the continuous pileup of matter in the center. Its slope is close to $\rho \propto r^{-2.7}$ (see also Fig.~\ref{fig:evol-var4} for this same snapshot in redshift), and is just as steep as the outer slope of the envelope! That outer slope is already close to the characteristic NFW behavior of $\rho \propto r^{-3}$.  

In the bottom panel of Fig.~\ref{fig:formation-var4}, the corresponding pressure profile \Psigma{} of the same halo at the same redshift is shown.  
We stress again that the pressure drops off steeply, as a result of the shock front, as outside of it there is no significant velocity dispersion left. In fact, this finding has been also reported for CDM halos in the fluid approximation by \citet{Ahn2005}.

As time progresses, the central region of the CDM halo makes another transition back to a shallower profile, which can be best seen in Fig.~\ref{fig:evol-var4}, which shows the evolution of the halo density profiles all the way down to $z=0$. As of $z \sim 0.9$, we see a flattening of the central profile, and at $z=0$ the central profile goes, indeed, like $\rho \propto r^{-1.15}$, thus very similar to the characteristic NFW cusp of $\rho \propto r^{-1}$. 
In fact, this transition from the steeper cusp to the shallower $\sim r^{-1}$ cusp can be also inferred by looking at Fig.~1 of \CPaperSDR{}: at the halo formation time ($a=a_f$; $a$ being the scale factor), the profile is steeper than at the later time of $5a_f$, also shown in that figure.
We should also stress that the ``would-be bump'' in the density in the inner region of the CDM halo is the result of a mere dynamical fluctuation, i.e. it is temporary and not a persistent feature.

Figure~\ref{fig:evol-var4} now clearly displays how the collapse of the initial ($\epsilon=1/6$) profile is followed by an increase in density in the center with time, establishing a cuspy profile. Once the density and pressure are high enough to withstand the pressure of the infalling matter, it is reflected and a shock wave forms, which moves outward and transports material back into the outskirts of the halo. This shock front then separates the halo from the surrounding environment, providing a proxy for the virial radius of the halo. At the final snapshot $z=0$, its location is around $300$~kpc, i.e. a reasonable number for the virial radius of this halo of $10^{12} $~M$_{\odot}$.

Overall, our findings confirm previous results by \citet{Ahn2005}, \citet{Dawoodbhoy2021} and \CPaperSDR{} for the CDM regime, in terms of halo structure. This way, our 3D fluid simulations of CDM in RAMSES also confirm the usefulness and robustness of previous results obtained for 1D. Finally, our fluid approximation reveals present-day CDM halo density profiles which are in good accordance with the analytic NFW profile, originally devised from fits to CDM N-body simulation data.  \par
%

%

\section{Formation and evolution of SFDM-TF halos}\label{sec:simsRAMSES}

\subsection{Comparing the two fluid approximations}\label{sec:resultV1V2}

In the first step of our SFDM-TF halo simulations, we compare both sets of fluid approximations, the 3D version of Eqs.~\eqref{eq:sfdmSDR} from \CPaperSDR{} versus Eqs.~\eqref{eq:sfdmHEA} from \CPaperHWM{}. In the forthcoming, we call the simulations based on \CPaperSDR{} as ``Var1'' (``variant 1''). Remember that in these equations, the pressure \PSI{} appears only in the momentum equation, but not in the energy equation which deals solely with \Psigma{}. Also, the source terms from self-gravity of the fluid are ignored. On the other hand, we call the simulations based on \CPaperHWM{} as ``Var2'' (``variant 2''). Here, \PSI{} contributes to the momentum equation and to the energy equation. Moreover, the source terms due to gravity are not neglected.

We followed the procedure described in \CPaperSDR{} and simulated the collapse of a single SFDM-TF halo with an initial, spherically symmetric ($\epsilon=1/6$) profile, placed in the center of the box. However, whereas \CPaperSDR{} established their cosmological environment in their 1D simulations by adopting the universal CDM MAH found by  \citet{Wechsler2002}, our RAMSES simulations apply ICs according to the description in Sec.~\ref{sec:ICsBox}; the cosmological environment (i.e. the expanding background universe, etc.) is handled by RAMSES in a standard way.  
On the other hand, the RAMSES simulations by \CPaperHWM{} were not initialized from a spherical infall model, but their ICs were generated  with MUSIC. As a result, when \CPaperHWM{} analyzed their halos at their final snapshot of $z=0.5$, these halos have undergone minor and major mergers.

\begin{figure*} [!htb]		
	{\includegraphics[width=1.0\columnwidth]{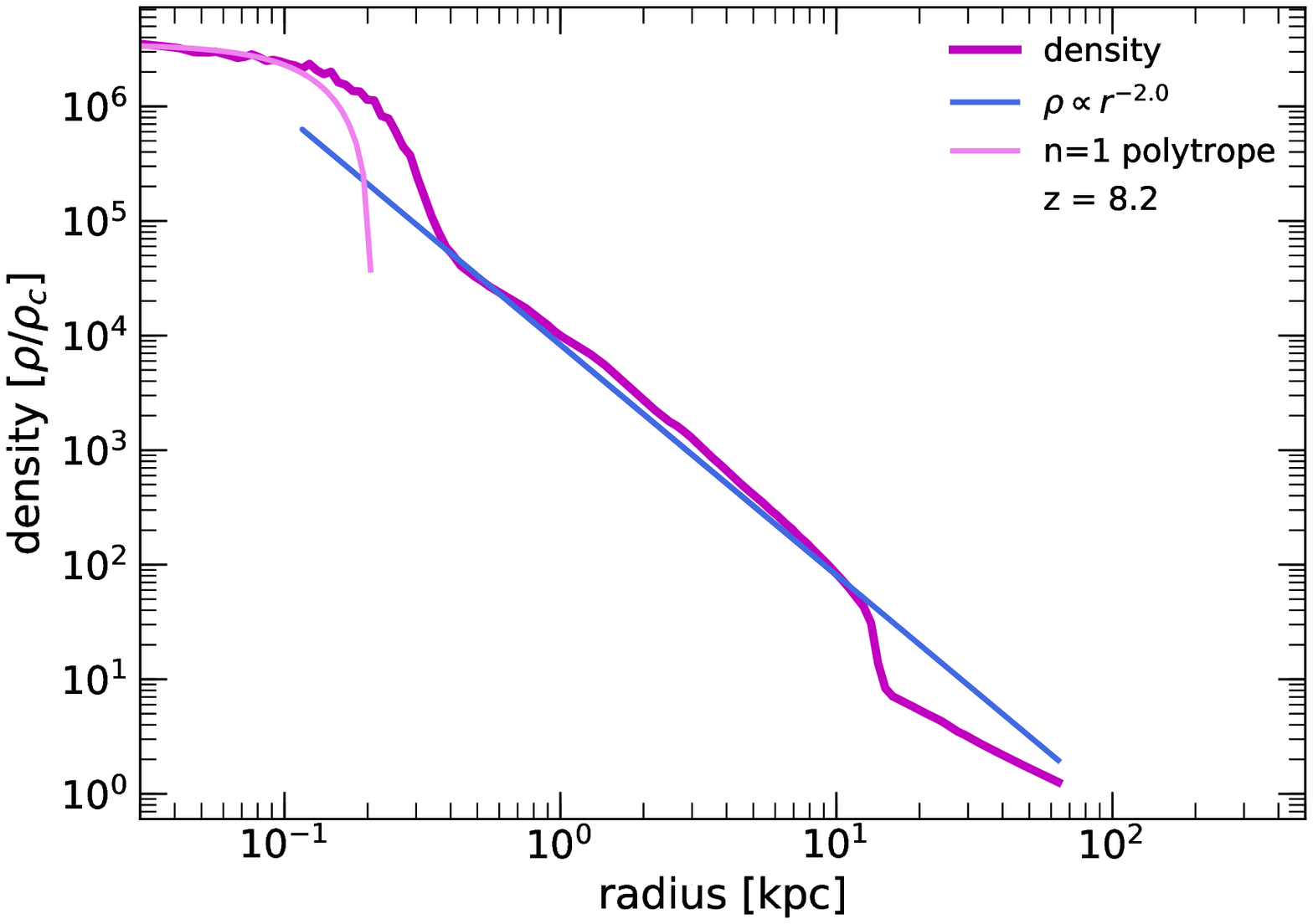}}
	\includegraphics[width=1.0\columnwidth]{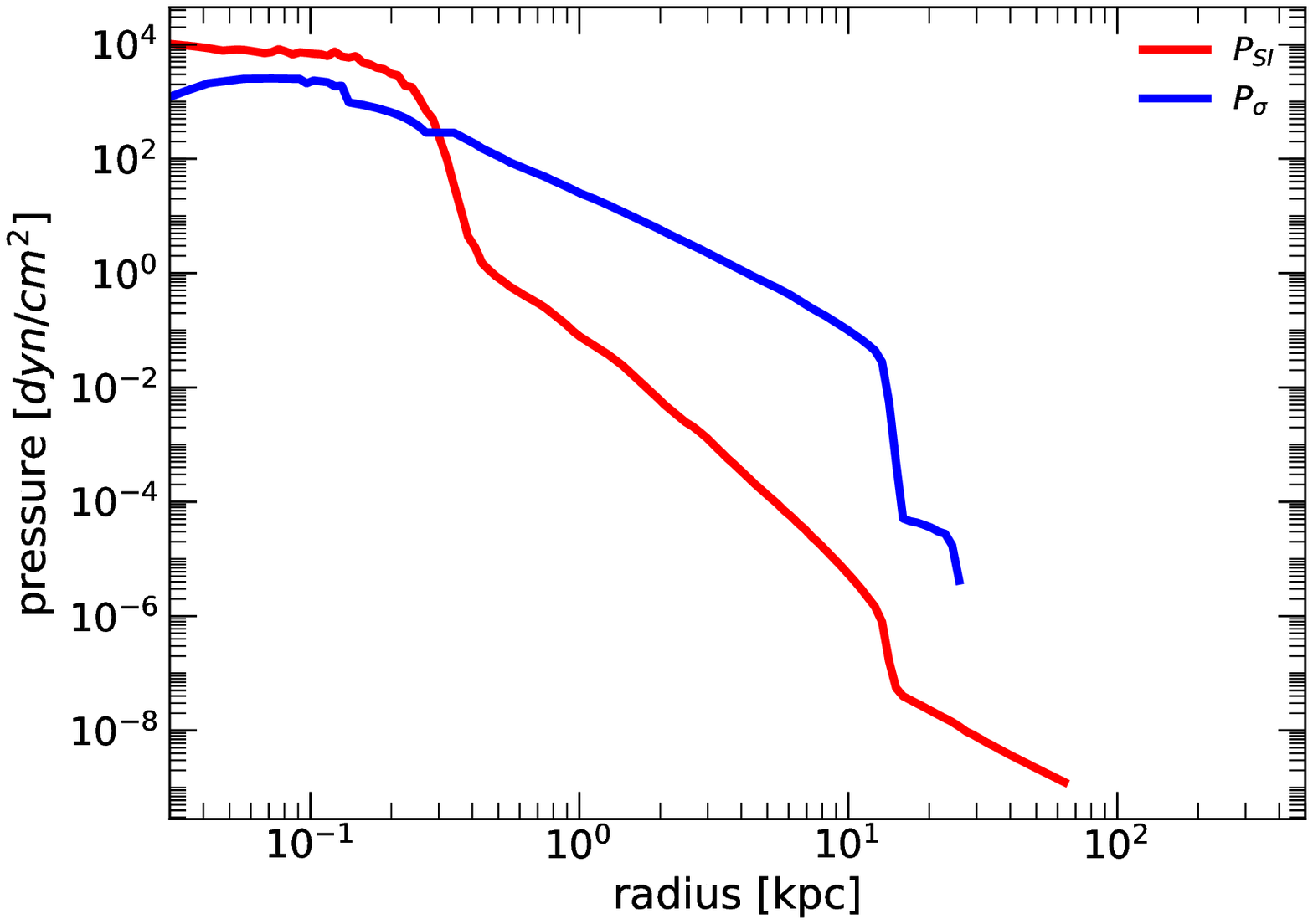}	
	\caption[Formation of a SFDM Halo Var1]
	{\textbf{Formation of a SFDM-TF halo of mass $10^{12} $~M$_{\odot}$ using Var1}. The left-hand panel displays the density profile of the halo at its formation time ($z \sim 8$), using the 3D version of the fluid approximation of \CPaperSDR{}; we call it Var1. The central profile is similar to a $(n=1)$-polytropic core. At the edge of this core occurs a transition to a nearly isothermal profile, which is terminated by the outward moving shock front, formed by the external pressure of the accreting DM and background pressure. The right-hand panel displays the two pressure contributions, \PSI{} (red solid line) and \Psigma{} (blue solid line), again at the formation time. We can see that \PSI{} dominates in the core by a factor of $\sim 5$, whereas \Psigma{} dominates in the envelope by $\sim 3$ orders of magnitude. At the edge of the core, the pressures are nearly equal. The central fluctuations in the density and pressures are a numerical artifact, caused by limited spatial resolution in the central region in the AMR mechanism of RAMSES. As a result, at this redshift the core size exceeds \RTF{} by a factor of $\sim 3$. (However, this ``core expansion'' must not be confused with the expansion of halos, and their cores described in Sec.~\ref{sec:resultEVOL}.)}
	\label{fig:formation-var1}
\end{figure*}

\begin{figure*} [!htb]		
	{\includegraphics[width=1.0\columnwidth]{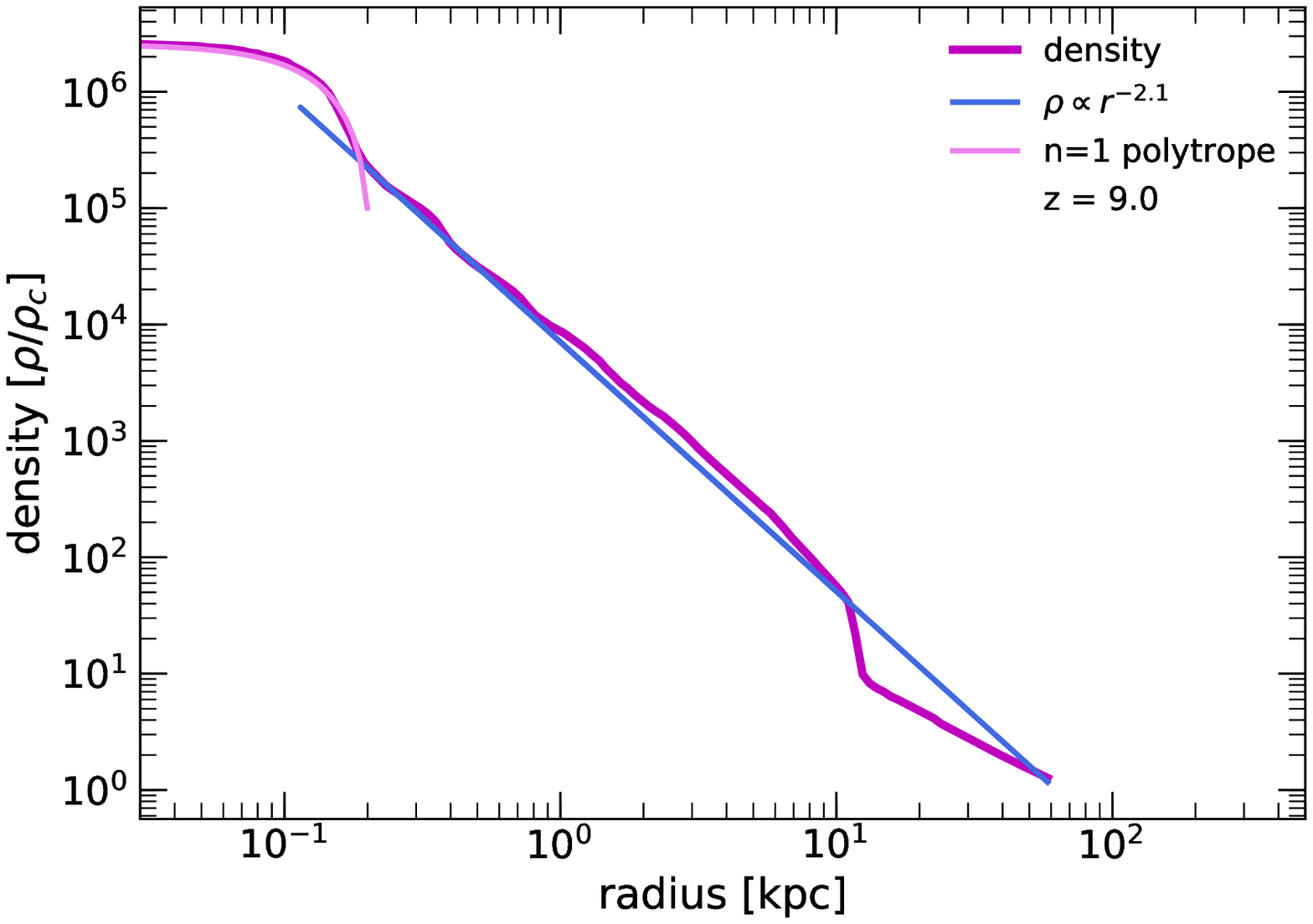}}
	\includegraphics[width=1.0\columnwidth]{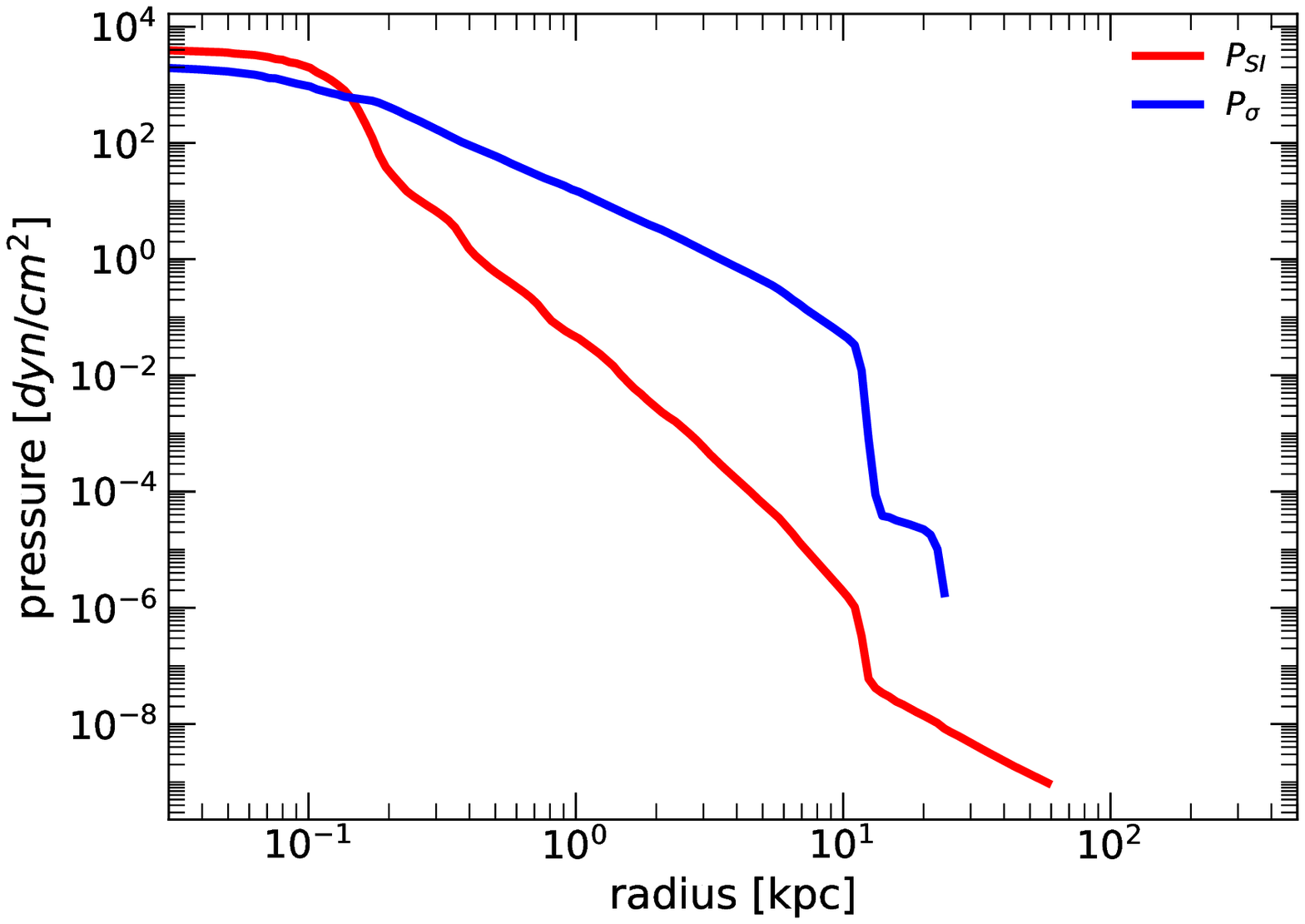}	
	\caption[Formation of a SFDM Halo of mass $10^{12} $~M$_{\odot}$ using Var2]
	{\textbf{Formation of a SFDM-TF halo of mass $10^{12} $~M$_{\odot}$ using Var2}. The left-hand panel displays the density profile of the halo at its formation time ($z \sim 9$), using the equations of \CPaperHWM{}; we call it Var2. We recognize a similar establishment of a core-envelope structure, with the same characteristics as in Fig.~\ref{fig:formation-var1} (see caption there). However, in contrast to the case of Var1, the core density is very close to a ($n=1$)-polytrope, with a radius close to \RTF{}, as well. This difference has to do with resolution issues in Var1, discussed in the main text. The right-hand panel displays the profiles of the pressures, \PSI{} (red solid line) and \Psigma{} (blue solid line), again at the formation time. We see the same features as in the Var1 run: \PSI{} dominates in the core by a factor of $\sim 3$, whereas \Psigma{} dominates in the envelope by $\sim 3$ orders of magnitude. The pressures are nearly equal at the edge of the core. Hence, we conclude that the use of the different set of fluid equations, Var1 vs Var2, is not the reason for the discrepancies between the results in \CPaperSDR{} vs those in \CPaperHWM{}, concerning the pressure profiles. 
	}
	\label{fig:formation-var2}
\end{figure*}

We show the results for the density and pressure profiles of our simulated halo of mass $10^{12} $~M$_{\odot}$, at its formation time of $z \sim 8$ in Fig.~\ref{fig:formation-var1} for Var1, and at $z \sim 9$ in Fig.~\ref{fig:formation-var2} for Var2, respectively. This difference in formation time amounts to less than $\sim 10$ \% of the cosmic time. We think it stems from the numerical resolution issues that arise in Var1, as discussed shortly.\par

The left-hand panel of Fig.~\ref{fig:formation-var1} displays the density profile of the SFDM-TF halo at formation time, which is also the time when the shock front has moved outward, separating the almost isothermal envelope of the halo from that region, where infall of DM onto the newly formed halo is taking place. Toward the center, the density profile develops a close resemblance to a ($n=1$)-polytrope. The right-hand panel displays both pressure contributions: the red solid line shows \PSI{} and the blue solid line \Psigma{}. Within the core, \PSI{} dominates over \Psigma{}, whereas at the edge of the core both pressures are nearly equal. The fluctuations of \Psigma{} in the center of the core are artificial and caused by the AMR mechanism, in combination with the chosen spatial resolution of the simulation. As a result of this limited resolution, the core size exceeds the expected \RTF{} predicted by the ($n=1$)-polytrope. We tolerated this behavior in our comparison, because our main interest was to clarify the characteristics of the pressure contributions. Otherwise, the CPU requirements would have been too demanding, an issue that was also pointed out in \CPaperHWM{}. 

The results of Fig.~\ref{fig:formation-var1} using Var1 basically confirm the earlier finding in 1D presented by \CPaperSDR{}, concerning the core-envelope structure and the dominance of \PSI{} in the core.   \par
Figure~\ref{fig:formation-var2} displays the results of Var2 at the formation time of the SFDM-TF halo. Remember, in this variant \PSI{} contributes to both momentum equation and energy equation. Again, the left-hand panel displays the density profile. In contrast to Var1, we see less fluctuations in the central region. This difference in the resolution between Var1 and Var2 is caused by the fact that the modified gradients in the energy equation of Var2 force the AMR mechanism to create a higher spatial resolution in the early stage of halo formation; this effect is also noticeable by the significantly increased CPU resources for Var2. In other words, the spatial (and time) resolution in Var2 is higher than in Var1, because the additional gradient of the former -- thanks to the additional pressure term in the energy equation --  leads to an enforced better resolution in the AMR mechanism.
As a result, the artificial fluctuations in the density as well as in the pressures go away when Var2 is used. Therefore, the core size is also smaller than its counterpart in Var1, and the core density now follows very closely a ($n=1$)-polytrope, in terms of slope and radius \RTF{}. On the other hand, the envelope displays the same nearly isothermal profile, ending at the shock front, which isolates the halo from the infalling background matter. However, due to the numerical fluctuations in Var1, the pressures stabilize at a later time for this case, which we think explains the small difference in the formation time/redshift of the halos between Var1 and Var2.\par
Now interestingly, in the right-hand panel of Fig.~\ref{fig:formation-var2}, we can see that \PSI{} also dominates over \Psigma{} within the core and out to the edge of it, using Var2. Therefore, we conclude that the use of the different set of fluid equations, Var1 vs Var2, is not the reason for the discrepancies between the results in \CPaperSDR{} vs those reported in \CPaperHWM{} concerning the overall run of the pressure profiles. We will get back to this point in the next subsections.
\par
In Sec.~\ref{sec:shockwave} we discussed the properties of shock waves in the formation of a CDM halo. In the early stages of the formation of the CDM halo, a similar core-envelope structure, as seen in the formation of the SFDM-TF halos in Figs.~\ref{fig:formation-var1} and \ref{fig:formation-var2}, is present in the CDM halo; see Fig.~\ref{fig:shock-front}. However, while CDM halos eventually develop a central cusp, SFDM-TF halos preserve their core-envelope structure over time. Here, \PSI{} balances gravity, as \PSI{} dominates over \Psigma{} in the core, during the late phases of the evolution of the SFDM-TF halo, which is in accordance with \CPaperSDR{}. On the other hand, \Psigma{} dominates over \PSI{} in the envelope, which is also consistent with \CPaperSDR{}. The findings of Sec.~\ref{sec:shockwave} can be applied to SFDM-TF halos, as well, but have to be seen in the context of the mutual effect of the two contributions to the pressure, \Psigma{} and \PSI{}. In the very early stages of halo collapse, \Psigma{} dominates while the polytropic core is about to build up. Once this process is completed, the shock front forms and begins to move outward, which can be seen in Fig.~\ref{fig:evol-var1-2}, where the shock front appears at $z \sim 20$ (see also Sec.~\ref{sec:evolpressure}).\par
In contrast, \CPaperHWM{} report a similar resulting core-envelope halo structure, but they find that the cores, too, are dominated by \Psigma{} by the time of their final shapshots. The authors interpret this difference to \CPaperSDR{} as a consequence of mixing, which leads to a dynamical heating of the core as the shock-heated outer layers mix with the core.
Since the spherically symmetric 1D simulations of \CPaperSDR{} are blind to such mixing effects, it can explain the discrepancies.\par
However, we have seen that our 3D simulations produce similar results between Var1 and Var2, thus mixing may not be the final explanation for the differing results. We will discuss this in more detail in the next subsections, but we should stress that, unfortunately, the y-axis of Fig.~5 in \CPaperHWM{}, which is comparable to our Figs.~\ref{fig:formation-var1} and \ref{fig:formation-var2}, does not extend all the way down to lower numbers, in order to reveal the location of the shock front at larger halo-centric distance. At least in the late stages of halo evolution, there should be a shock front, and these outer regions are too far away from the core in order to mix with the matter in the core.
In fact, as we will see shortly, we attribute the difference in the results between \CPaperHWM{} vs \CPaperSDR{} and ours here to the limited simulation run-time of \CPaperHWM{}, i.e. they start their simulations much too late in terms of $z_{ini}$. Also, they analyze their halos at a snapshot quite earlier than $z=0$.\par

\subsection{Evolution of SFDM-TF halos}\label{sec:resultEVOL}

In Sec.~\ref{sec:differences} we already noted the differences in the simulation setup of \CPaperSDR{} vs \CPaperHWM{}, where we emphasized that \CPaperSDR{} analyzed the structure of their SFDM-TF halos at the formation time, whereas \CPaperHWM{} analyzed their halos at a final snapshot at redshift $z=0.5$. 

\begin{figure} [!htb]		
	{\includegraphics[width=1.0\columnwidth]{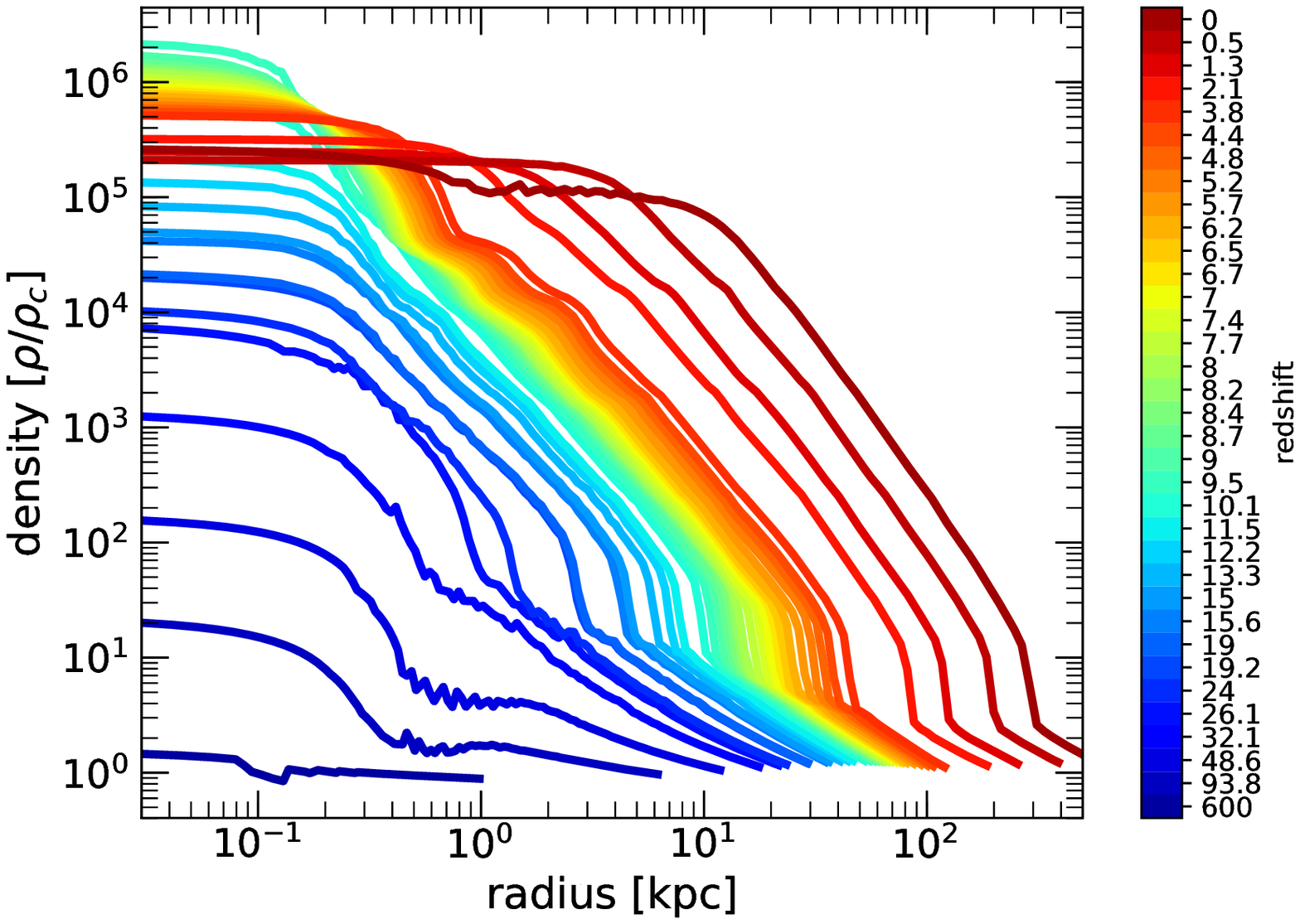}}
	{\includegraphics[width=1.0\columnwidth]{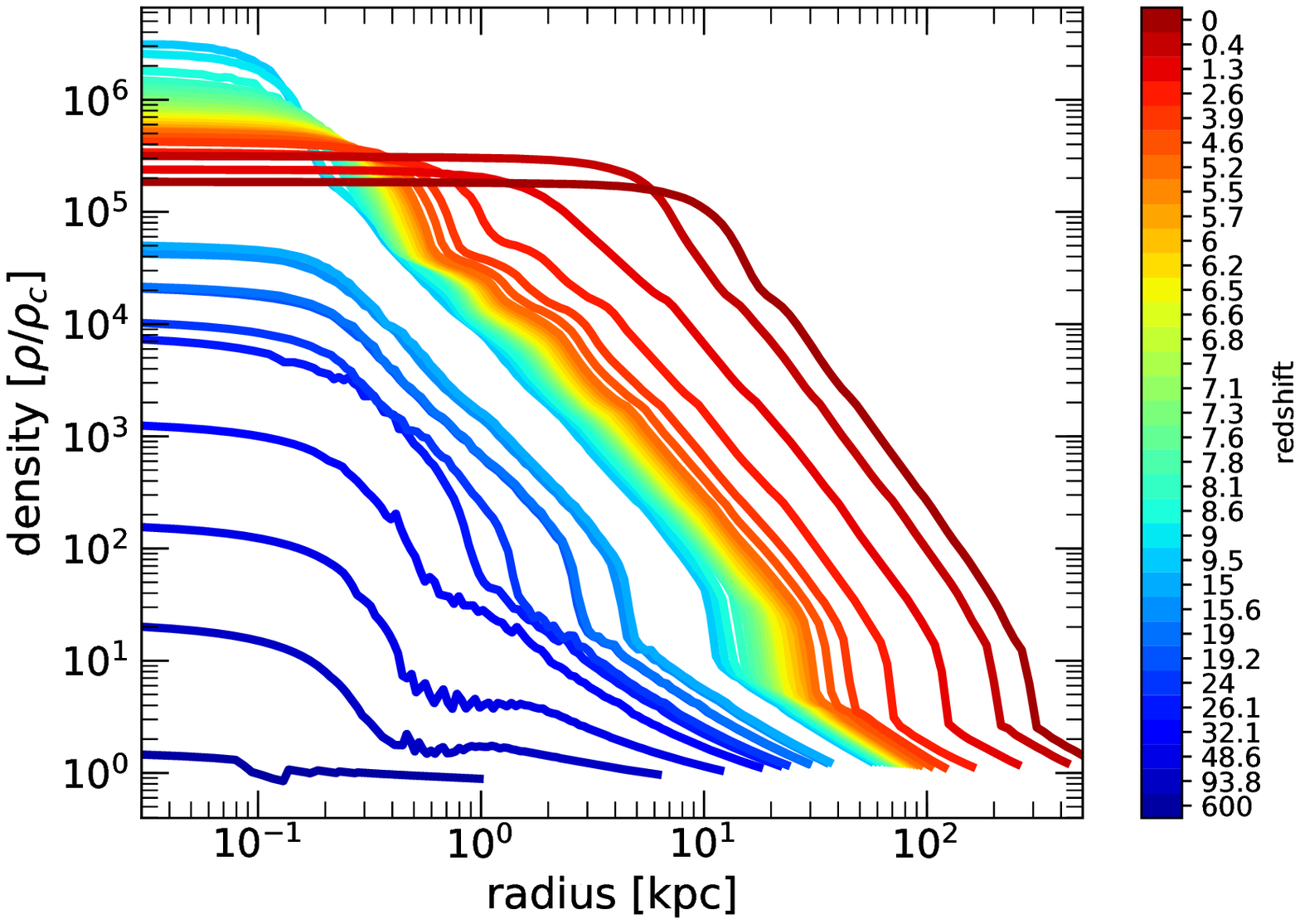}}
	\caption[Evolution of SFDM Halos]
	{\textbf{Evolution of a SFDM-TF halo of mass $10^{12} $~M$_{\odot}$ from $z_{ini}=600$ to $z=0$.} The color-coded solid lines display the evolution of the density profile, beginning with the collapse of the initial ($\epsilon=1/6$) profile down to $z=0$ within a $4$~Mpc simulation box (top panel: Var1; bottom panel: Var2). The halo density continuously grows in the center and at $z \sim 24$ the shock wave forms and begins to move outward, establishing a clear core-envelope structure. The core is close to a ($n=1$)-polytrope at the formation time ($z \sim 8$ in the top panel, $z \sim 9$ in the bottom panel), surrounded by a nearly isothermal envelope. After the formation of the halo, the halo keeps growing in mass and size. Eventually, the core expands and flattens out by $z \sim 4$, losing its polytropic shape. There are no substantial differences in the results between Var1 and Var2.
	}
	\label{fig:evol-var1-2}
\end{figure}

In the last subsection, we focused on the density and pressure profiles of our simulated halos, which resulted from different fluid approximations, Var1 and Var2, respectively, at their formation time. Now, we show the overall evolution of these halos, from the initial collapse redshift of $z_{ini}=600$ all the way to redshift $z=0$. We display the results in Fig.~\ref{fig:evol-var1-2}.\par
Comparing our results for Var1 and Var2, we see no substantial differences between them. We can see that our forming halos conform to the results of \CPaperSDR{}. Moreover, we can see the transformation from the $(n=1.5)$-polytropic core, in the early stages of the evolution, to the final $(n=1)$-polytropic core, beginning at $z \sim 20$. After some period of almost constant size, the polytropic core flattens out and both core and envelope expand. We interpret this effect to be a consequence of the expansion of the background universe, as follows. It is a well-known fact that the size of an isothermal sphere is determined by the surrounding pressure. This external pressure has two contributions: the density of the background universe and the pressure of the infalling matter. As the density of the background universe decreases with its expansion, the envelope grows and consecutively the core also expands. We will put this result into perspective in Sec.~\ref{sec:summary}.
Furthermore, we can see that the density at the outer edge of the shock front steadily decreases from $\rho/\rho_c \sim 10^2$, beginning with the formation of the shock front at $z \sim 24$, to $\rho/\rho_c \sim 3$ at $z \sim 4$. Subsequently, the density at the outer edge of the shock front remains constant. \par
\begin{figure} [!htb]		
	{\includegraphics[width=1.0\columnwidth]{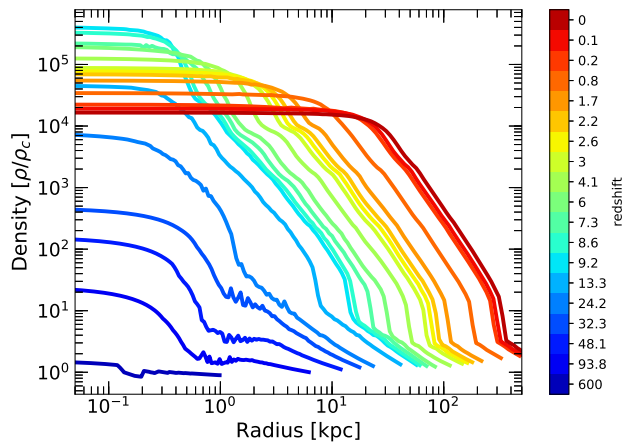}}
	\caption[Evolution of a SFDM Halo Var2 box size 12 Mpc]
	{\textbf{Evolution of a SFDM-TF halo of mass $10^{12} $~M$_{\odot}$ from $z_{ini}=600$ to $z=0$, using Var2 and a 12 Mpc simulation box.} 
		The color-coded solid lines display the evolution of the density profile, beginning with the collapse of the initial ($\epsilon=1/6$) profile down to $z=0$ in a $12$~Mpc box. The overall evolution of the density follows the same pattern as in the case shown in Fig.\ref{fig:evol-var1-2} (see caption there). Likewise, as seen in Fig.~\ref{fig:evol-var1-2}, we recognize an expanding envelope and core. But in contrast to the $4$~Mpc box, the transition seen here, from the polytropic core to the extended core, does not happen as abruptly as before. Instead, there is a smooth transition beginning at $z \sim 4$. 
	}
	\label{fig:evol-var2-12}
\end{figure}

Given the fact that only the density of the background universe seems to determine the size of the envelope, we think that there is no pressure originating from the infall of matter onto the halo anymore by and after the time of about $z \sim 4$. In contrast, the simulations of \CPaperSDR{} enforce their adopted MAH throughout their entire simulation, and they do not see an expansion of core nor envelope. Given this difference, we conclude that at the time when the halo begins to expand and the density at the outer edge of the shock front remains constant, there is not enough matter left in our simulation box, thus the infall of matter onto the halo ceases. In order to test this assumption, we repeat the simulation of the same SFDM-TF halo but in a $12$~Mpc simulation box with an accordingly increased spatial resolution. Also, we use Var2 for this simulation, because of its better overall resolution characteristics. The result is shown in Fig.~\ref{fig:evol-var2-12}.\par
We can see that the overall evolution of the halo density follows the same pattern as in the previous case with the smaller simulation box.
However, in contrast to the smaller box, where the density of the polytropic core decreases abruptly ($z \sim 7$), and envelope and core begin to expand, this transition proceeds more smoothly in the larger box. The reason is as follows: in the smaller box, the infall of matter stops more or less abruptly, which leads to a sudden expansion of the envelope, as a reaction to the decreased pressure and the expansion of the background universe. On the contrary, in the larger box the infall of matter diminishes more slowly, as the density decreases more slowly with the expansion of the background universe. It seems that by redshift $z \sim 4$ the infall of matter does not anymore dominate the ``external pressure'' exposed on the isothermal halo envelope. 
Finally, at $z=0$, the location of the shock front is at $\sim 300$~kpc, which is a reasonable value for the virial radius of the halo. A similar value was found in the CDM case in the previous section. In fact, this size of the halo agrees well with observations; e.g. \citet{Posti2019} found $R_{vir} = 287^{+22}_{-25}$~kpc for the Milky Way. 

We note that this whole evolutionary trend in the simulations remains the same, if we pick different initial redshifts of $z_{ini} = 3400;1000$. There are differences, however, if halo collapse is initiated at substantially later $z_{ini}$, as seen in Fig.~\ref{fig:evol-var2-z50} to be discussed later.\par
Now, we finally turn our attention to the simulation of a lower-mass halo, where we choose $10^{9}$~M$_{\odot}$ as a typical host halo for ultra-faint dwarf galaxies. These constitute the smallest galaxies known, and are DM-dominated systems.
We use again the fluid approximation of Var2 for its better resolution characteristics in the central region. Also, we pick the same value for $R_\text{{TF}} = 110$~pc, as before.

\begin{figure} [!htb]		
	{\includegraphics[width=1.0\columnwidth]{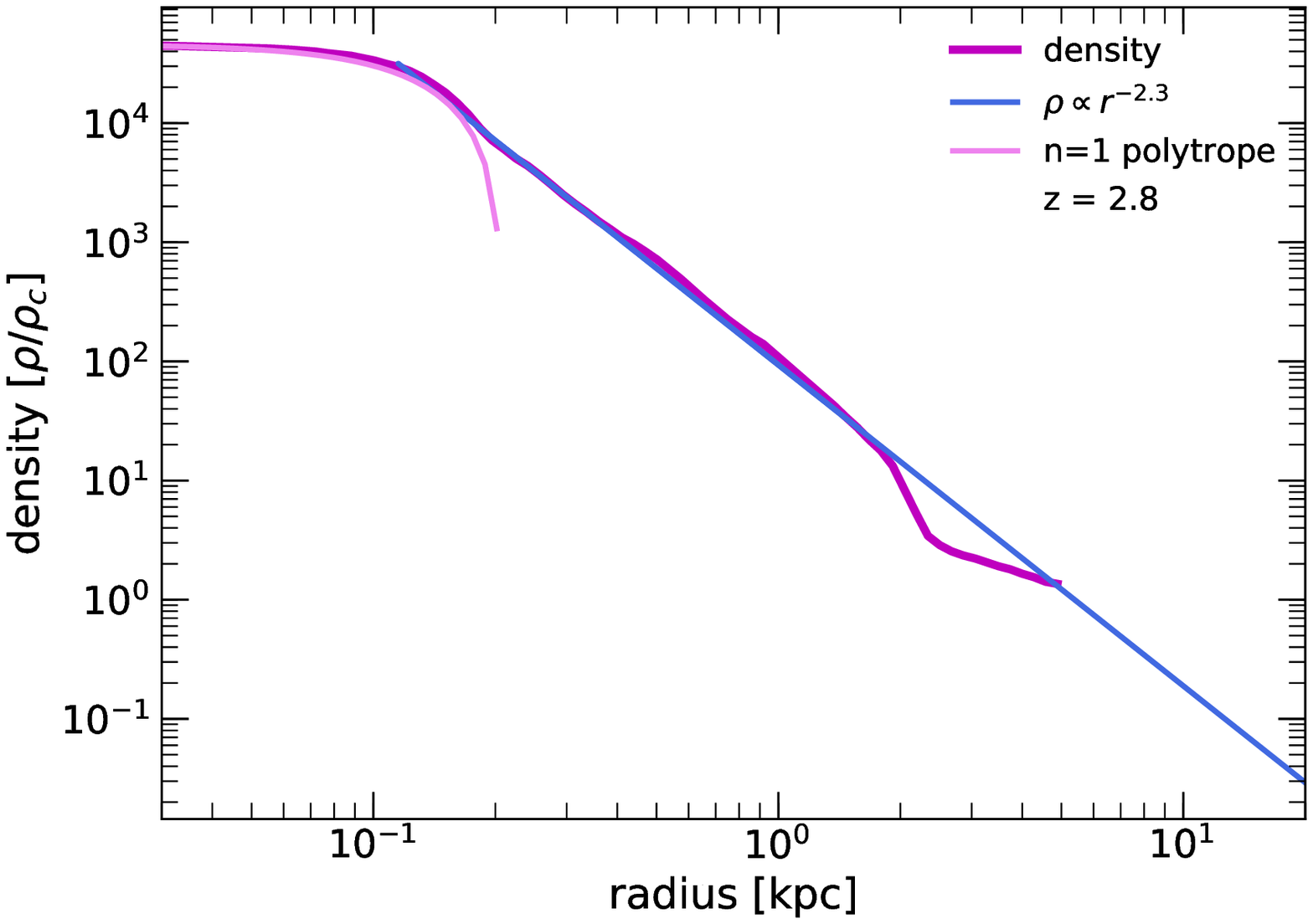}}
	\includegraphics[width=1.0\columnwidth]{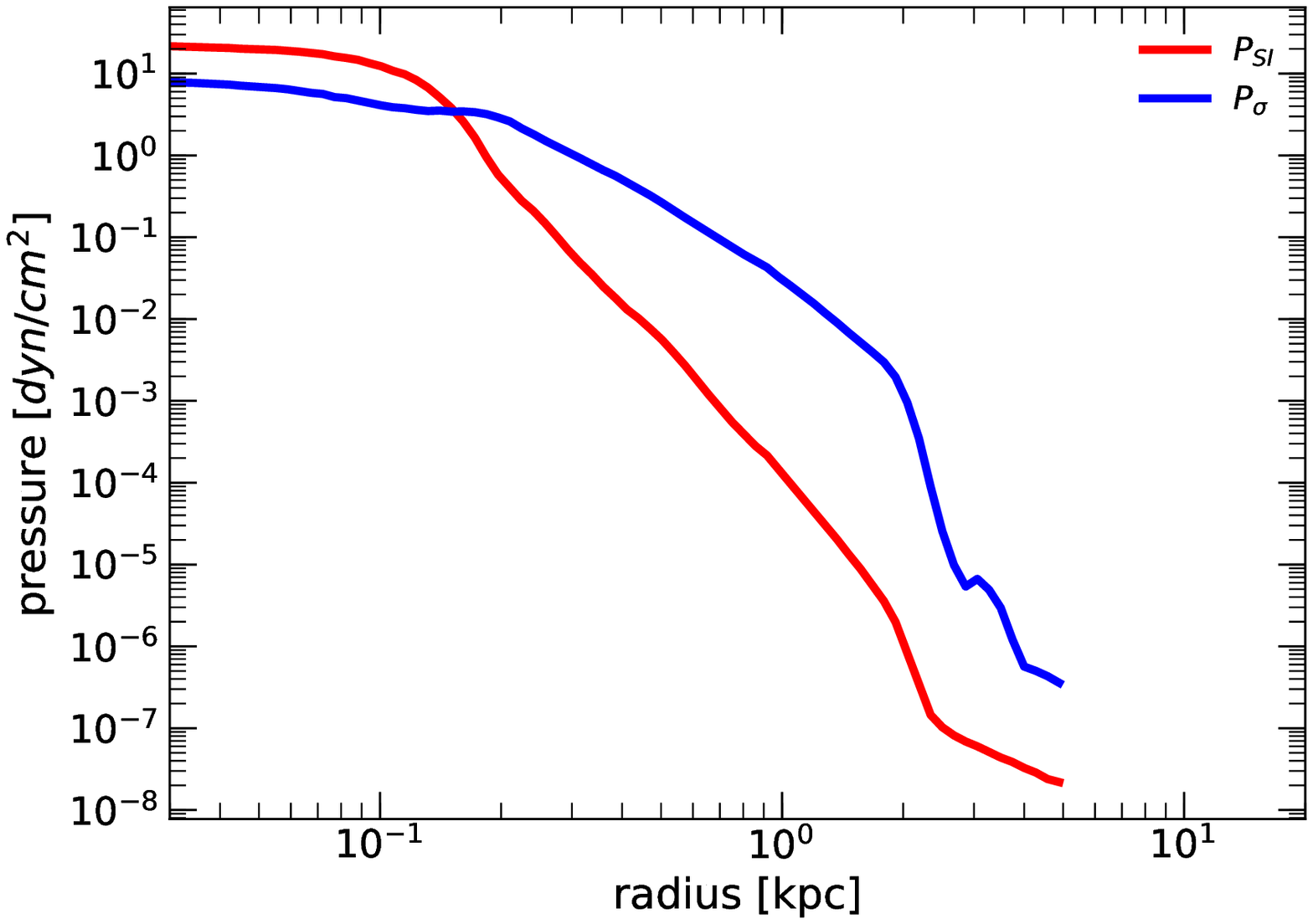}	
	\caption[Formation of a small SFDM Halo Var2]
	{\textbf{Formation of a SFDM-TF halo of mass $10^{9} $~M$_{\odot}$ using Var2}. The top panel displays the density profile at the formation time $z = 2.8$; it shows the same characteristic shape as the previous case of the Milky-Way-sized halo (see Figs.~\ref{fig:formation-var1} and \ref{fig:formation-var2}). The core follows closely a ($n=1$)-polytrope, in terms of shape and size \RTF{}. The shock front separates the halo from the background environment. The bottom panel displays the pressures, \PSI{} (red solid line) and \Psigma{} (blue solid line), respectively. They also show the same features as seen in Figs.~\ref{fig:formation-var1} and \ref{fig:formation-var2}, and are equal at approximately the radius \RTF{}.
	}
	\label{fig:formation-var2s}
\end{figure}

\begin{figure} [!htb]		
	{\includegraphics[width=1.0\columnwidth]{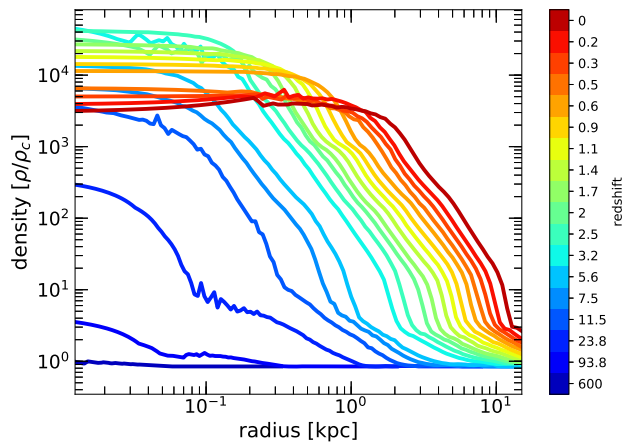}}
	\caption[Evolution of a small SFDM Halo]
	{\textbf{Evolution of a SFDM-TF halo of mass $10^{9} $~M$_{\odot}$ from $z_{ini}=600$ to $z=0$, using Var2 and a 2 Mpc simulation box.} The color-coded solid lines display the evolution of the density profiles. Their shapes are very similar to the halo profiles shown in Fig.~\ref{fig:evol-var2-12}. 
	}
	\label{fig:evol-var2-M09}
\end{figure}

The results for the density and pressure profiles at the halo formation time are shown in Fig.~\ref{fig:formation-var2s}, whereas Fig.~\ref{fig:evol-var2-M09} displays the evolution of the density profiles over the entire simulation run. \par

In order to provide the necessary spatial resolution for the lower-mass halo, we reduced the size of the box to $2$~Mpc and adapted the ICs accordingly.
Although smaller halos form typically at earlier times than larger halos, we initiated our simulation at the same $z_{ini}=600$, as in our previous runs, in order to get comparable results. In fact, this choice of $z_{ini}$ is consistent for both halo masses, according to \citet{Klyping2016_Massrelation}.
Nevertheless, the mass infall onto the lower-mass halo is slower, given the lower mass-accretion rates, hence the formation of the lower-mass halo is delayed compared to its more massive brethren, and its formation redshift is at $z \sim 3$. Likewise, the development of the shock front\footnote{See also Fig.~\ref{fig:initial-profiles}.} also happens later ($z \sim 6$) compared to the higher-mass halo.

Apart from the later formation time, the formation of the halo with mass $10^{9}$~M$_{\odot}$ shows, however, the same characteristic properties as the Milky-Way-size halo with mass $10^{12}$~M$_{\odot}$, see Fig.~\ref{fig:formation-var2s}. The central core profile displays a shape close to a ($n=1$)-polytrope. It is enshrouded by a nearly isothermal envelope, which is separated from the background environment by a shock front; however, the envelope profile is now not as close to the isothermal sphere as for the larger halo before. But as previously, \PSI{} dominates in the core, here by a factor of $\sim 2$ to $\sim 3$ over \Psigma{}, and at the core radius close to \RTF{} both pressure contributions are nearly equal. Again, in the envelope, \Psigma{} dominates over \PSI{}, here by $\sim 2$ orders of magnitude. 
Overall, the central densities and hence the pressures (due to the polytropic equation of states $P \propto \rho^{\gamma}$) are significantly lower for the lower-mass halo than for the Milky-Way-sized halo. The main reason is the later formation time of the lower-mass halo, which forms at a time when the background density has reduced, compared to the epoch when the higher-mass halo formed.

Furthermore, we show the evolution of the density profile in Fig.~\ref{fig:evol-var2-M09}. Again, we recognize a similar behavior as reported before: while the central density of the halo increases over time, at some point the core will expand beyond the polytropic shape that it acquired during the evolution. However, the central pileup of density during the evolution is not as pronounced, as in the previous cases for the higher-mass halo.
Although we reduced the size of the box, we can see the same smooth decrease in density in the core, as indicated in Fig.~\ref{fig:evol-var2-12} before which showed the re-simulation of the $10^{12}$~M$_{\odot}$ halo in a $12$~Mpc box. This similarity in the outcome is explained by the fact that the mass is reduced by $3$ orders of magnitude, whereas the volume of the box has been reduced by a factor of $8$. 
%

\subsection{Evolution of pressure}\label{sec:evolpressure}
\begin{figure} [!h]
	\begin{tabular}{c c } 
		{\includegraphics[width=0.49\columnwidth]{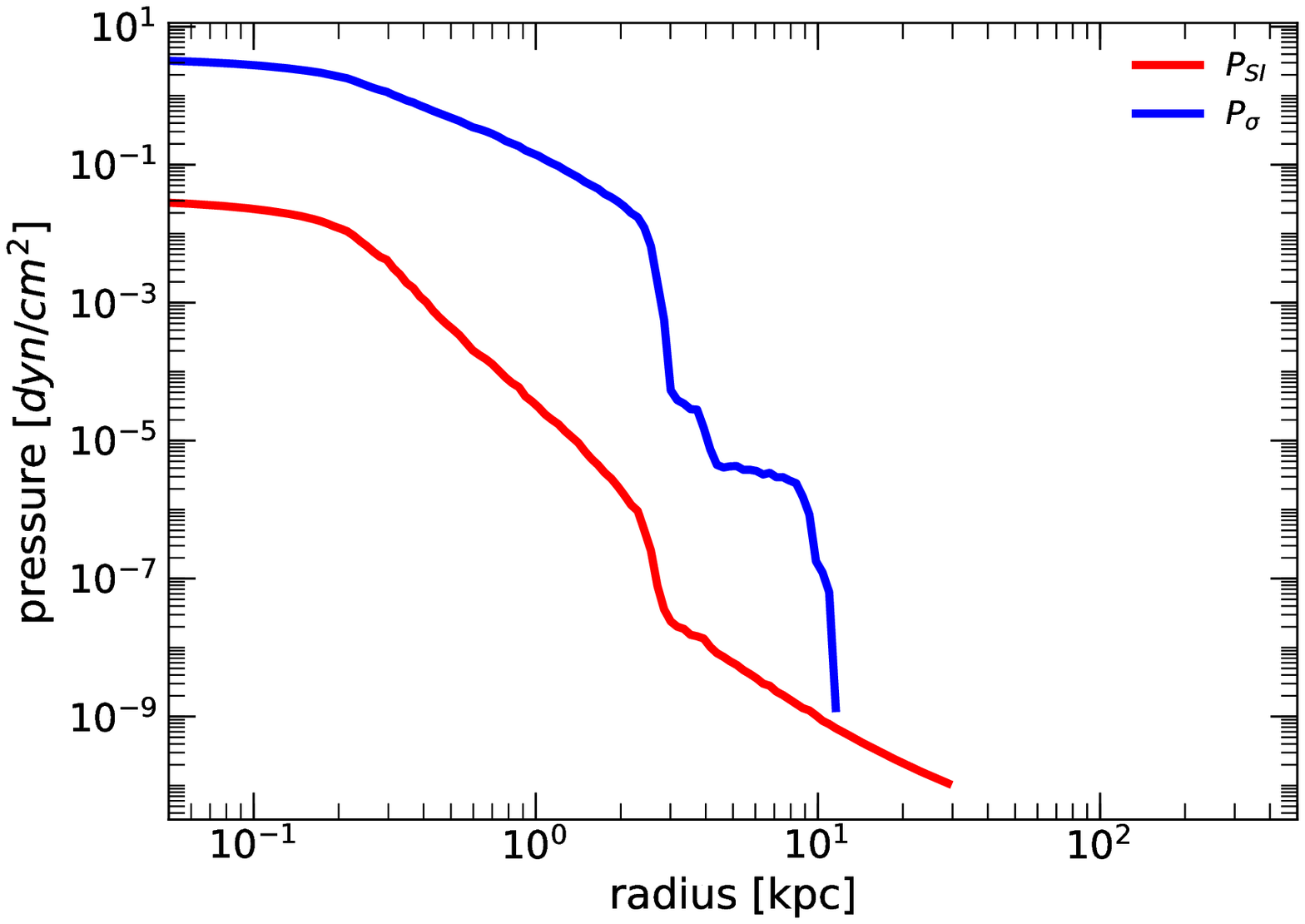}} &
		{\includegraphics[width=0.49\columnwidth]{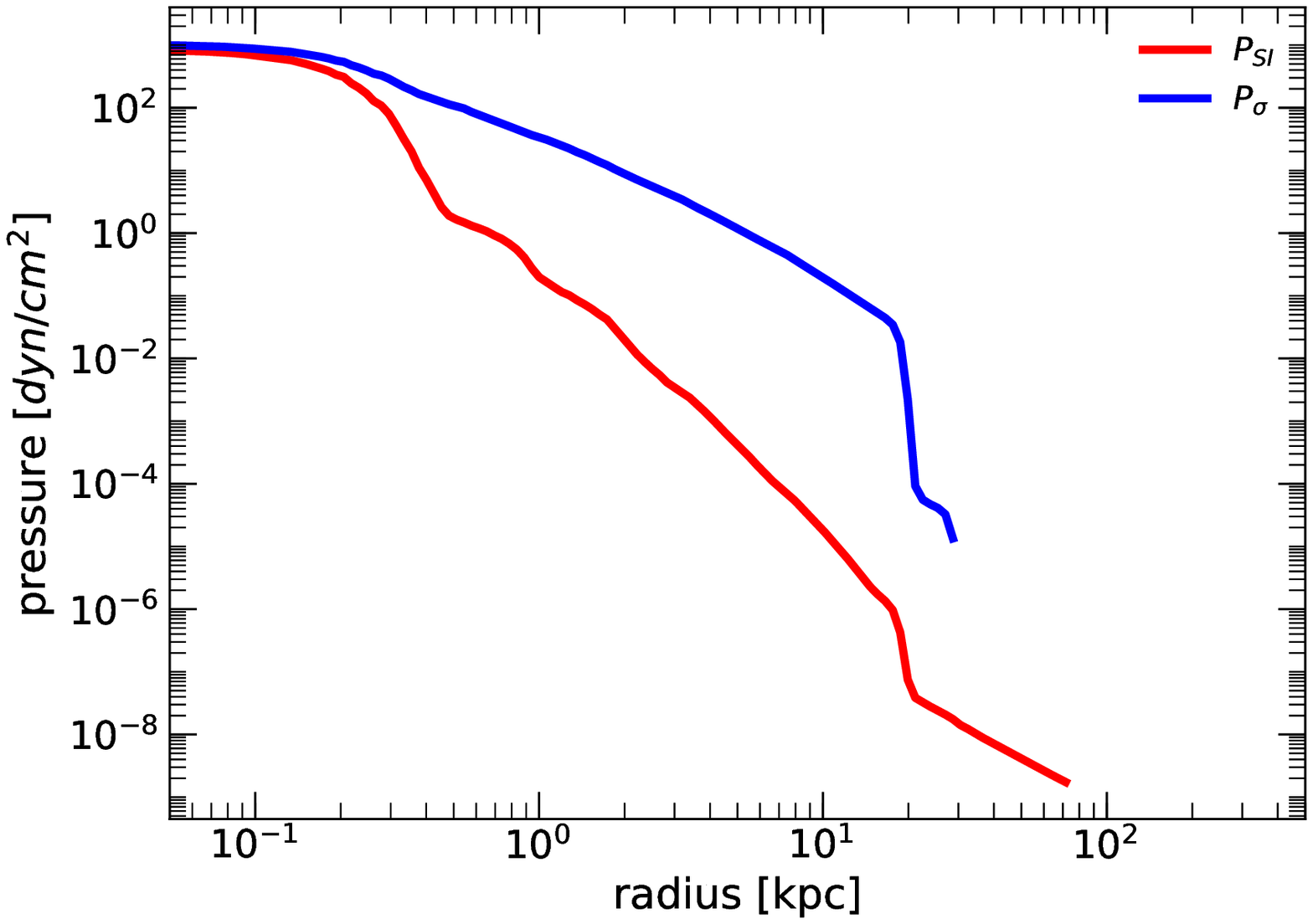}}\\ 
		\vspace{-10ex} \\
		(a) & (b) \\
		\vspace{1ex} \\
		{\includegraphics[width=0.49\columnwidth]{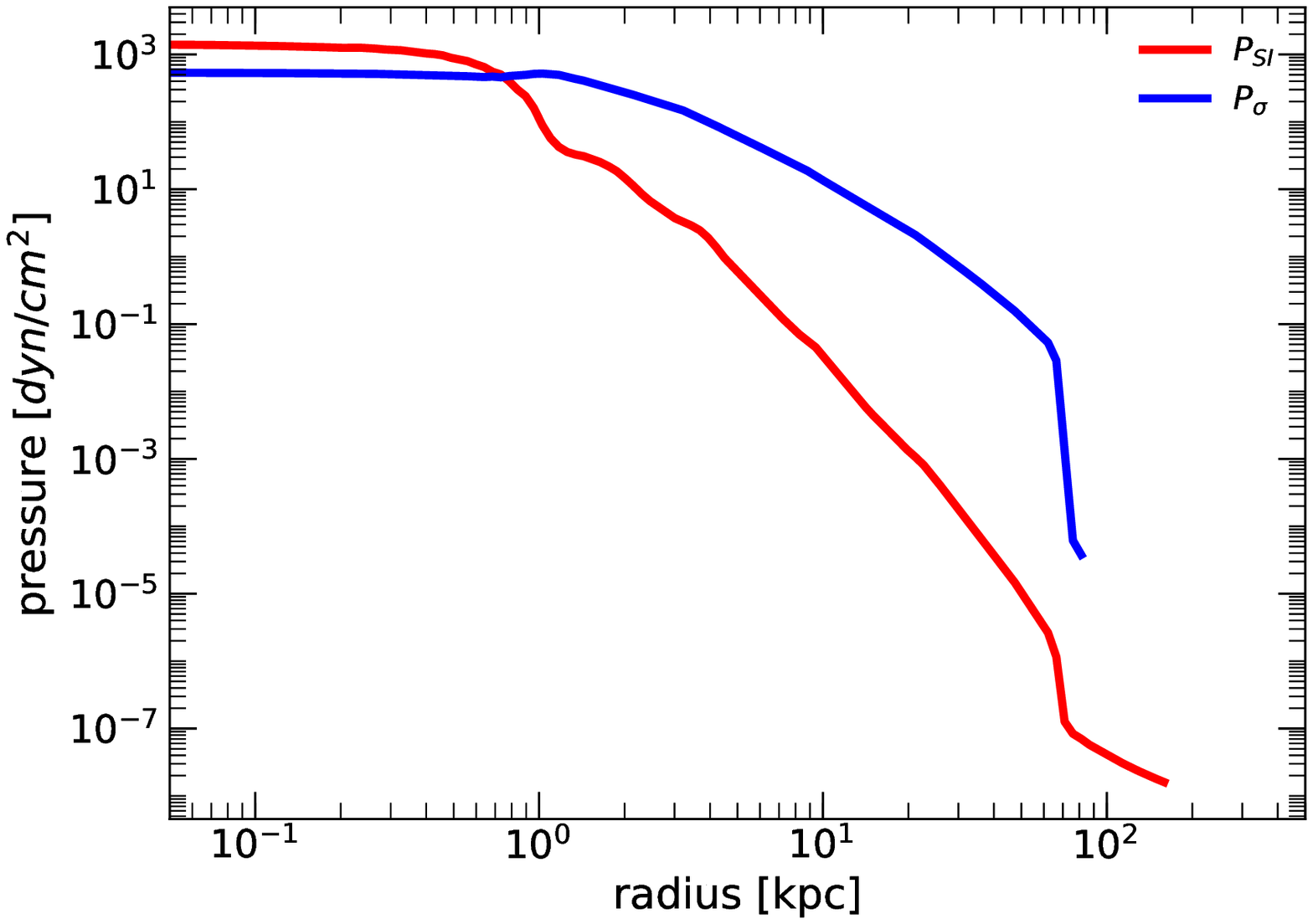}} &
		{\includegraphics[width=0.49\columnwidth]{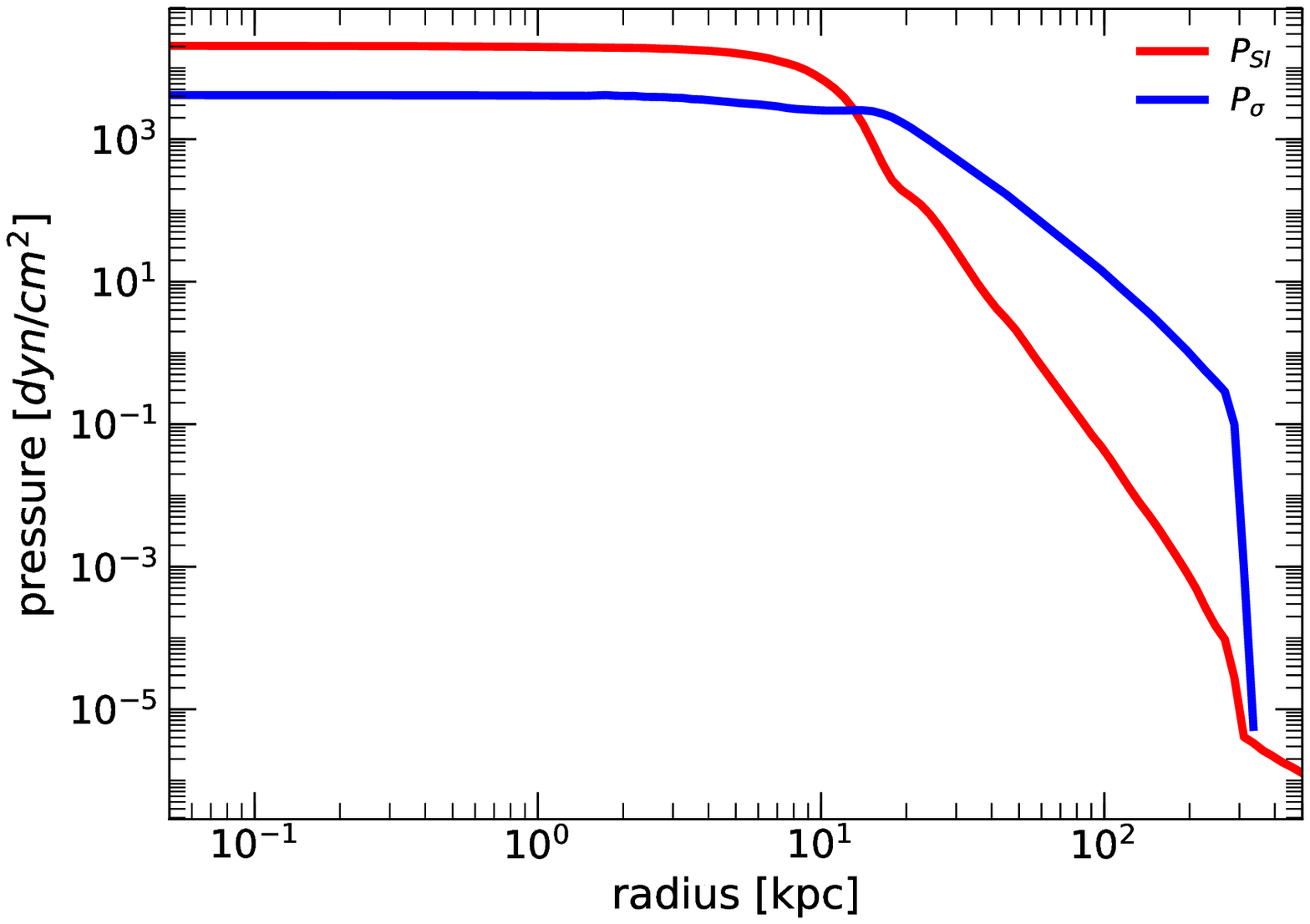}}\\ 
		\vspace{-10ex}\\
		(c) & (d) \\
		\vspace{1ex} \\
	\end{tabular}
	\caption[Evolution of pressure]
	{\textbf{Evolution of pressure.} The panels shown in this figure display the evolution of the pressure contributions of the SFDM-TF halo of Fig.~\ref{fig:formation-var2}, \PSI{} (red solid line) and \Psigma{} (blue solid line). Panels (a),(b),(c) and (d) display their evolution at the respective redshifts of $z \sim 20; 7; 2.7; 0$ from the top left-hand panel to the bottom right-hand panel.  Panel (a) displays the very early stages of the evolution, where \Psigma{} dominates the $(n=1.5)$-polytropic core. Around the formation time ($z \sim 9$; see Fig.~\ref{fig:formation-var2}), \PSI{} dominates in the core. This is followed by a short period of fluctuations in the central region. Panel (b) displays the situation at $z \sim 7$, where \PSI{} has decreased significantly, which coincides with the decrease of density in the core, seen in the bottom panel of Fig.~\ref{fig:evol-var1-2}. At $z \sim 3$ displayed in panel (c), \PSI{} begins to increase again and eventually dominates over \Psigma{}, until $z = 0$, displayed in panel (d). We can see that the importance of \PSI{} is steadily increasing with time. This can be also seen in the shape of the density profile of the bottom panel of Fig.~\ref{fig:evol-var1-2}, where the central density profile evolves into a shape close to a ($n=1$)-polytrope at the formation time, followed by an expansion of the core thereafter.
	}
	\label{fig:evol-var2-pressure}
\end{figure}

%
In light of the reported differences in the pressure runs within halos in \CPaperHWM{}, compared to \CPaperSDR{}, we explore in more detail the evolution of the pressure contributions \PSI{} and \Psigma{}, following the formation of the halo of $10^{12}~M_{\odot}$. We show their profiles in Fig.~\ref{fig:evol-var2-pressure}.\par

In the early evolutionary stages of the SFDM-TF halo, depicted in panel (a), the density in the central region is low and, therefore, the repulsive SI is not significant. This can be seen by the fact that \PSI{} is lower than \Psigma{}, although its polytropic exponent $\gamma$ is higher than that of \Psigma{}. In this stage, the central structure of the halo has a shape close to a ($n=1.5$)-polytrope, stabilized by \Psigma{}. We find thus the same behavior as in the CDM case, reported in the previous section. 
Once the density increases sufficiently, it reaches a point where \Psigma{} cannot balance gravity anymore and the central structure transitions into a core described by a higher polytropic exponent $\gamma$, i.e. \PSI{} becomes dominant in the core. Because of this pressure, the evolution differs now from the CDM regime.
This increasing importance of \PSI{} is accompanied by a short period of fluctuations in the central region, during which the central density decreases and \Psigma{} temporarily reaches the magnitude of \PSI{}. In panel (b) 
we see this decrease of pressure compared to the pressure seen in the right-hand panel of Fig.~\ref{fig:formation-var2}. This is approximately at the time when the density of both the envelope and the core starts to decrease\footnote{We think this effect is merely caused by the fluctuations that arise in the $4$ Mpc box, due to the sudden stop of matter infall. These fluctuations neither appear in the larger $12$ Mpc box, nor in the case of the lower-mass halo simulation.} at redshift $z \sim 7$. During the subsequent evolution, the importance of \PSI{} rises again in the core, and it gets finally dominant again by redshift $z \sim 2.7$ [panel (c)]. This increase in \PSI{} then continues to redshift $z=0$, where we see a clear dominance of \PSI{} in the core [panel (d)]. This evolution of pressures is accompanied by the central density profile establishing a shape close to a ($n=1$)-polytrope at the formation time. We interpret this evolution as a consequence of establishing a new equilibrium of the halo, after the infall of matter has stopped in our simulations, and as a result of the decrease of external pressure due to Hubble expansion; an effect that we have also seen in the larger box of Fig.~\ref{fig:evol-var2-12}.
In contrast, the dominance of \PSI{} is maintained throughout the entire simulation in scenarios with ongoing mass accretion, which is the case of the choice of MAH in \CPaperSDR{}. However, the fact that we find the central dominance of \PSI{} established at late times in the evolution of the halo, especially at $z=0$, which conforms to the results of \CPaperSDR{}, we conclude that their interpretation is correct, after all, namely that \PSI{} dominates over \Psigma{} in the cores of virialized SFDM-TF halos.\par

\subsection{Importance of the time of collapse}

The conclusion of the last subsection, however, stands in contrast to the conclusion of \CPaperHWM{}, who do not find that \PSI{} dominates in their halo cores at $z=0.5$. We think that their halos have not yet established the ($n=1$)-polytropic core, due to the deferred collapse as their simulations start at $z_{ini}=50$. It seems the halos are rather in the stage, where the ($n=1.5$)-polytropic core is still in place, thus prior to the point when the halo core is finally dominated by \PSI{}. This could also explain why the core radii in \CPaperHWM{} are rather large, showing a correlation with the halo mass and the halo central density, in a way seen in their Fig.~5. 
Unfortunately, we cannot quite make a one-to-one comparison between our halo density profiles and the profiles in \CPaperHWM{}, because the segment of the y-axis displayed in their Fig.~5 is too short to include the location of the expected shock front at larger halo-centric radius -- and there ought to be a shock front in their halos, as well! 

\begin{figure} [!htb]		
	{\includegraphics[width=1.0\columnwidth]{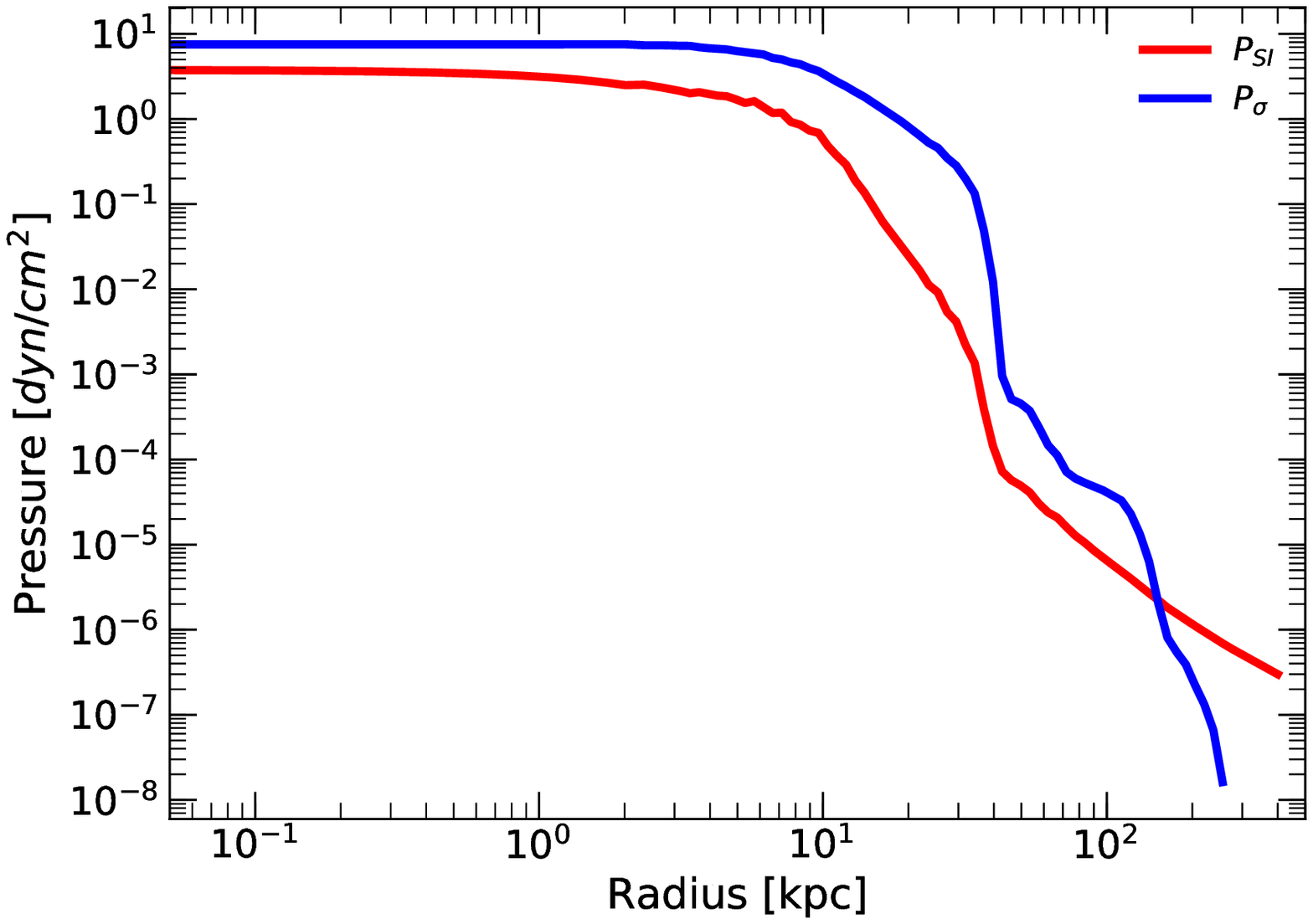}}
	{\includegraphics[width=1.0\columnwidth]{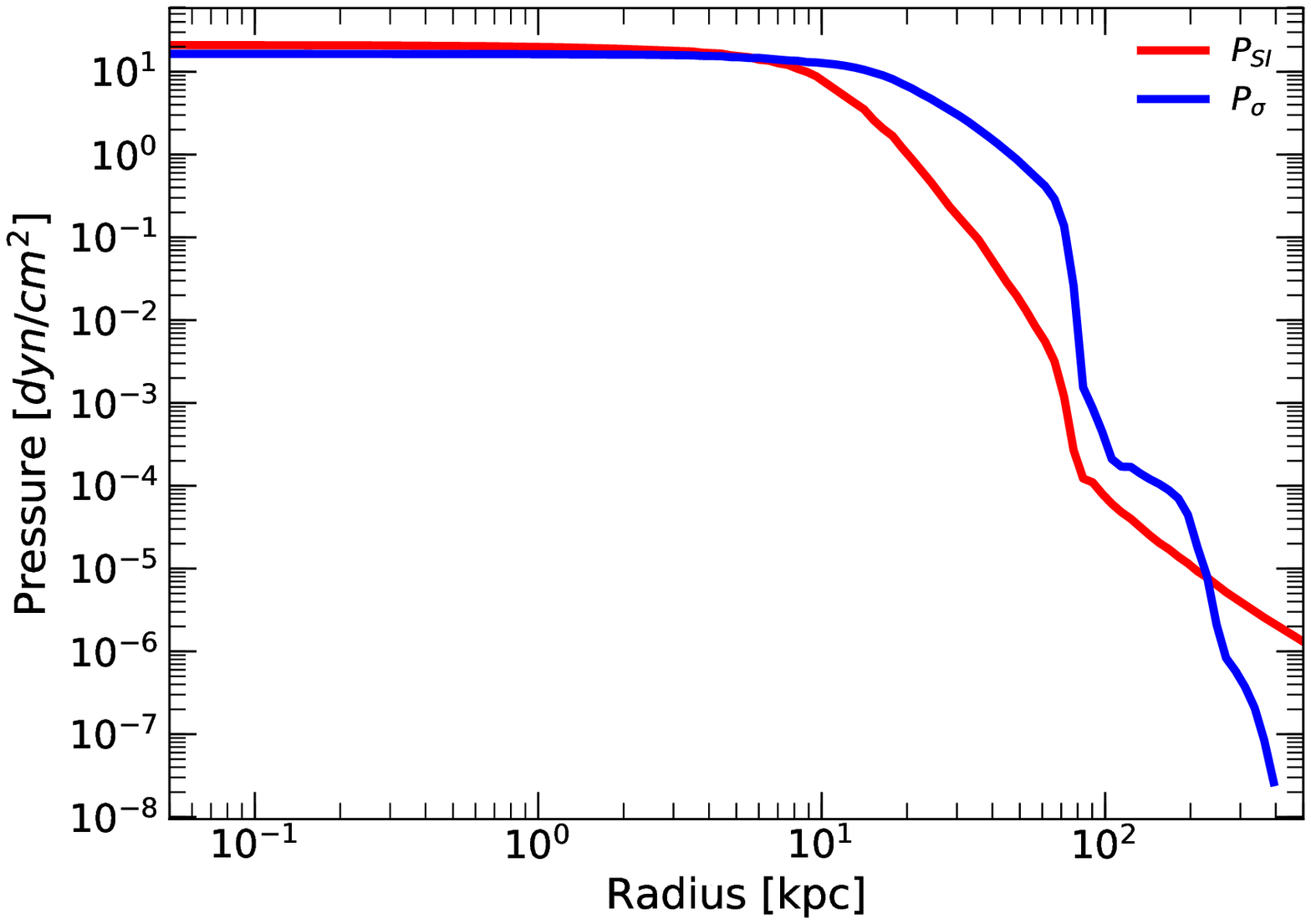}}
	\caption[Evolution SFDM Halo $\bm{z_{ini}=50}$]
	{\textbf{Evolution of a SFDM-TF halo of $10^{12}~M_{\odot}$ using Var2, but initializing the simulation at $\bm{z_{ini}=50}$.} Shown are the pressure profiles, \PSI{} (red solid line) and \Psigma{} (blue solid line), at $z=0.5$ (top panel) and at $z=0.0$ (bottom panel). We can see that at $z=0.5$ the halo is still in the stage of formation, prior to the point where \PSI{} gets dominant when the formation is completed. 
	}
	\label{fig:evol-var2-pressure-z50}
\end{figure}

\begin{figure} [!htb]		
	{\includegraphics[width=1.0\columnwidth]{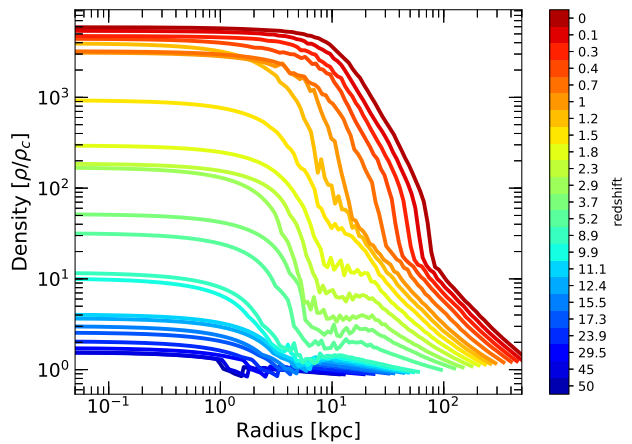}}
	\caption[Evolution SFDM Halo Var2 with $\bm{z_{ini} = 50}$]
	{\textbf{Evolution of a SFDM-TF halo of $10^{12}~M_{\odot}$ using Var2, from $\bm{z_{ini}=50}$ to $\bm{z=0}$}. The color-coded solid lines display the evolution of the density profiles, beginning with the collapse of the initial ($\epsilon=1/6$) profile until $z=0$ in a $4$~Mpc box. We can see a steadily growing density in the center. Around $z \sim 10$, the core begins to build up with almost equal contributions from \PSI{} and \Psigma{}. At $z = 0.4$, the formation of the envelope begins, where \Psigma{} dominates over \PSI{}, and a shock front forms which moves outward. The remaining time until $z=0$ is not sufficient to establish a clear core-envelope structure.
	}
	\label{fig:evol-var2-z50}
\end{figure}

Therefore, we performed a comparison simulation, where we re-simulated the Milky-Way-size halo of $10^{12}~M_{\odot}$ using Var2, but choose the same ``late'' initial redshift of $z_{ini}=50$ for our ICs as in \CPaperHWM{}. We show our results in Figs.~\ref{fig:evol-var2-pressure-z50} and~\ref{fig:evol-var2-z50}.\par
\par

Figure ~\ref{fig:evol-var2-pressure-z50} displays the profiles of the two pressure contributions, \PSI{} and \Psigma{}, at $z=0.5$ (top panel) and at $z=0$ (bottom panel). At $z=0.5$ the halo is still in the stage of formation, see also Fig.~\ref{fig:evol-var2-z50}, prior to the point where \PSI{} becomes dominant in the core, once the formation is complete. Even at $z=0$, the halo had not enough time to completely evolve to its final stage, and \PSI{} is just about to get the dominant contribution to the pressure in the core (compare also to Fig.~\ref{fig:evol-var2-pressure}). In this sense, our result is in agreement with \CPaperHWM{}, where \PSI{} is not dominant in the core at $z=0.5$. However, since our results show that \PSI{} should dominate in the later stages of the evolution, we conclude that the halos in \CPaperHWM{} did not have enough time to evolve, given the late $z_{ini}=50$, and so the transition to \PSI{}-dominated cores has not yet taken place.

\section{SFDM-TF halos from cosmological ICs}\label{sec:cosmosims}

\begin{figure*} [!htb]		
	{\includegraphics[width=1.0\columnwidth]{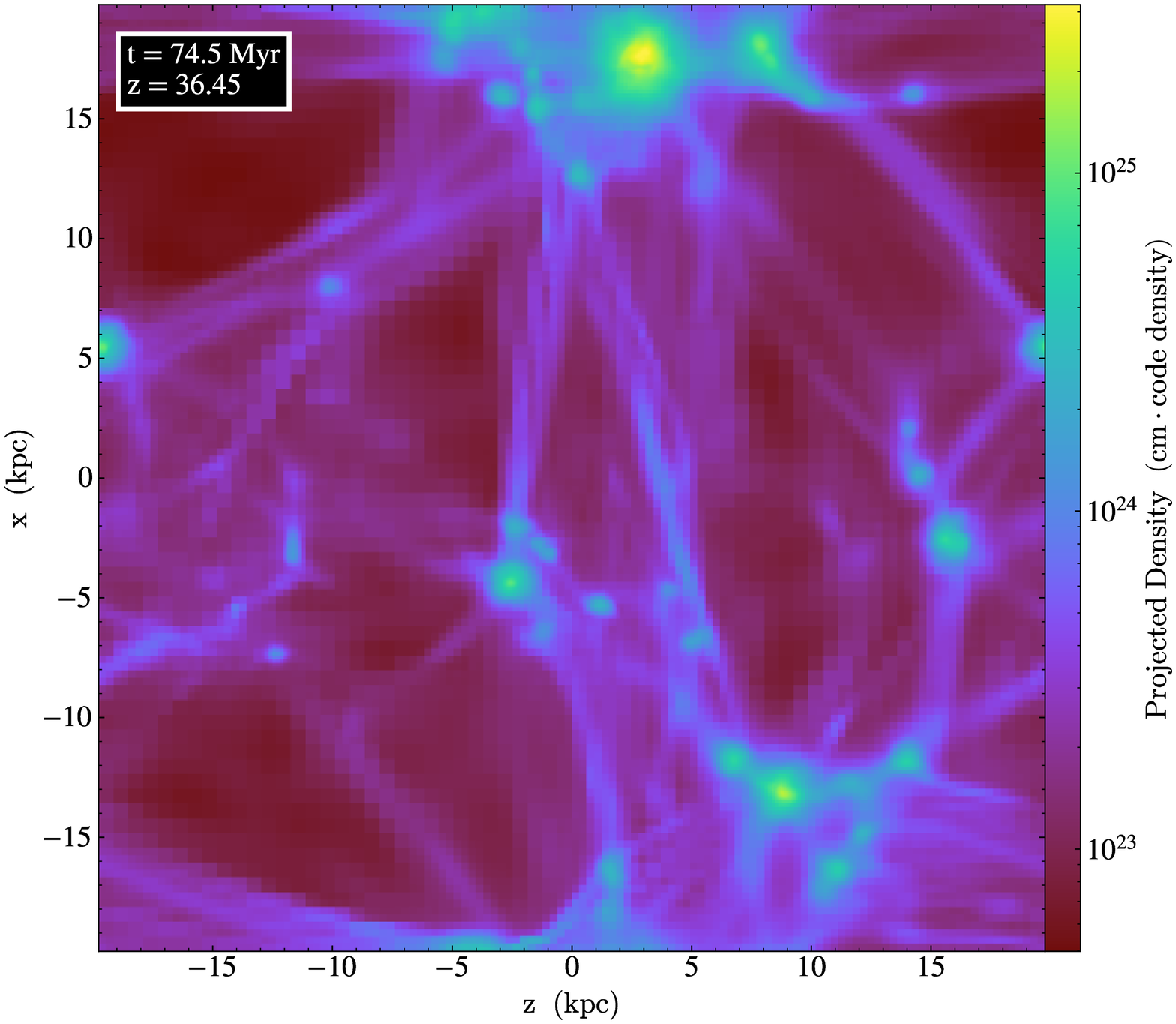}}
	\includegraphics[width=1.0\columnwidth]{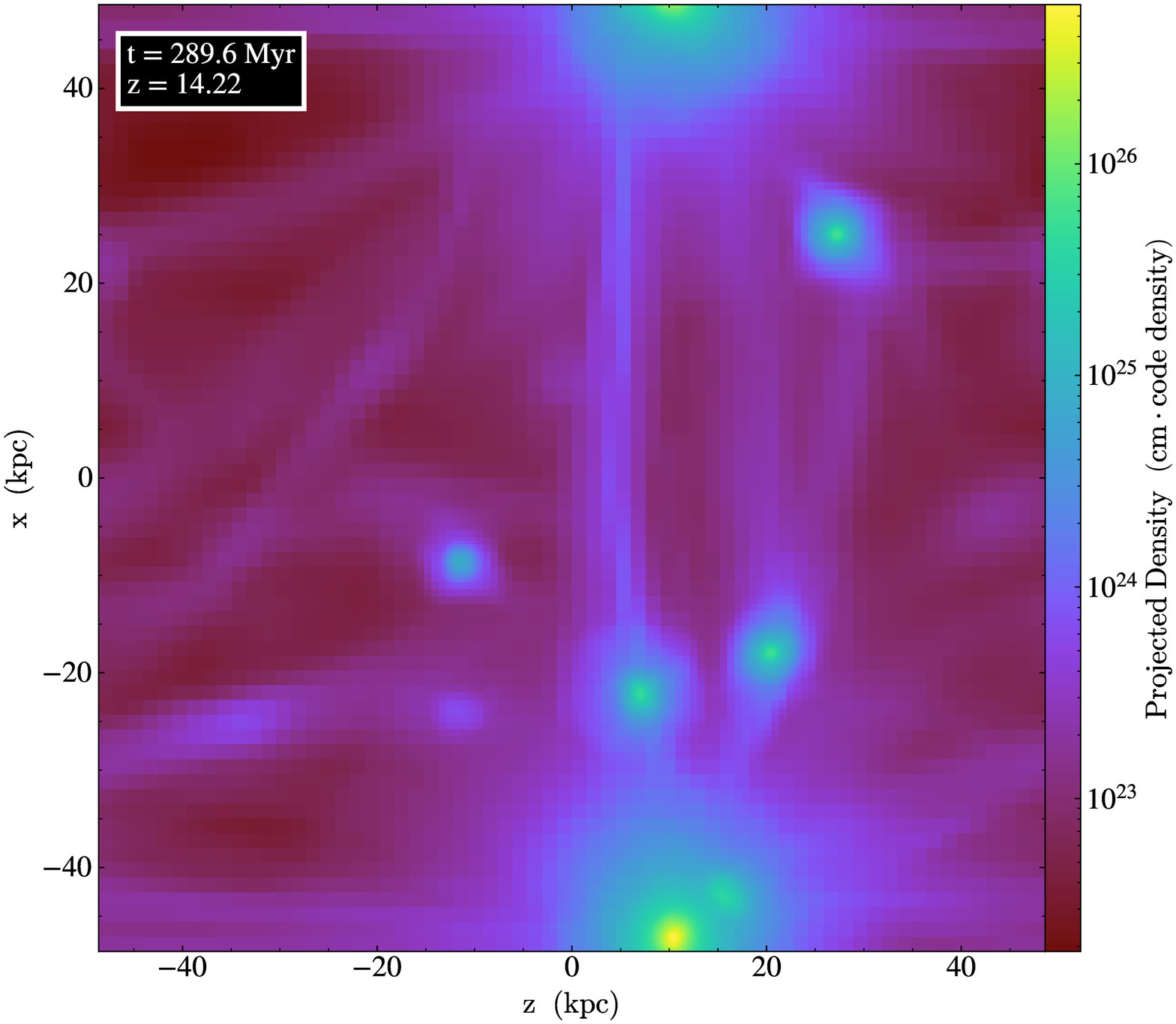}
	\includegraphics[width=1.0\columnwidth]{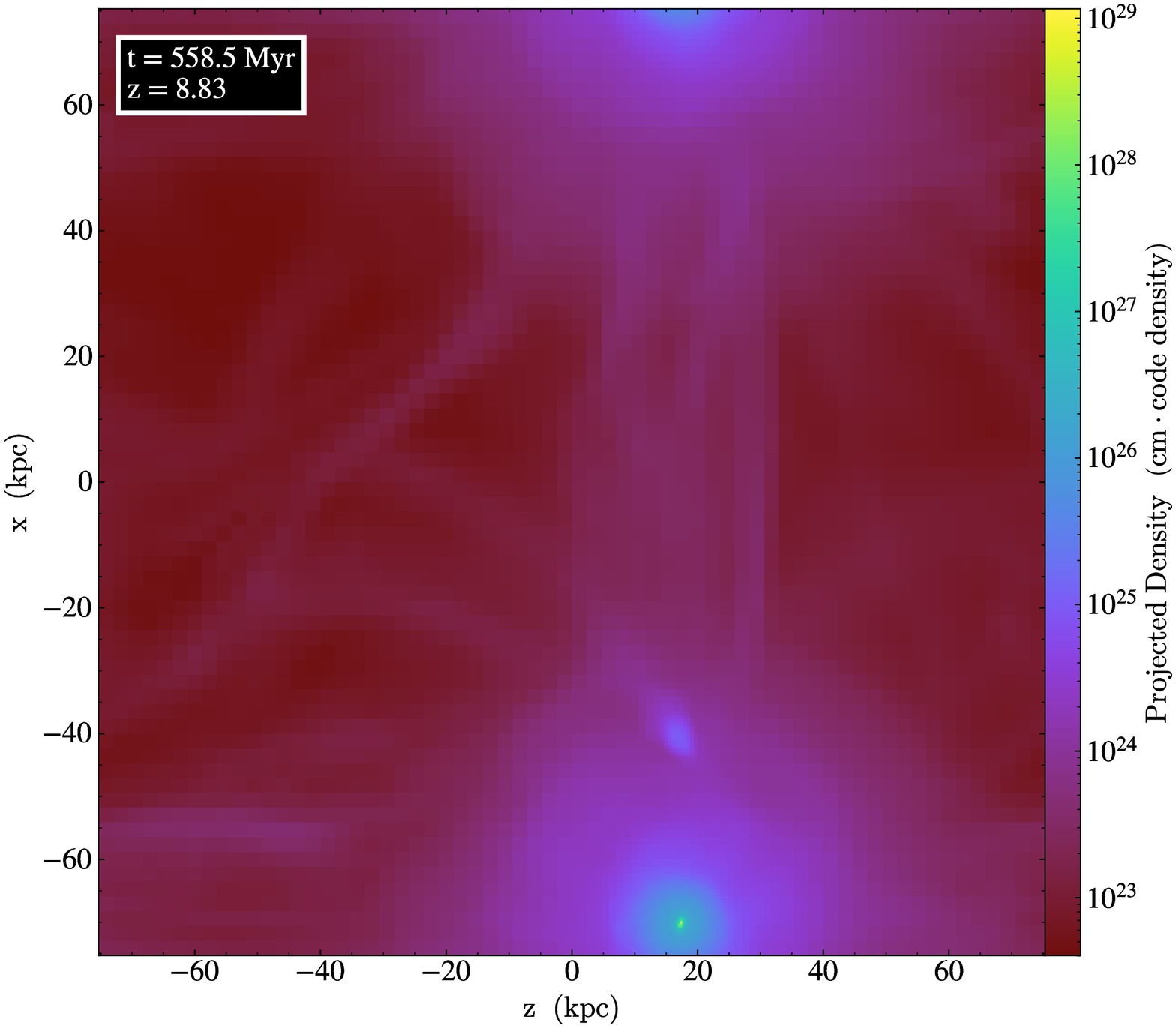}	
	\includegraphics[width=1.0\columnwidth]{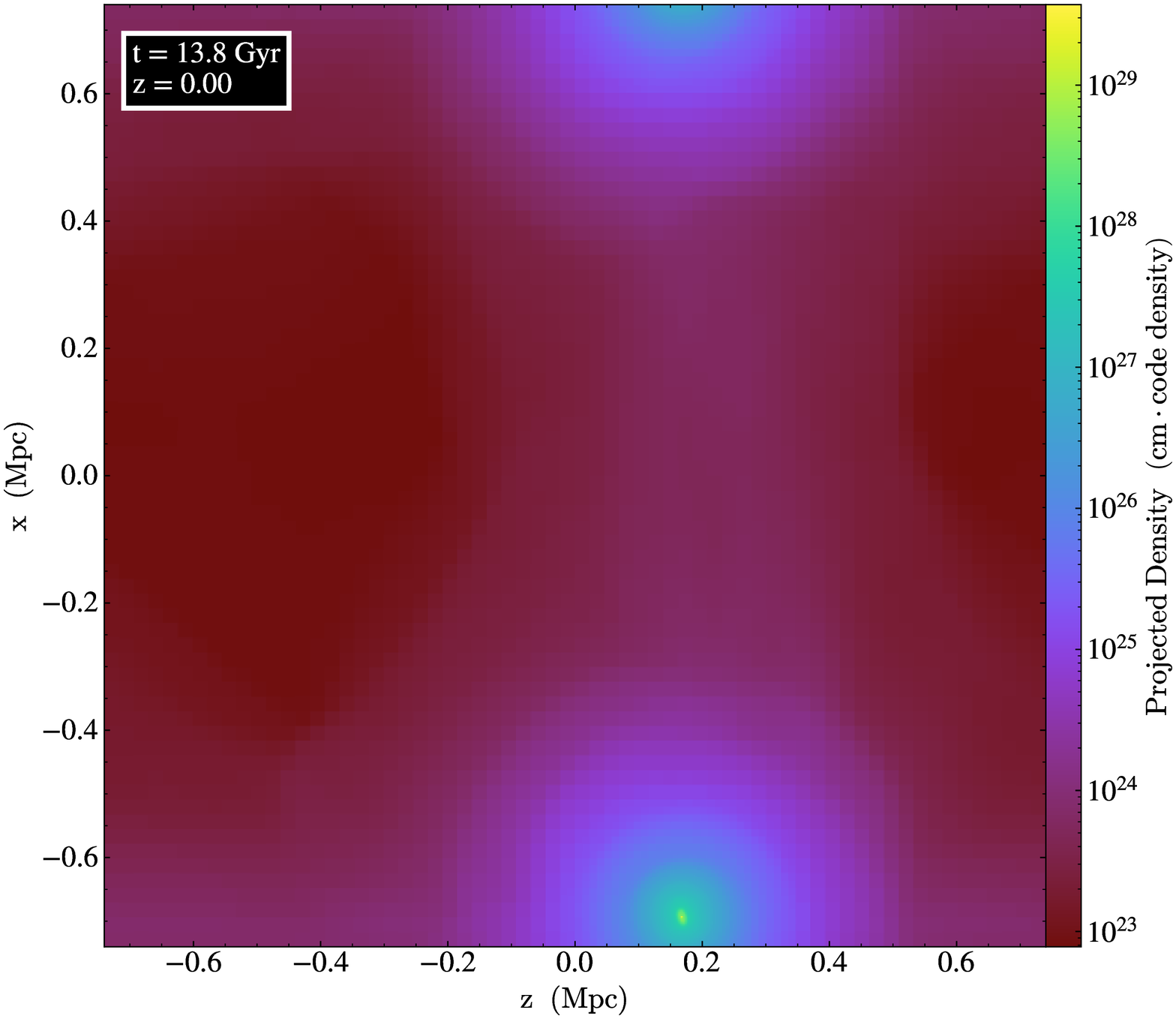}	
	\caption[Mergers of SFDM Halos]
	{\textbf{Mergers of SFDM-TF halos}. SFDM-TF halo formation from ICs generated with MUSIC. In the first stage, a number of halos form from nearly spherical infall of matter. Soon after, the regions between halos show the typical filamentary structure of the cosmic web (see top left-hand panel). The infall of matter proceeds then mainly via the filaments. The top right-hand panel illustrates an evolutionary stage, where some halos have already merged, with a dynamically dominant massive halo that affects the motions of the smaller halos around it. In the bottom left-hand panel, only one massive halo is left, with some matter being expelled again, but which spherically falls back onto the halo consecutively. A small halo starts to merge with the massive halo. Finally, the bottom right-hand panel displays a stage where a single halo remains in a nearly virialized state. (Owing to the periodic BCs, the halo at the bottom of the box coincides with the one at the top.)
	}
	\label{fig:form-MUSIC}
\end{figure*}

In Sec.~\ref{sec:compareNBody}, we compared the fluid approximation for CDM to an N-body simulation of cosmological structure formation. We used MUSIC to create our ICs in order to provide a realistic background environment. On the other hand, in our simulations of the single-halo infall problem in SFDM-TF in the previous section, we placed a single spherical perturbation in the center of a box and investigated the formation and evolution of the halo that results upon the almost spherical infall of matter onto it.  
However, the simulations run by \CPaperHWM{} formed halos out of cosmologically realistic ICs generated with MUSIC. To test this scenario with our own modified version of RAMSES, we add in this section a proof-of-concept simulation, using also ICs generated by MUSIC, in a comoving box of $1$ Mpc. 
Thereby, we also confirm that our implementation yields comparable outcomes to \CPaperHWM{}, in the sense that our fluid implementation also produces a cosmic-web-like structure for SFDM-TF. The difference here is the smaller box size that we use, in order to show a less computationally expensive simulation, than those performed in \CPaperHWM{}.  Our results are shown in Fig.~\ref{fig:form-MUSIC}.  \par
We can see that, in the beginning, the evolution of halos is dominated by nearly spherical infall of matter. Soon after, the typical filamentary structure of the cosmic web forms, where the infall of matter onto the collapsed halos is mostly via the filaments. In the box, a number of halos have formed which enter a stage dominated by mergers. In this stage, a significant amount of matter is expelled from halos, but matter falls back predominantly onto the dominant halo in a nearly spherical manner, i.e. the late stage of halo mass accretion is yet again similar to spherical infall. Hence, we conclude that the results of the single-halo infall calculations are a good approximation to the more complicated reality.
\par
\section{Summary and Conclusions}\label{sec:summary}
%

In this paper, we studied SFDM in the Thomas-Fermi regime, also known as ``SFDM-TF'', as a valuable alternative model to CDM. We were particularly motivated by the cusp-core problem of CDM, which describes the discrepancy between predicted central density cusps in halos vs the more core-like profiles determined from observations of DM-dominated dwarf galaxies.  
The novel dynamics of SFDM as a quantum fluid poses challenges to realistic structure formation simulations in SFDM. Although much progress has been made over the past years, numerical simulations still lack resolution and/or sufficiently long run-times in order to make a fair comparison with observations of galaxies. This problem is exacerbated for SFDM-TF which has basically two important length scales, the TF radius \RTF{} and the de Broglie length \ldb{} where $\mRTF \gg \mldb$. While the characteristic SI pressure of SFDM-TF is at play on scales of \RTF{}, the genuine quantum dynamics acts at scales down to $\lambda_{deB}$. Therefore, it is useful to find computational approaches which average over the dynamics on these very small scales, without neglecting its physical effects on macroscopic scales.  
Appropriate fluid approximations suggest themselves as a way to accomplish this, and previous literature has come up with solutions. \citet{Dawoodbhoy2021} and \CPaperSDR{} have devised a fluid approximation for SFDM-TF and performed singe-halo infall calculations using a 1D Lagrangian fluid code, in order to investigate the formation and evolution of such halos. The results agreed with earlier analytic expectations, in that such halos form a core-envelope structure, with a central core close to a ($n=1$)-polytrope with radius \RTF{}, surrounded by a CDM-like (i.e. NFW-like) halo envelope. Within the cores, the SI pressure, $\mPSI \propto R_\text{{TF}}^2 \rho^2$, dominated, whereas the envelope was stabilized by an ``effective'' velocity-dispersion pressure due to the small-scale quantum dynamics, $P_{\sigma} \propto \rho^{5/3}$. 

Subsequently, \CPaperHWM{} have performed 3D structure formation simulations of SFDM-TF with the cosmological code RAMSES, using a somewhat different fluid approximation in 3D. They confirmed many of the previous results of \CPaperSDR{}, but also reported discrepancies, notably that the cores of their halos would not be dominated by \PSI{}. Instead, the two pressure contributions, \PSI{} and \Psigma{}, were roughly equal in the core, whereas \Psigma{} dominated in the envelope. They attributed this difference to the possibility of mixing and energy exchange between inner and outer halo parts in 3D, a phenomenon that the previous 1D simulations could not model.

Our aim in this paper was to settle this issue by performing more dedicated 3D cosmological simulations and to compare our findings with those of \CPaperSDR{} and \CPaperHWM{}. Like in these papers, we neglect baryons in our simulations.

In order to make a fair comparison, we also used RAMSES and modified it such that we implemented both versions of fluid approximations, the 3D version of \CPaperSDR{} (we call it Var1) and the version by \CPaperHWM{} (we call it Var2). In contrast to Var1, the fluid equations of Var2 contain the source terms due to self-gravity of the SFDM fluid. Also, Var1 includes the SI pressure \PSI{} in the momentum equation, but not in the energy equation. On the other hand, in Var2 the pressure \PSI{} appears in both momentum and energy equation.  

We performed proof-of-concept structure formation simulations in a box of comoving $1$ Mpc, using initial conditions from MUSIC, in order to convince ourselves that our implementation worked correctly. Our results agreed with those of \CPaperHWM{}, in that we also see the formation of a cosmic web in SFDM-TF. 
In the process of this test, we also studied the CDM regime, setting $\mPSI = 0$ which reduces the equations to those of the ``CDM fluid approximation''. We compared our result with a standard N-body structure formation simulation of CDM, both calculated with RAMSES, and found very good overall agreement. 

However, since our focus was the infall problem of single-halo formation, we did tests of halo collapse, using several initial density profiles as input. We found no significant impact onto the formed halo, and decided to generate the initial conditions for our forthcoming simulations with the ($\epsilon=1/6$) profile of a scale-free, spherical perturbation; also in order to compare to the results of \CPaperSDR{}. Apart from some test cases, we always chose an initial redshift of $z_{ini} = 600$. \par

Most results in this paper concern halo hosts of Milky-Way mass of $10^{12}~M_{\odot}$, but we also simulated a $10^{9}~M_{\odot}$ halo which is a typical host mass for ultra-faint galaxies. In all simulations of SFDM-TF, we used the same choice of $R_\text{{TF}} = 110$~pc.  While this core size seems too small to resolve the cusp-core problem (or other ``small-scale structure problems'' for that matter), it is motivated by the findings of \CPaperSDR{}, \citet{Foidl2022} and \citet{Hartman2022}. It was shown in these papers that linear structure growth constrains the allowed parameter space of SFDM-TF more severely than previously thought. As a result, associated \RTF{} of kpc size and beyond are highly disfavored, because such a choice would suppress structure formation on too large scales, in conflict with observations.

Now, in order to test our code modifications, and for the sake of our understanding, we first performed 3D single-halo infall simulations in the CDM regime, where $\mPSI = 0 = \mRTF$, using a comoving $4$~Mpc cosmological box with periodic boundary conditions. 
In the CDM regime, the only pressure contribution that helps CDM to oppose gravitational collapse is the pressure due to the random motion of collisionless CDM particles. This pressure is formally the same as \Psigma{} above, never mind that it derives from a different phase-space averaging procedure. 

While our results for CDM halo formation and evolution conform very well with previous works that investigated the CDM fluid approximation, we found some interesting details, as follows.  We were able to follow the evolution of the halo density and pressure profiles over time, including the development of the expected shock front in CDM halos. In the fluid picture, the location of this shock front at $z=0$ is a good proxy for the virial radius of the halo, because it separates the inner evolved part from the outer low-density background. In fact, at the present it is located at $\sim 300$~kpc for the CDM halo of $10^{12}~M_{\odot}$, in good agreement with estimates of the size of the Milky Way.  

During its evolution, the central density in the CDM halo rises, but in its early stage it follows a ($n=1.5$)-polytropic core. By the formation time of the halo, this core has transitioned into a steep central cusp whose slope is almost the same as the outer density slope, which is close to $\rho \propto r^{-3}$ as expected from NFW. Finally, the central profile makes another transition to a ``shallower cusp'', very close to the NFW behavior of $\rho \propto r^{-1}$, which remains in place until $z=0$. These transitions have to do with the respective search for equilibrium between gravity and pressure.

Next, we performed 3D simulations of the formation of a single SFDM-TF halo, again using a $4$~Mpc cosmological box with periodic boundary conditions. The first simulation was configured to use the fluid approximation of \CPaperSDR{}, i.e. Var1, where we could confirm their results concerning the core-envelope structure as well as the runs of the pressures \PSI{} and \Psigma{} within the halo, at the formation time of the SFDM-TF halo, i.e. at the same time snapshot as in \CPaperSDR{}. However, since we calculated the entire evolution down to $z=0$, we encounter a new effect in our simulation, namely the polytropic core and the envelope both start to expand at some point. In the second simulation, we used the fluid approximation of \CPaperHWM{}, i.e. Var2. We found that our results in this case \textit{also} agreed with those of \CPaperSDR{} at the formation time of the halo, although the same phenomenon of ``core/envelope expansion'' occurs as in the previous case. In any case, we found no significant discrepancy in the results between Var1 and Var2, except for the fact that Var2 exhibits better resolution characteristics. As such, a choice of Var2 is preferred over Var1. Furthermore, for both Var1 and Var2, we found that the evolution of SFDM-TF halos goes through a two-stage process, where the central profile at early times is close to a ($n=1.5$)-polytrope, which only later transitions to a ($n=1$)-polytrope when $P_\text{SI}$ finally dominates in the halo core. On the other hand, the halo envelopes show the same characteristics as CDM halos. 

Before we discuss the comparison between \CPaperSDR{} and \CPaperHWM{} in more detail, let us elaborate on the ``core expansion'' that we see in our simulations, as follows.
As the envelope of the SFDM-TF halo is almost isothermal, the size of the halo is determined by the external pressure, which has two contributions: the pressure of the infalling matter and the density of the background universe. For the simulations in a $4$~Mpc box, we saw a rapid decrease in the density of the core at $z\sim 7$ and a consecutive expansion of the envelope starting at $z \sim 4$.
Therefore, we performed a further simulation, using a $12$~Mpc box with accordingly increased spatial resolution. In this case, we found that the transition from the polytropic core to the expanded core is much smoother in the larger box than in the smaller box. We explain this behavior with the smooth decrease in the infall of matter onto the halo caused by the expansion of the background universe, compared to the immediate stop of the infall due to the lack of matter in the smaller box.

Now, with respect to the discrepancies in the results of \CPaperSDR{} and \CPaperHWM{}, concerning the pressure runs, we explain them as a result of systematic differences in the setup of the simulations. The most significant difference is that \CPaperSDR{} analyzes the structure of SFDM-TF halos at the formation time of the halo, and they start their simulations at $z_{ini} = z_{eq} \sim 3000$, whereas \CPaperHWM{} start their simulations at $z_{ini}=50$ and do the analysis of halo structure at a redshift of $z=0.5$, i.e. later than the formation time. So, the main issue is the fact that in \CPaperHWM{} the collapse of the halo takes place too late, thus it had not enough time to form and virialize. Our assessment is confirmed by a test simulation that we performed, where we also used a late $z_{ini}=50$, in order to show the incomplete evolution of the halo.

A further difference is as follows. \CPaperSDR{} investigate the formation of SFDM halos by spherical infall in a cosmological environment that maintains a predefined MAH according to \citet{Wechsler2002}, whereas \CPaperHWM{} use MUSIC in their simulations, where a number of several halos collapse and form over time. The MAH for each individual halo in the simulations by \CPaperHWM{} is not maintained throughout the entire simulation for the following reason: after the first stage of halo collapse with spherical infall, filaments form rapidly and the infall onto halos goes via these filaments; the remaining 3D space in the vicinity of the halos is diluted dramatically; but the shock waves following mergers transport a significant part of the material into the outskirts of the halo; at the final stages, the material falls back onto the halo in a nearly spherical manner.\par

To re-iterate: in our 3D simulations of the formation and evolution of SFDM-TF halos, we have seen that both fluid approximations in \CPaperSDR{} vs \CPaperHWM{} reveal results with no significant differences. With regard to the evolution of SFDM-TF halos we identified two important processes: the MAH that is determined by the amount of matter in the vicinity of the halo and the expansion of the background universe. The amount of matter determines how ``fast'' the transition from the polytropic core to the expanded core evolves. This could imply that the core size of a halo might be impacted by its environment, as it determines the MAH onto the halo.
This is an interesting finding, because it implies that core size is not necessarily determined only by the parameters of the DM model. Even without the inclusion of baryons and feedback effects, our simulations reveal that an initial primordial DM core of sub-kpc size, $\sim 100$~pc -- demanded by power spectrum constraints from linear theory  -- can evolve into a larger core of $\gtrsim 1$~kpc, after all, during DM-only halo evolution.

In fact, in \CPaperHWM{} a similar, but less pronounced, effect is mentioned and seen in their Figs.~3 and 5, where they display halos with masses between $10^{7}$ and $10^{11}$~M$_{\odot}$. Their halo cores extend to a radius beyond $\mRTF = 1$~kpc, where more massive halos display a more pronounced effect than the less massive halos. This effect supports the aforementioned argument, that their halos are in an early stage of formation, where \Psigma{} is still dominant and responsible for the central density that follows closely a ($n=1.5$)-polytrope, which extends beyond the ($n=1$)-polytrope. These findings agree with ours, as more massive halos are impacted stronger from the lack of outer pressure onto the terminating shock front of the halos's envelope. Again, given that \CPaperHWM{} started their simulations at $z_{ini}=50$ and stopped them at $z=0.5$ may explain why the effect is less pronounced compared to our simulations that started at $z_{ini}=600$ and ended at $z=0$.

Of course, future simulations should include baryons and should be preferably performed in larger cosmological volumes with realistic initial conditions. The great computational challenge of such simulations has prevented faster progress in this field so far. We expect that baryons will modify previous and our own results here, in that their presence will lead to multiple feedback processes that affect central halo densities and the size of cores. 
While previous constraints from linear theory have shown that cores in SFDM-TF should be of sub-kpc size, our results here have shown that primordial cores can expand upon halo evolution, independent of baryonic feedback. 

The amount of expansion depends on the mass of the halo, and the latter is predefined in our ICs (i.e. in the chosen mass of the initial overdensity), which is a good approximation for low-mass halos, but less so for massive halos of Milky-Way size which are believed to acquire much of their mass by multiple mergers during their lifetime. 
Now, using a primordial core radius of $R_\text{{TF}} = 110$~pc, we find that the core of the $10^9~M_{\odot}$ low-mass halo expands up to a radius of $\sim 2$~kpc by $z=0$, which is a value that is still in accordance with observations of ultra-faint galaxies; see e.g. \citet{Bernal2018,MartinezMedina2014} for a comparison. On the other hand, for the $10^{12}~M_{\odot}$ Milky-Way-like halo, that same primordial core expands up to a radius of $\sim 20$~kpc by $z=0$, and this value also depends more strongly on the box size of the simulation, given its impact onto the mass accretion history. Such large cores might be in conflict with observations of galaxies, which are dominated by baryons in the central regions such as the Milky Way, M31 or similar galaxies. However, there are massive low-surface-brightness spiral galaxies which do have large cores of $\gtrsim 10$~kpc, and these cores might be possibly explained by expanding DM-only primordial cores. In any case, the astrophysics of baryons impacts the appearance of galaxies on various scales. Hence, future work is desired, especially with respect to the inclusion of baryons, in order to determine whether SFDM in general, or SFDM-TF in particular, may yet resolve the ``cusp-core problem'', after all.


\begin{acknowledgments}
	The authors are grateful for helpful discussions with Paul Shapiro, Taha Dawoodbhoy and Stian Hartman.
	T.R.-D. acknowledges the support by the Austrian Science Fund FWF through a FWF Elise Richter Fellowship Grant No. V656-N28, and through the FWF Single-Investigator Grant (FWF-Einzelprojekt) No. P36331-N. 
\end{acknowledgments}

\end{document}